# On the Use of Walsh Domain Equalizer for Performance Enhancement of MIMO-OFDM Communication Systems

**Khaled Ramadan**

Department of Communications and Computer Engineering,

The Higher Institute of Engineering at Al-Shorouk City, Cairo, Egypt

Email: ramadank637@gmail.com

*Abstract*— **The purpose of this article is to investigate the viability of Multi-Carrier Modulation (MCM) systems based on the Fast Walsh Hadamard Transform (FWHT). In addition, a nonlinear Joint Low-Complexity Optimized Zero Forcing Successive Interference Cancellation (JLCOZF-SIC) equalizer is proposed. To that end, general equations for the number of flops of the proposed equalizer and various other equalizers are given. This article discusses the use of Banded Matrix Approximation (BMA) as a technique for reducing complexity. The proposed equalizer uses BMA to accomplish both equalization and co-Carrier Frequency Offset (co-CFO) corrections. In addition, three cases involving the proposed equalizer were investigated. In the first case, diagonal compensation is used. In the second case, BMA compensation is used. In the third case, complete matrix compensation is used. In the presence of frequency offset, noise, and frequency-selective Rayleigh fading environments, analysis and simulation results show that the OFDM-FWHT system with the proposed equalizer outperforms the conventional OFDM system with various linear and nonlinear equalizers.**

**Nomenclature**

| | |
|---|---|
| $\mathrm{X}^{i} \in \mathbb{R}^{N\times 1}$ , $i \in \{1, 2, \dots\}$ | The $i^{\text{th}}$ polar Non-Return-to-Zero (NRZ) vector |
| $\mathrm{x}^{i} \in \mathbb{C}^{N\times 1}$, $i \in \{1, 2, \dots\}$ | The modulated vector via IFFT associated to $i^{\text{th}}$ stream |
| $\mathrm{x}^{i} \in \mathbb{R}^{N\times 1}$, $i \in \{1, 2, \dots\}$ | The modulated vector associated to $i^{\text{th}}$ stream in the transform domain |
| $\mathrm{x}^{i}_{\mathrm{CP}} \in \mathbb{C}^{(N+N_{\mathrm{CP}})\times 1}$, $i \in \{1, 2, \dots\}$ | The modulated vector with CP associated to $i^{\text{th}}$ stream |
| $\mathrm{X}_{i} \in \mathbb{C}^{N\times 1}$, $i \in \{1, 2, \dots\}$ | The $i^{\text{th}}$ polar NRZ vector |
| $\mathrm{y}^{j} \in \mathbb{C}^{(N+N_{\mathrm{CP}})\times 1}$ | The received modulated vector at the $j^{\text{th}}$ receiving antenna |
| $\bar{\mathrm{y}}^{j} \in \mathbb{C}^{N\times 1}$ | The output of the CP removal block at the $j^{\text{th}}$ receiving antenna |
| $\tilde{y}^j \in \mathbb{C}^{N\times 1}$ | The $j^{\text{th}}$ pre-equalized data stream vector |
| $\bar{\bar{y}}^j \in \mathbb{C}^{N\times 1}$ | The output of the transform at the $j^{\text{th}}$ receiving antenna |
| $\hat{y}^j \in \mathbb{C}^{N\times 1}$ | The estimated polar NRZ vector at the $j^{\text{th}}$ receiving antenna |
| $\dot{y}^j \in \mathbb{C}^{N\times 1}$ | The output of the SIC at the $j^{\text{th}}$ receiving antenna |
| $\mathbf{P}_{\text{CP}^+} \in \mathbb{R}^{(N+N_{\text{CP}})\times N}$ | The Cyclic Prefix (CP) insertion matrix |
| $\mathbf{P}_{\text{CP}^-} \in \mathbb{R}^{N\times(N+N_{\text{CP}})}$ | The CP removal matrix |
| $\mathbf{I}_{a\times b}$ | An identity matrix of dimension $a \times b$ |
| $\mathbf{0}_{a\times b}$ | A zero matrix of dimension $a \times b$ |
| $(\#)^T$ | The vector/matrix transpose of # |
| $(\#)^H$ | The vector/matrix Hermitian of # |
| $\mathbf{\Gamma}_N^{-1} \in \mathbb{C}^{N\times N}$ | The inverse transform matrix |
| $\mathbf{\Gamma}_N \in \mathbb{C}^{N\times N}$ | The transform matrix |
| $\mathbf{F}_N^{-1} \in \mathbb{C}^{N\times N}$ | The Inverse Fast Fourier Transform (IFFT) matrix |
| $\mathbf{F}_N \in \mathbb{C}^{N\times N}$ | The FFT matrix |
| $\mathbf{W}_N^{-1} \in \mathbb{C}^{N\times N}$ | The IFWHT matrix |
| $\mathbf{W}_N \in \mathbb{C}^{N\times N}$ | The FWHT matrix |
| $\mathbf{K}_{N_j} \in \mathbb{C}^{iN\times \boldsymbol{i}N}, i \in \{1,2,\ldots.\}$ | The covariance matrix associated to the $j^{\text{th}}$ stream |
| $\boldsymbol{\varphi}(2^k) \in \mathbb{R}^{2^k\times 2^k}$ | The basic real Walsh Hadamard matrix |
| $\boldsymbol{\mathcal{A}}, \boldsymbol{\mathcal{B}}, \boldsymbol{\mathcal{U}} \in \mathbb{C}^{N\times N}$ | Each of them is an arbitrary complex matrix of dimension $N \times N$ |
| $\boldsymbol{\mathcal{H}}^{j,i} \in \mathbb{C}^{(N+N_{\text{CP}})\times(N+N_{\text{CP}})}$ | The channel Impulse Response Matrix (IRM) between the $i^{\text{th}}$ transmitting antennas, and $j^{\text{th}}$ receiving antenna |
| $h^{j,i} \in \mathbb{C}^{L\times 1}$ | The complex channel impulse response coefficients between the $i^{\text{th}}$ transmitting antennas, and $j^{\text{th}}$ receiving antenna |
| $\boldsymbol{\pi}^{j,i} \in \mathbb{C}^{N\times N}$ | The channel plus co-CFO full matrix |
| $\mathbf{\Pi}^{j,i} \in \mathbb{C}^{N\times N}$ | The banded-matrix of $\boldsymbol{\pi}^{j,i}$ |
| $\text{z}^j \in \mathbb{C}^{(N+N_{\text{CP}})\times 1}$ | The additive noise at the $j^{\text{th}}$ receiving antenna |
| $\otimes$ | The Kronecker product |
| $\varepsilon_{j,i}$ | The normalized co-CFO |
| $N$ | The transform length (i.e., the sub-carrier number) |
| $N_{\text{CP}}$ | The CP length |
| $[\boldsymbol{\eta}_{m,p}]_{\text{Fourier}}$ | The Fourier transform interference matrix |
| $[\boldsymbol{\eta}_{m,p}]_{\text{Walsh}}$ | The Walsh transform interference matrix |
| $\text{P}_i$ | The average power per symbol associated with the $i^{\text{th}}$ data stream vector |
| $\text{S}_\text{o}$ | The Power Spectral Density (PSD) of white Gaussian noise |
| $\mathcal{O}(\#)$ | Operations of order # |
| $\boldsymbol{\psi}_i \in \mathbb{C}^{iN\times 1}, i \in \{1,2,\ldots.\}$ | The composite matrix that contains different processing between $i^{\text{th}}$ transmitting antennas and specific receiving antenna including the channel |
| $\mathbf{\Xi}_i \in \mathbb{C}^{iN\times 1}, i \in \{1,2,\ldots.\}$ | The banded matrix of $\boldsymbol{\psi}_i$ |
| $f$ | The frequency shift |
| $\Delta f$ | The sub-carrier spacing |
| $\tau$ | The banded-matrix bandwidth |
| $\Re(\#)$ | Real part of # |
| $\Im(\#)$ | Imaginary part of # |
| $\Re(\xi)$ | The covariance matrix optimization parameter |
| $\mathfrak{J}(\xi)$ | The Joint Low Complexity Optimized Linear Zero Forcing (JLCOLZF) regularization parameter |
| $\rho$ | A real exponent term, and its value less than unity |
| $\delta$ | Number of covariance matrix terms |
| $\mathbf{C}_i \in \mathbb{C}^{N\times N}$ | The $i^{\text{th}}$ JLORLZF equalizer solution matrix |
| $\boldsymbol{\mathcal{C}}_{\text{LZF}}, \boldsymbol{\mathcal{C}}_{\text{LMMSE}} \in \mathbb{C}^{jN\times iN}$ | The solution matrix of the LZF, and LMMSE equalizers |
| $\boldsymbol{\Omega}_i \in \mathbb{C}^{iN\times N}$ | A composite matrix that contains different processing between $i^{\text{th}}$ transmitting antennas and specific receiving antenna including the channel with BMA |
| $\boldsymbol{\beta}_{jN\times N} \in \mathbb{C}^{iN\times N}$ | An arbitrary complex matrix of dimension $jN \times N$ |
| $\mathbb{E}\{\#\}$ | The expectation of # |
| $\sigma_{\boldsymbol{X}}^2$ | The transmitted signal power |
| $\mathbf{R}_{\text{z}}$ | The covariance matrix of AWGN |
| $\Gamma$ | The computational complexity efficiency |
| $\mathcal{N}_1$ | The number of flops of the proposed equalizer |
| $\mathcal{N}_2$ | The number of flops of the compared equalizer |

## I. Introduction

The frequency-selective transmission channel in Multi-Carrier Modulation (MCM) systems is partitioned into a group of flat fading channels [1], the impacts of which can be corrected with a one-tap per subcarrier equalizer. Channel partitioning techniques commonly used in cellular and wireline communication networks include Orthogonal Frequency Division Multiplexing (OFDM) and Discrete Multi-Tone (DMT) modulation [2], [3], [4], [5]. The benefits of OFDM technology extend its use to future wireless communications (5G) [6], [7], [8], [9], [10].

MCM can be applied in a variety of ways, but the most common version is based on the Fast Fourier Transform (FFT) due to various benefits such as its ease or efficacy against frequency-selective fading [11], [12]. However, the OFDM system can be implemented using orthogonal pairs for modulation and demodulation at the transmitter and receiver sides, respectively [13], [14], [15], [16]. Also, several authors [17], [18] have suggested the use of the Fast Walsh Hadamard Transform (FWHT), which has recently garnered increased interest in several fields [19], [20], [21].

The Walsh Hadamard transform has lately gained popularity in a wide range of engineering applications [22], [23], [24]. Because this transform is most readily implemented on digital computers [19] a hunt for the class of applications where the Walsh Hadamard transform can supplant or potentially augment the conventional function of the Fourier transform has begun. Despite their similarities and shared qualities, these transforms are traits of two distinct topological groups and will not be interchangeable in general. The Walsh transform is the natural representation of systems with dyadic symmetry, whereas the Fourier base (exponential functions) is the natural representation of systems with translational symmetry.

One of the majors that can be considered for future study is the use of Deep Reinforcement Learning (DRL) principal-based resource allocation [25]. Similarly, the authors of [26] suggest a novel strategy for combining optimal power allocation and user association methods in which cells are powered by a common grid network and other energy sources to achieve effective resource management. The authors of [7] proposed a novel adaptive backhaul topology that can respond to changing traffic patterns. The authors of [6] propose a dynamic optimization approach to lower the overall energy consumption of fifth-generation (5G) heterogeneous networks while maintaining the required coverage and capacity.

The goal of this article is to use the Walsh transform for OFDM, in addition to a proposal for a nonlinear equalizer. For the entire transmission and receiving procedure, a general matrix formulation is presented. We investigate the application of a proposed nonlinear equalizer known as the JLCOZF-SIC, which uses the concept of the BMA. Throughout our study, we also looked at three different cases linked to equalization and co-CFO compensation [27]. In the first case, only the diagonal compensation is taken into account. In the second case, the idea of BMA is viewed as a method for reducing complexity. The third case involves the use of complete matrix compensation. To rectify the channel effects in these instances, both time-domain and transform-domain equalizations are required. In comparison to the proposed, we deduce the formula for the number of flops, which enables us to find the estimated number of arithmetic operations for various linear and nonlinear equalizers. To the best of our understanding, the above FWHT-MCM system research is still ongoing.

The contribution of this paper can be summarized in the following points:

1. The conventional OFDM system is modulated via Inverse FWHT (IFWHT) instead of Inverse FFT (IFFT).
2. A general matrix formulation of the transceiver structure is presented.
3. A proposal for a nonlinear Walsh domain equalizer called the JLCOZF-SIC equalizer is presented.
4. The proposed equalizer reduces computational complexity by utilizing the BMA idea as well as an optimization parameter that prevents estimating both the signal power of each vector and the Signal-to-Noise Ratio (SNR).
5. The proposed equalizer takes into account the effects of the Rayleigh fading channel, noise, and the co-CFO.
6. The number of flops of various linear and nonlinear equalizers is given and compared to that of the suggested equalizer.

The organization of this paper is as follows: Section II describes the general model of the MCM system, and then a detailed mathematical formulation of the proposed equalizer is given in section III. Simulation results and analysis is discussed in section IV. Section V covers a detailed analysis of the complexity evaluation of the proposed equalizer with respect to different equalizers. Finally, the concluded marks are presented in section VI.

## II. The MCM System Model

Figures 1a and 1b depict the general block diagram of an $i \times j$ MIMO-OFDM multicarrier transceiver. Figure 1c shows the corresponding flowchart of a typical an $i \times j$ MIMO-OFDM multicarrier transceiver. Let's consider that the polar Non-Return-to-Zero (NRZ) bits are given by:

$$\mathrm{X}^{i} = [X_0^i \quad X_1^i \quad \ldots\ldots \quad X_{N-2}^i \quad X_{N-1}^i]^T \tag{1}$$

where $\mathrm{X}^{i} \in \mathbb{R}^{N\times 1}$ is the polar NRZ vector that modulated via Binary Phase Shift Keying (BPSK). The transmitted data vector in the frequency or Walsh domain can be formulated as:

$$\mathrm{x}^{i} = \mathbf{\Gamma}_N^{-1}\, \mathrm{X}^{i} = [x_0^i \quad x_1^i \quad \ldots\ldots \quad x_{N-2}^i \quad x_{N-1}^i]^T \tag{2}$$

where $\boldsymbol{\Gamma}_N^{-1}$ is an $N$-point inverse transform. In the case of IFFT, the entries of the $\boldsymbol{\Gamma}_N^{-1}$ matrix is given as:

$$\boldsymbol{\Gamma}_N^{-1} = \mathbf{F}_N^{-1} = \frac{1}{\sqrt{N}} e^{j\frac{2\pi f m n}{N}} \tag{3}$$

where $m, n \in \{0,1,2,..,N-1\}$, $\mathrm{x}^i \in \mathbb{C}^{N\times 1}$ is the modulated frequency-domain vector associated to the $i^{\text{th}}$ data stream vector. In the case of IFWHT, the entries of the $\boldsymbol{\Gamma}_N^{-1}$ matrix is given as [28], [29]:

$$\boldsymbol{\Gamma}_N^{-1} = \mathbf{W}_N^{-1}(2^k) = \frac{1}{\sqrt{N}} \begin{bmatrix} \boldsymbol{\varphi}(2^{k-1}) & \boldsymbol{\varphi}(2^{k-1}) \\ \boldsymbol{\varphi}(2^{k-1}) & -\boldsymbol{\varphi}(2^{k-1}) \end{bmatrix} = \boldsymbol{\varphi}(2) \otimes \boldsymbol{\varphi}(2^{k-1}) \tag{4}$$

where $\otimes$ denotes the Kronecker product, $2 \leq k \in N$, and $\boldsymbol{\varphi}(2^1)$ is given as:

$$\boldsymbol{\varphi}(2^1) = \begin{bmatrix} 1 & 1 \\ 1 & -1 \end{bmatrix} \tag{5}$$

In the same manner, $\boldsymbol{\varphi}(2^2)$ is given as:

$$\boldsymbol{\varphi}(2^2) = \begin{bmatrix} 1 & 1 & 1 & 1 \\ 1 & -1 & 1 & -1 \\ 1 & 1 & -1 & -1 \\ 1 & -1 & -1 & 1 \end{bmatrix} \tag{6}$$

In the same manner, $\mathrm{x}^i \in \mathbb{R}^{N\times 1}$ is the modulated Walsh vector associated to the $i^{\text{th}}$ data stream vector. The CP can be inserted to the $i^{\text{th}}$ modulated vector as:

$$\mathrm{x}_{\mathrm{CP}}^i = \mathbf{P}_{\mathrm{CP}^+} \boldsymbol{\Gamma}_N^{-1} \mathrm{X}^i = \left[ x_{0,\mathrm{CP}}^i \quad x_{1,\mathrm{CP}}^i \quad \ldots.. \quad x_{N-2,\mathrm{CP}}^i \quad x_{N-1,\mathrm{CP}}^i \right]^T \tag{7}$$

with

$$\mathbf{P}_{\mathrm{CP}^+} = \left[ [\mathbf{0}_{N_{\mathrm{CP}}\times(N-N_{\mathrm{CP}})}; \mathbf{I}_{N_{\mathrm{CP}}\times N_{\mathrm{CP}}}]^T, \mathbf{I}_{N\times N} \right]^T \tag{8}$$

where $\mathbf{P}_{\mathrm{CP}^+} \in \mathbb{R}^{(N+N_{\mathrm{CP}})\times N}$ is the CP insertion matrix. This is followed by Parallel-to-Serial (P/S) converter then transmitted over the Rayleigh fading channel.

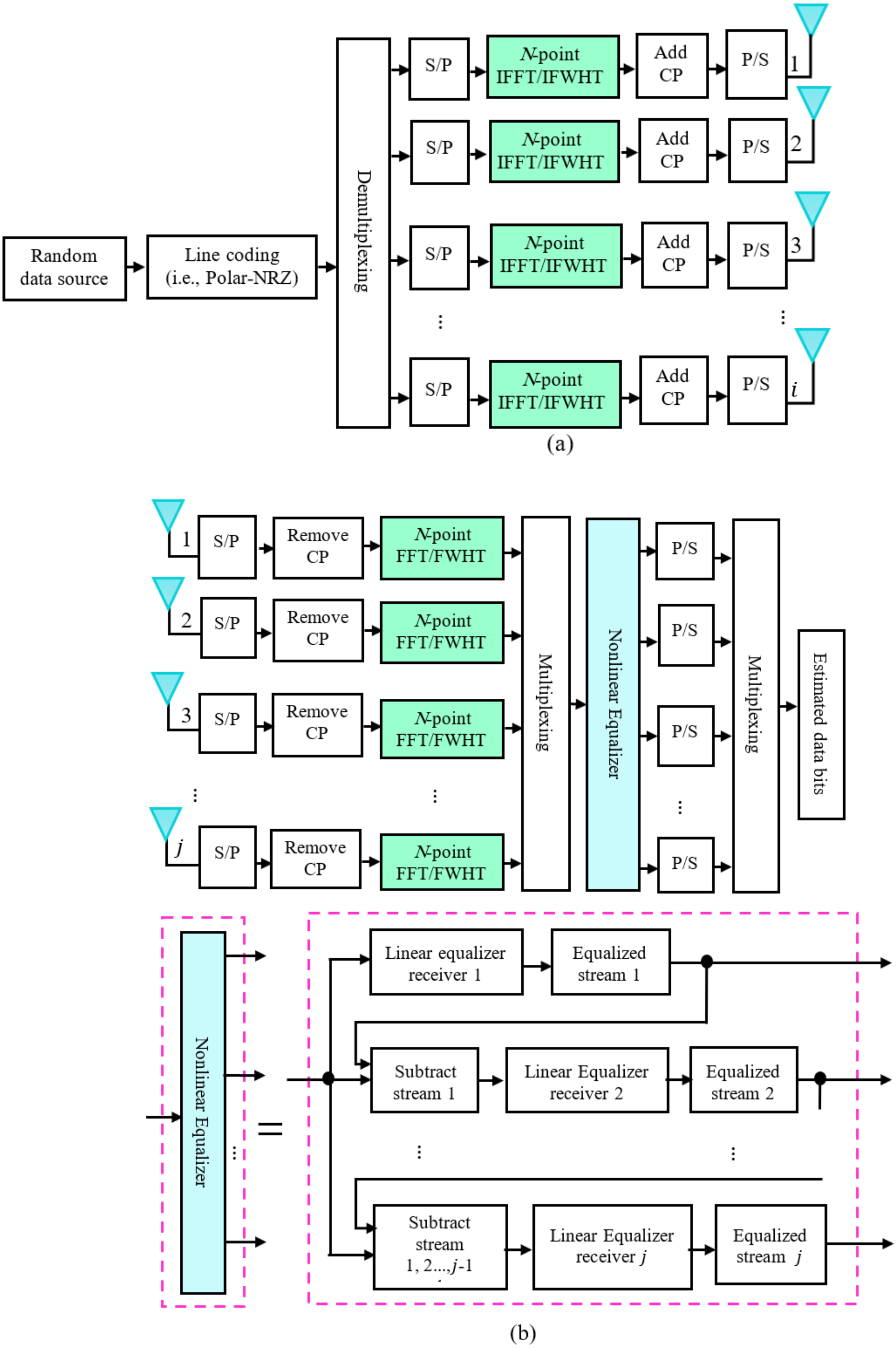

**Figure 1**. The transceiver block diagram of an $i \times j$ MIMO-OFDM communication system.

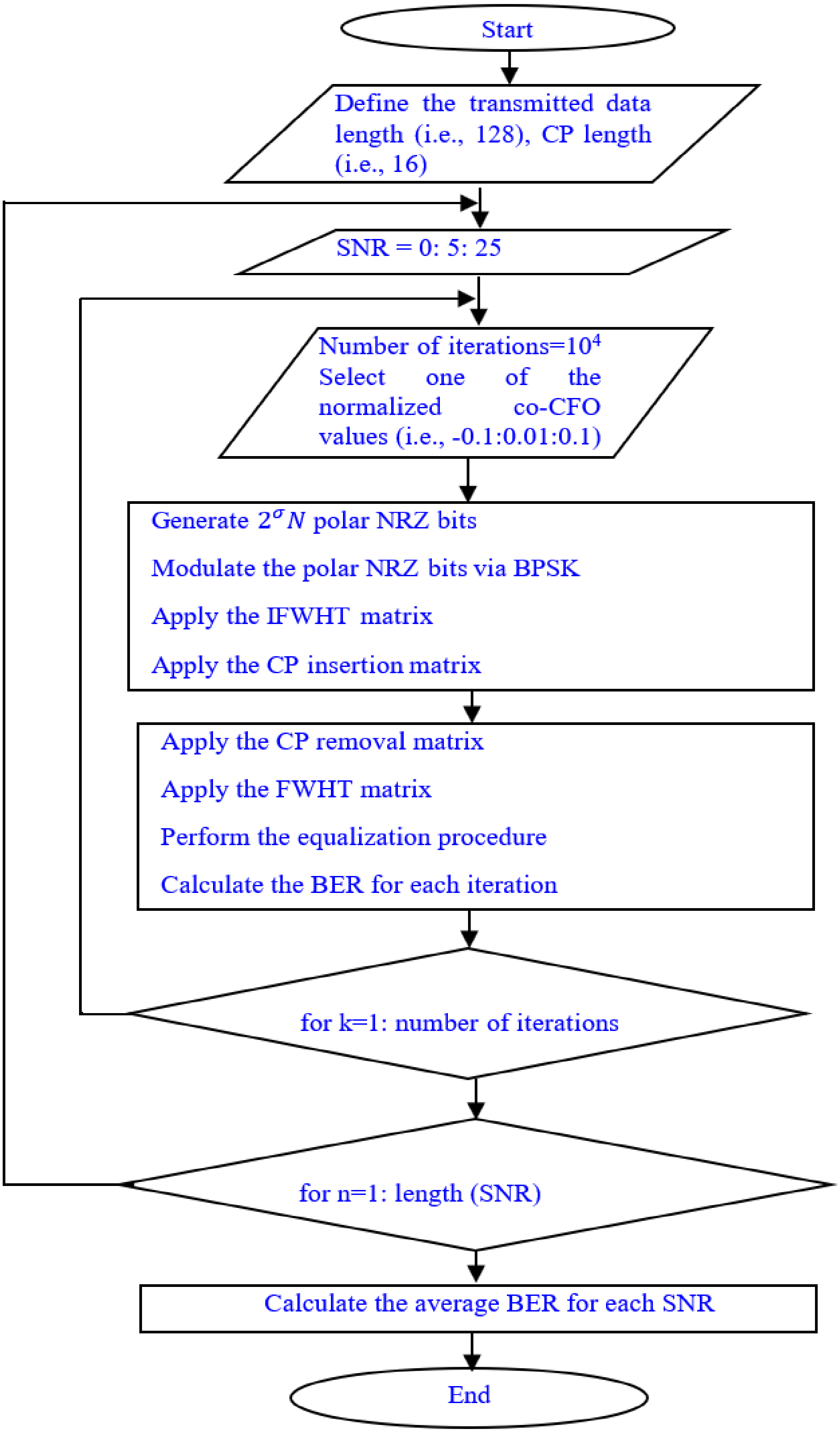

**Figure 1c**. Flowchart of an $i \times j$ MIMO-OFDM transceiver system.

At the received side, the received data stream vector via the $j^{\text{th}}$ receiving antenna can be formulated as:

$$\mathrm{y}^{j} = \boldsymbol{\psi}^{j,i}\, \boldsymbol{\mathcal{H}}^{j,i} \mathbf{P}_{\mathrm{CP}^{+}}\, \boldsymbol{\Gamma}_{N}^{-1}\, \mathrm{X}^{i} + \mathrm{z}^{j} \tag{9}$$

where $\mathrm{z}^{j} \in \mathbb{C}^{(N+N_{\mathrm{CP}})\times 1}$ represents the complex AWGN vector, complex Gaussian $\backsim \mathcal{CN}(0, \sigma_n^2)$ and $\sigma_n^2$ is the noise power, $\boldsymbol{\psi}^{j,i} \in \mathbb{C}^{(N+N_{\mathrm{CP}})\times(N+N_{\mathrm{CP}})}$ is the co-CFO matrix between the $i^{\text{th}}$ transmitting antenna and $j^{\text{th}}$ receiving antenna. The co-CFO matrix is diagonal matrix defined as:

$$\boldsymbol{\psi}^{j,i} = \mathrm{Diag}\left\{1, e^{j\frac{2\pi\varepsilon_{j,i}}{N}}, e^{j\frac{4\pi\varepsilon_{j,i}}{N}}, \ldots, e^{j\frac{2\pi\varepsilon_{j,i}(N+N_{\mathrm{CP}}-2)}{N}}, e^{j\frac{2\pi\varepsilon_{j,i}\,(N+N_{\mathrm{CP}}-1)}{N}}\right\} \tag{10}$$

where $\varepsilon_{j,i}$ denotes the normalized co-CFO between the $i^{\text{th}}$ transmitting antenna and $j^{\text{th}}$ receiving antenna. The normalized co-CFO can be expressed as:

$$\varepsilon_{j,i} = \frac{f}{\Delta f} \tag{11}$$

where $f$ is the amount of the frequency shift and $\Delta f$ describes the amount of the sub-carrier spacing. The value of the normalized co-CFO depends on some factors. These factors are the length of the transform, the value of the carrier frequency, the stability of the oscillator, the user equipment speed, and the transmitted signal bandwidth. Finally, the $\mathcal{H}^{j,i} \in \mathbb{C}^{(N+N_{\mathrm{CP}})\times(N+N_{\mathrm{CP}})}$ matrix describes the channel IRM between the $i^{\text{th}}$ transmitting antenna and the $j^{\text{th}}$ receiving antenna, which can be expressed as:

$$\boldsymbol{\mathcal{H}}^{j,i} = \begin{bmatrix} h^{j,i}(0) & 0 & 0 & 0 & \ldots & 0 & 0 & 0 \\ h^{j,i}(1) & h^{j,i}(0) & 0 & 0 & \ldots & 0 & 0 & 0 \\ h^{j,i}(2) & h^{j,i}(1) & h^{j,i}(0) & 0 & \ldots & 0 & 0 & 0 \\ \vdots & h^{j,i}(2) & h^{j,i}(1) & h^{j,i}(0) & \vdots & 0 & 0 & 0 \\ h^{j,i}(L-1) & \vdots & h^{j,i}(2) & h^{j,i}(1) & \ddots & 0 & 0 & 0 \\ 0 & h^{j,i}(L-1) & \vdots & h^{j,i}(2) & \vdots & h^{j,i}(0) & 0 & 0 \\ 0 & 0 & h^{j,i}(L-1) & \vdots & \ldots & h^{j,i}(1) & h^{j,i}(0) & 0 \\ 0 & 0 & 0 & h^{j,i}(L-1) & \ldots & h^{j,i}(2) & h^{j,i}(1) & h^{j,i}(0) \end{bmatrix} \tag{12}$$

where $h^{j,i} \in \mathbb{C}^{L\times 1}$ is the complex channel impulse response coefficients between the $i^{\text{th}}$ transmitting antennas, and $j^{\text{th}}$ receiving antenna, and $L$ is the channel length. The distorted head of each vector due to the channel delay spread, which represented by the CP vector is discarded at the $j^{\text{th}}$ antenna as:

$$\bar{y}^{j} = \mathbf{P}_{\text{CP}^-}\, \mathrm{y}^{j} = \mathbf{P}_{\text{CP}^-} \boldsymbol{\psi}^{j,i}\, \boldsymbol{\mathcal{H}}^{j,i} \mathbf{P}_{\text{CP}^+}\, \boldsymbol{\Gamma}_N^{-1}\, \mathrm{X}^{i} + \mathbf{P}_{\text{CP}^-} \mathrm{z}^{j} \tag{13}$$

where $\mathbf{P}_{\text{CP}^-} \in \mathbb{R}^{N\times(N+N_{\text{CP}})}$ indicates the removal matrix of the CP. Because the Walsh matrix is symmetric, the forward and backward transformations are similar processes with the exception of the $\frac{1}{\sqrt{N}}$ scaling factor. Hence, the FWHT matrix can be expressed using Eq. (4) as:

$$\boldsymbol{\Gamma}_N = \mathbf{W}_N(2^k) = \sqrt{N}\begin{bmatrix} \boldsymbol{\varphi}(2^{k-1}) & \boldsymbol{\varphi}(2^{k-1}) \\ \boldsymbol{\varphi}(2^{k-1}) & -\boldsymbol{\varphi}(2^{k-1}) \end{bmatrix} = \boldsymbol{\varphi}(2)\otimes\boldsymbol{\varphi}(2^{k-1}) \tag{14}$$

The transform output at the $j^{th}$ receiving antenna can be formulated as:

$$\bar{\bar{y}}^{j} = \boldsymbol{\Gamma}_N\, \mathbf{P}_{\text{CP}^-} \boldsymbol{\psi}^{j,i}\, \boldsymbol{\mathcal{H}}^{j,i} \mathbf{P}_{\text{CP}^+}\, \boldsymbol{\Gamma}_N^{-1}\, \mathrm{X}^{i} + \boldsymbol{\Gamma}_N\, \mathbf{P}_{\text{CP}^-} \mathrm{z}^{j} \tag{15}$$

where $\boldsymbol{\pi}^{j,i} = \boldsymbol{\Gamma}_N\, \mathbf{P}_{\text{CP}^-} \boldsymbol{\psi}^{j,i}\, \boldsymbol{\mathcal{H}}^{j,i} \mathbf{P}_{\text{CP}^+}\, \boldsymbol{\Gamma}_N^{-1}$, Equation (15) can be expressed in two forms as:

$$\begin{bmatrix} \bar{\bar{y}}^{1} \\ \bar{\bar{y}}^{2} \\ \vdots \\ \bar{\bar{y}}^{j} \end{bmatrix} = \begin{bmatrix} \boldsymbol{\pi}^{1,1} & \boldsymbol{\pi}^{1,2} & \dots & \boldsymbol{\pi}^{1,i} \\ \boldsymbol{\pi}^{2,1} & \boldsymbol{\pi}^{2,2} & \dots & \boldsymbol{\pi}^{2,i} \\ \vdots & \vdots & \ddots & \vdots \\ \boldsymbol{\pi}^{j,1} & \boldsymbol{\pi}^{j,2} & \dots & \boldsymbol{\pi}^{j,i} \end{bmatrix} \begin{bmatrix} \mathrm{X}^{1} \\ \mathrm{X}^{2} \\ \vdots \\ \mathrm{X}^{i} \end{bmatrix} + \begin{bmatrix} \boldsymbol{\Gamma}_N\, \mathbf{P}_{\text{CP}^-} \mathrm{z}^{1} \\ \boldsymbol{\Gamma}_N\, \mathbf{P}_{\text{CP}^-} \mathrm{z}^{2} \\ \vdots \\ \boldsymbol{\Gamma}_N\, \mathbf{P}_{\text{CP}^-} \mathrm{z}^{j} \end{bmatrix} \tag{16}$$

or:

$$\begin{bmatrix} \bar{\bar{y}}^{1} \\ \bar{\bar{y}}^{2} \\ \vdots \\ \bar{\bar{y}}^{j} \end{bmatrix} = \underbrace{\begin{bmatrix} \boldsymbol{\pi}^{1,1} \\ \boldsymbol{\pi}^{2,1} \\ \vdots \\ \boldsymbol{\pi}^{j,1} \end{bmatrix}}_{\boldsymbol{\psi}_1} [\mathrm{X}^{1}] + \underbrace{\begin{bmatrix} \boldsymbol{\pi}^{1,2} \\ \boldsymbol{\pi}^{2,2} \\ \vdots \\ \boldsymbol{\pi}^{j,2} \end{bmatrix}}_{\boldsymbol{\psi}_2} [\mathrm{X}^{2}] + .. + \underbrace{\begin{bmatrix} \boldsymbol{\pi}^{1,i} \\ \boldsymbol{\pi}^{2,i} \\ \vdots \\ \boldsymbol{\pi}^{j,i} \end{bmatrix}}_{\boldsymbol{\psi}_i} [\mathrm{X}^{i}] + \begin{bmatrix} \boldsymbol{\Gamma}_N\, \mathbf{P}_{\text{CP}^-} \mathrm{z}^{1} \\ \boldsymbol{\Gamma}_N\, \mathbf{P}_{\text{CP}^-} \mathrm{z}^{2} \\ \vdots \\ \boldsymbol{\Gamma}_N\, \mathbf{P}_{\text{CP}^-} \mathrm{z}^{j} \end{bmatrix} \tag{17}$$

Now, let's check the magnitude of the $\boldsymbol{\pi}^{j,i}$ matrix. Consider the case of $2 \times 2$ MIMO-OFDM communication systems based on the Fourier and Walsh transforms. Using a six-tap Rayleigh fading channel based on Jake's model [30]. Figure 2 shows the magnitude of the $\boldsymbol{\pi}^{j,i}$ $, j, i \in \{1,2\}$ matrix in the case of a $2 \times 2$ MIMO-OFDM communication system based on the Fourier transform. This figure indicates that higher sub-carrier indices have a lower amount compared to that of the diagonal elements, which builds the consideration of BMA that may be taken into account [31]. The concept of the BMA reduces the number of multiplications and additions needed for matrix-by-matrix additions, subtraction, division, and inversion [31]. The BMA takes some elements with respect to its diagonal and ignores the remaining elements in this row.

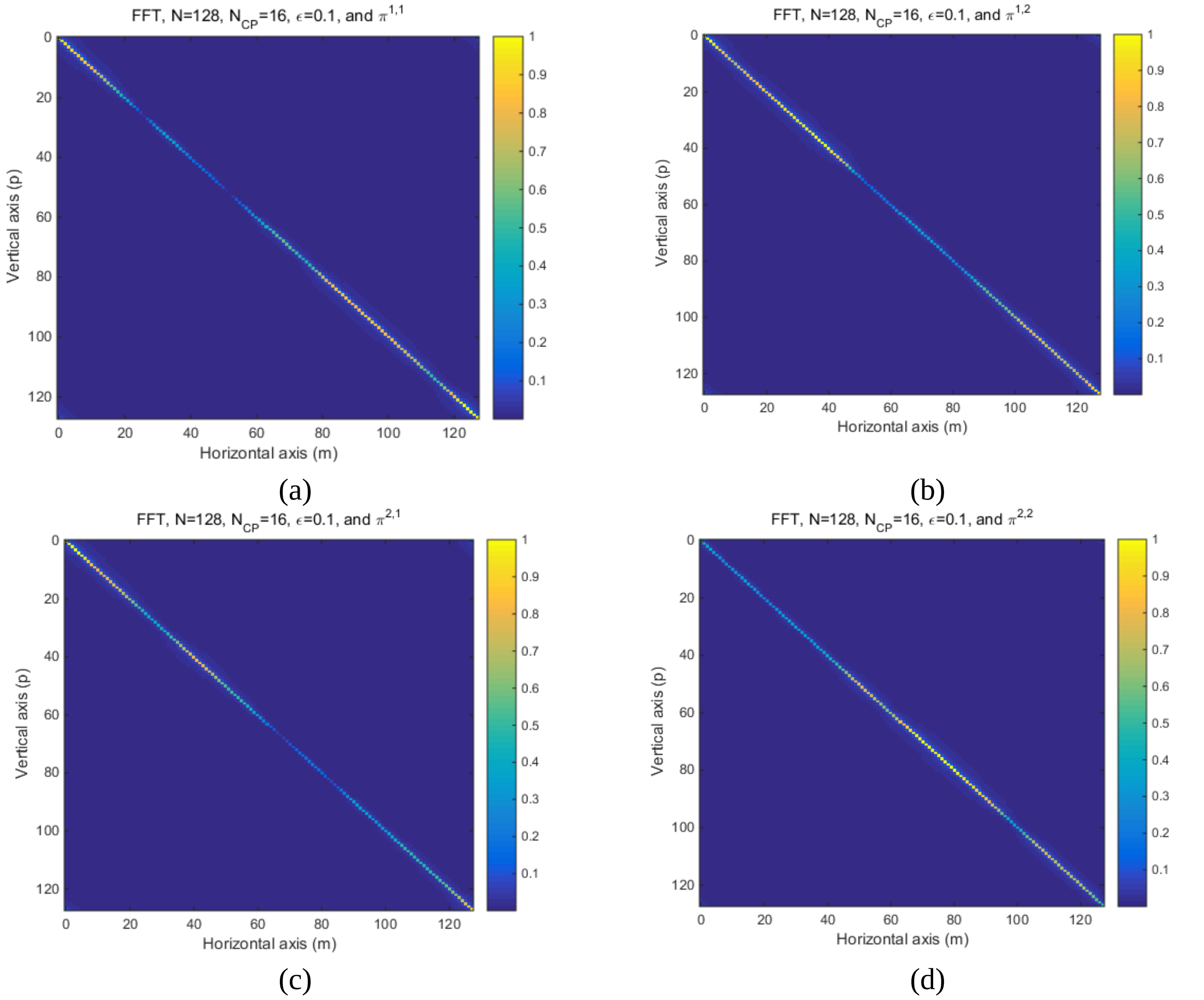

**Figure 2**. The normalized magnitude of the interference matrix vs. sub-carrier indices using Fourier transform.

Figure 3 shows the magnitude of the $\boldsymbol{\pi}^{j,i}$ , $j, i \in \{1,2\}$ matrix in the case of a $2 \times 2$ MIMO-OFDM communication system based on the Walsh transform. With respect to the diagonal elements, the sub-carriers with $N/2$ away have a considerable value that must be taken into account to reserve the BER performance. Thus, the BMA concept must take at least half of the sub-carriers, which increases the number of multiplications and additions compared to the case of the Fourier transform. Thus, the concept of the BMA in the case of the Walsh transform may be useful as that half of the sub-carrier or more is taken into account. This will be discussed and analyzed deeply through the simulation results and analysis section to accurately specify the optimum value of the BMA bandwidth. The bandwidth of the BMA specifies the number of elements with respect to the diagonal that will be taken into account.

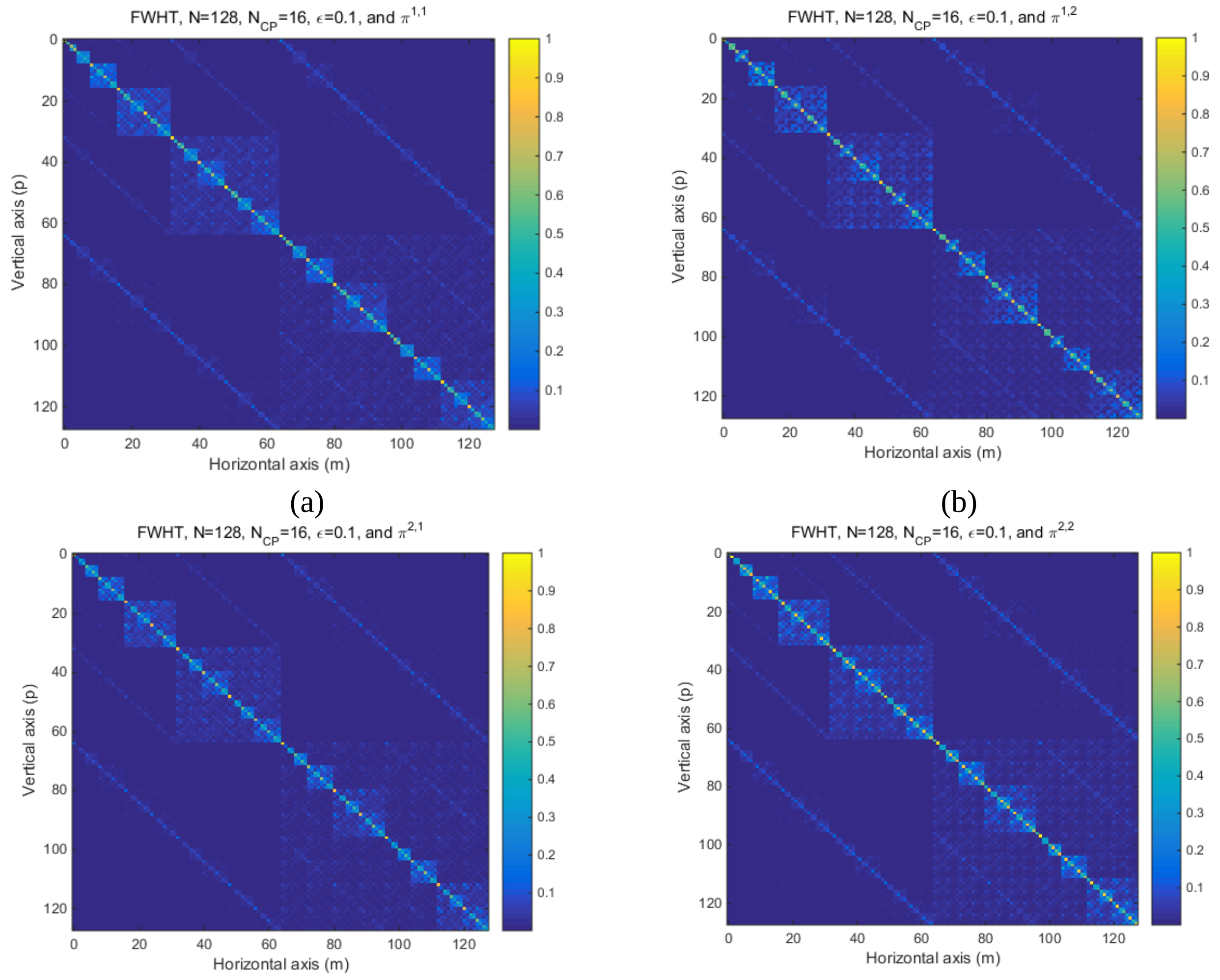

(a) (b) (c) (d)

**Figure 3**. The normalized magnitude of the $\boldsymbol{\pi}^{j,i}$ , $j, i \in \{1,2\}$ matrix vs. sub-carrier indices using Walsh transform.

Let's now give an overview of the interference matrix of that of based on the Fourier and Walsh transforms. In the case of Fourier transform the generated interference matrix can be formulated as:

$$[\boldsymbol{\eta}_{m,p}]_{\text{Fourier}} = \mathbf{F}_N^{-1}.\,\text{diag}\left\{e^{j\frac{2\pi\varepsilon n}{N}}\right\}.\,\mathbf{F}_N \qquad (18)$$

In the case of the Walsh transform, the interference matrix can be formulated as:

$$[\boldsymbol{\eta}_{m,p}]_{\text{Walsh}} = \mathbf{W}_N^{-1}.\,\text{diag}\left\{e^{j\frac{2\pi\varepsilon n}{N}}\right\}.\,\mathbf{W}_N \qquad (19)$$

Let's check the magnitude of the interference matrix in Eqs. (18), and (19). Figure 4 describes the amount of interference from all sub-carriers up on the desired sub-carrier (i.e., diagonal elements are the desired sub-carriers). It is clear that the interference amount on the desired sub-carrier gradually decreased with respect to the relative sub-carrier index in the case of Fourier. In the case of the Walsh transform, the interference amount on the desired sub-carrier is almost zero, except for some sub-carrier indices, which have a considerable value of 15.84% with respect to the desired (i.e., diagonal). This may have a bad effect on the Bit-Error-Rate (BER) performance using the principles of the BMA. Hence, according to the interference amount on the desired sub-carriers illustrated in Fig. 4, the OFDM communication system may be useful for the principal of the BMA in the case of FFT and discouraged in the case of FWHT with a bandwidth less than $N/2$. This will be discussed deeply in the simulation results and analysis section.

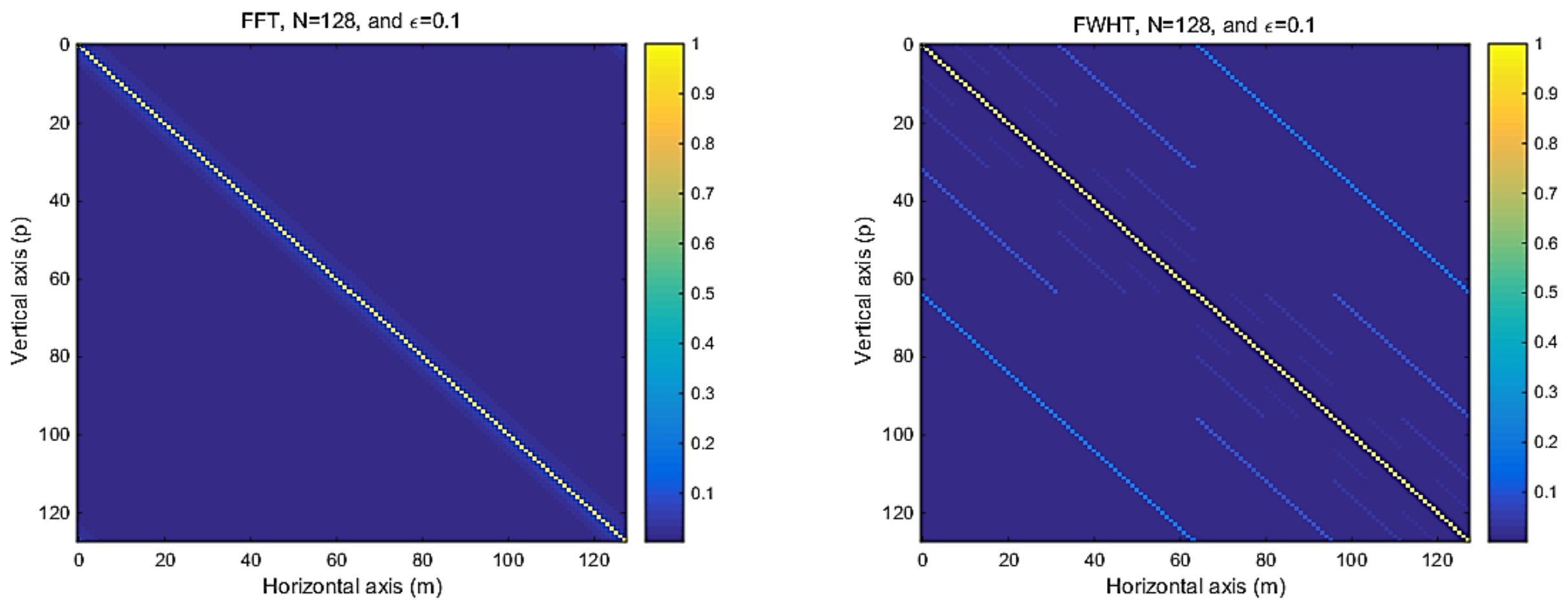

(a) (b)

**Figure 4**. The normalized magnitude of the interference matrix vs. sub-carrier indices using different transforms.

## III. The Proposed JLCOZF-SIC Equalizer

To demonstrate the major distinctions between the MMSE-SIC and the proposed equalizer's mathematical formulation process, we should first describe the MMSE-SIC process. The major benefits of the proposed equalizer over the conventional equalizer are included.

- *The conventional MMSE-SIC equalizer*

The conventionally accurate operation of the MMSE-SIC requires estimation of the average power per symbol in addition to computation of the SNR [32], [33]. Although the MMSE-SIC equalizer outperforms all linear algorithms in terms of BER efficiency, its computation cost is significantly higher (i. e., $\mathcal{O}(N^3)$ operations). The following stages describe the MMSE-SIC equalizer operation process for an $i \times j$ MIMO configuration as follows:

1) The construction of the covariance matrix is performed as [34]:

$$\mathbf{K}_{\boldsymbol{N}_j} = \mathrm{S_o}\mathbf{I}_{jN\times jN} + \sum_{a\neq j}^{\boldsymbol{i}} \mathrm{P}_a\ \boldsymbol{\psi}_{\boldsymbol{a}}\ \boldsymbol{\psi}_{\boldsymbol{a}}^{*} \tag{20}$$

where $\mathrm{S_o}$ is the PSD of white Gaussian noise, $\mathrm{P}_a$ describes the $a^{\text{th}}$ average power per symbol. Let's consider that we're intended to recover the 1st data stream vector. This requires the covariance matrix to be in the form:

$$\mathbf{K}_{\boldsymbol{N}_1} = \mathrm{S_o}\mathbf{I}_{jN\times jN} + \mathrm{P}_2\ \boldsymbol{\psi}_2\ \boldsymbol{\psi}_2^{*} + \mathrm{P}_3\ \boldsymbol{\psi}_3\ \boldsymbol{\psi}_3^{*} + \cdots + \mathrm{P}_i\ \boldsymbol{\psi}_i\ \boldsymbol{\psi}_i^{*} \tag{21}$$

2) The covariance matrix in Eq. (21) is processed to get $\left(\mathbf{K}_{\boldsymbol{N}_1}\right)^{-\frac{1}{2}}$.

3) The $\left(\mathbf{K}_{\boldsymbol{N}_1}\right)^{-\frac{1}{2}}$ matrix is multiplied by the received vector expressed in Eq. (17).

4) The 1st Linear Minimum Mean Square Error (LMMSE) equalizer is applied to recover the 1st data stream vector.

5) The estimated vector is multiplied by the $\boldsymbol{\psi}_{\mathbf{1}}$ matrix, then subtracted from Eq. (17). This is known by the concept of the SIC.

6) Excepting the 2nd term in Eq. (21), another covariance matrix is built to recover the 2nd data stream vector as:

$$\mathbf{K}_{\boldsymbol{N}_2} = \mathrm{S_o}\mathbf{I}_{jN\times jN} + \mathrm{P}_3\ \boldsymbol{\psi}_3\ \boldsymbol{\psi}_3^{*} + \mathrm{P}_4\ \boldsymbol{\psi}_4\ \boldsymbol{\psi}_4^{*} + \cdots + \mathrm{P}_i\ \boldsymbol{\psi}_i\ \boldsymbol{\psi}_i^{*} \tag{22}$$

7) The $\left(\mathbf{K}_{\boldsymbol{N}_2}\right)^{-\frac{1}{2}}$ matrix is multiplied by the output of 6th step.

8) The 2nd LMMSE equalizer is applied to recover the 2nd data stream vector.

9) The 1st, and 2nd estimated vectors are multiplied by the corresponding $\boldsymbol{\psi}_1$, $\boldsymbol{\psi}_2$ matrices, then subtracted from Eq. (17).

10) Excepting the 2nd term in Eq. (22), another covariance matrix is built to recover the 3rd data stream vector as:

$$\mathbf{K}_{\boldsymbol{N_3}} = \mathrm{S_o}\mathbf{I}_{jN\times \boldsymbol{j}N} + \mathrm{P}_4\, \boldsymbol{\psi}_4\, \boldsymbol{\psi}_4^* + \mathrm{P}_5\, \boldsymbol{\psi}_5\, \boldsymbol{\psi}_5^* + \cdots + \mathrm{P}_i\, \boldsymbol{\psi}_i\, \boldsymbol{\psi}_i^* \quad (23)$$

11) The $\left(\mathbf{K}_{\boldsymbol{N}_3}\right)^{-\frac{1}{2}}$ matrix is multiplied by the output of 10th step.

12) The 3rd LMMSE equalizer is applied to recover the 3rd data stream vector.

13) These processing are repeated according to the number of transmitted vectors.

14) In the final step, the 1st, and 2nd up to the ($i$-1)th estimated vectors are multiplied by the corresponding $\boldsymbol{\psi}_1$, $\boldsymbol{\psi}_2$, up to $\boldsymbol{\psi}_{i-1}$ matrices, then subtracted from Eq. (17).

15) The $i$th LMMSE equalizer is applied to recover the $i$th data stream vector.

- *The Proposed JLCOZF-SIC equalizer*

The main steps of the proposed equalizer are the same as those of the MMSE-SIC equalizer, with some differences. These steps can be listed as follows:

1) The construction of the covariance matrix is modified to be:

$$\mathbf{K}_{\boldsymbol{N_j}} = \mathbf{I}_{jN\times jN} + \Re(\xi) \sum_{a\neq j}^{\boldsymbol{i}} \boldsymbol{\Xi_a}\, \boldsymbol{\Xi_a^*} \quad (24)$$

where $\xi$ is constant value named as the optimization parameter, $\boldsymbol{\Xi}_i = [\boldsymbol{\Pi}^{1,i} \quad \boldsymbol{\Pi}^{2,i} \quad \boldsymbol{\Pi}^{3,i} \quad \ldots \quad \boldsymbol{\Pi}^{j,i}]$ is the banded-matrix of $\boldsymbol{\psi}_i$ that defined as:

$$\left(\boldsymbol{\Pi}^{j,i}\right)_{m,p} = \begin{cases} \left(\boldsymbol{\pi}^{j,i}\right)_{m,p} & \text{if } |m-p| \le \tau \\ 0 & \text{if } |m-p| > \tau \end{cases} \quad (25)$$

where $\tau$ is the banded-matrix bandwidth, and $|m-p|$ is the relative sub-carrier spacing.

2) The covariance matrix related to the 1st data stream vector expressed in Eq. (21) will be modified as:

$$\mathbf{K}_{\boldsymbol{N}_1} = \mathbf{I}_{jN\times jN} + \Re(\xi)\ \Xi_2\ \Xi_2^* + \Re(\xi)\ \Xi_3\ \Xi_3^* + \cdots + \Re(\xi)\ \Xi_i\ \Xi_i^* \quad (26)$$

3) The covariance matrix in Eq. (26) is processed to get $\left(\mathbf{K}_{\boldsymbol{N}_1}\right)^{-\frac{1}{2}}$.

4) The $\left(\mathbf{K}_{\boldsymbol{N}_1}\right)^{-\frac{1}{2}}$ matrix is multiplied by the received vector expressed in Eq. (17) as:

$$\tilde{y}^1 = \left(\mathbf{K}_{\boldsymbol{N}_1}\right)^{-\frac{1}{2}} \times \bar{\bar{y}}^1 = \left(\mathbf{K}_{\boldsymbol{N}_1}\right)^{-\frac{1}{2}}.\boldsymbol{\psi}_1 \mathrm{X}^1 + \underbrace{\Re(\xi).\left(\mathbf{K}_{\boldsymbol{N}_1}\right)^{-\frac{1}{2}} \sum_{i=2}^{N_t} \boldsymbol{\psi}_i \mathrm{X}^i + \boldsymbol{\Gamma}_N\, \mathbf{P}_{\mathrm{CP}}\text{-}\mathrm{z}^1}_{\text{White noise}} \quad (27)$$

5) The 1st JLCOLZF equalizer is applied to recover the 1st data stream vector as:

$$\hat{y}^1 = \mathbf{C}_1 \times \tilde{y}^1 \approx \mathrm{X}^1 + \underbrace{\mathbf{C}_1 \times \Re(\xi).\left(\mathbf{K}_{\boldsymbol{N}_1}\right)^{-\frac{1}{2}} \sum_{i=2}^{N_t} \boldsymbol{\psi}_i \mathrm{X}^i + \boldsymbol{\Gamma}_N\, \mathbf{P}_{\mathrm{CP}}\text{-}\mathrm{z}^1}_{\text{White noise}} \quad (28)$$

where $\mathbf{C}_1 \times \left(\mathbf{K}_{\boldsymbol{N}_1}\right)^{-\frac{1}{2}}.\boldsymbol{\psi}_1 \approx \mathbf{I}_{N\times N}$, and

$$\mathbf{C}_1 = \left(\boldsymbol{\Omega}_1{}^H.\ \boldsymbol{\Omega}_1 + \Im(\xi)\mathbf{I}_{N\times N}\right)^{-1} \boldsymbol{\Omega}_1{}^H \quad (29)$$

where $\boldsymbol{\Omega}_1 = \left(\mathbf{K}_{\boldsymbol{N}_1}\right)^{-\frac{1}{2}}.\Xi_1$, and $\Im(\xi)$ is the JLCOLZF optimization parameter.

6) The SIC process is applied by multiplying the estimated vector by the $\Xi_1$ matrix, then subtracted from Eq. (17) as:

$$\dot{y}^2 = \bar{\bar{y}}^j - \Xi_1.\hat{y}^1 \quad (30)$$

7) Excepting the 2nd term in Eq. (26), another covariance matrix is built to recover the 2nd data stream vector as:

$$\mathbf{K}_{\boldsymbol{N}_2} = \mathbf{I}_{jN\times jN} + \Re(\xi)\ \Xi_3\ \Xi_3^* + \Re(\xi)\ \Xi_4\ \Xi_4^* + \cdots + \Re(\xi)\ \Xi_i\ \Xi_i^* \quad (31)$$

8) The $\left(\mathbf{K}_{\boldsymbol{N}_2}\right)^{-\frac{1}{2}}$ matrix is multiplied by the output of 6th step.

9) The 2nd JLCOLZF equalizer is applied to recover the 2nd data stream vector, and its solution is given as:

$$\mathbf{C}_2 = \left(\boldsymbol{\Omega}_2{}^H.\ \boldsymbol{\Omega}_2 + \Im(\xi)\mathbf{I}_{N\times N}\right)^{-1} \boldsymbol{\Omega}_2{}^H \quad (32)$$

where $\boldsymbol{\Omega}_2 = \left(\mathbf{K}_{\boldsymbol{N}_2}\right)^{-\frac{1}{2}}.\Xi_2$.

10) The 1st, and 2nd estimated vectors are multiplied by the corresponding $\Xi_1$ , $\Xi_2$ matrices, then subtracted from Eq. (17) as:

$$\dot{y}^3 = \bar{\bar{y}}^j - \Xi_1 . \hat{y}^1 - \Xi_2 . \hat{y}^2 \qquad (33)$$

11) Excepting the 2nd term in Eq. (31), another covariance matrix is built to recover the 3rd data stream vector as:

$$\mathbf{K}_{\boldsymbol{N}_3} = \mathbf{I}_{jN\times jN} + \Re(\xi)\ \Xi_4\ \Xi_4^* + \Re(\xi)\ \Xi_5\ \Xi_5^* + \cdots + \Re(\xi)\ \Xi_i\ \Xi_i^* \qquad (34)$$

12) The $\left(\mathbf{K}_{\boldsymbol{N}_3}\right)^{-\frac{1}{2}}$ matrix is multiplied by the output of 10th step.

13) The 3rd JLCOLZF equalizer is applied to recover the 3rd data stream vector as:

$$\mathbf{C}_3 = \left(\boldsymbol{\Omega}_3{}^H .\ \boldsymbol{\Omega}_3 + \ \Im(\xi)\mathbf{I}_{N\times N}\right)^{-1} \boldsymbol{\Omega}_3{}^H \qquad (35)$$

where $\boldsymbol{\Omega}_3 = \left(\mathbf{K}_{\boldsymbol{N}_3}\right)^{-\frac{1}{2}} . \Xi_3$.

14) These processing are repeated according to the number of transmitted vectors.

15) In the final step, the 1st, and 2nd up to the ($i$-1)th estimated vectors are multiplied by the corresponding $\Xi_1$, $\Xi_2$, up to $\Xi_{i-1}$ matrices, then subtracted from Eq. (17).

16) The $i$th JLCOLZF equalizer is applied to recover the $i$th data stream vector as:

$$\mathbf{C}_i = \left(\Xi_i{}^H .\ \Xi_i + \ \Im(\xi)\mathbf{I}_{N\times N}\right)^{-1} \Xi_i{}^H \qquad (36)$$

Note that each covariance matrix raised to negative half power $\left(\text{i.e.,}\ \left(\mathbf{K}_{N_j}\right)^{-\frac{1}{2}}\right)$ is approximated as [35]:

$$\begin{aligned}\mathbf{K}_{N_j} &= \left[\mathbf{I}_{jN\times jN} + \boldsymbol{\beta}_{jN\times jN}\right]^{\rho} \\ &= \mathbf{I}_{jN\times jN} + \rho\boldsymbol{\beta}_{jN\times jN} + \frac{\rho(\rho-1)}{2!}\left(\boldsymbol{\beta}_{jN\times jN}\right)^2 + \frac{\rho(\rho-1)(\rho-2)}{3!}\left(\boldsymbol{\beta}_{jN\times jN}\right)^3 + \cdots\end{aligned} \qquad (37)$$

where, $\rho < 1$, $\boldsymbol{\beta}_{jN\times \mathrm{j}N}\mathbb{C}^{N_tN\times N}$ is an $jN \times jN$ dimensional arbitrary complex matrix. As the terms of Eq. (37) were stretched to infinity, the computational complexity increased rapidly. As a result, let us verify the approximate number of terms that will be considered with respect to the case of MMSE-SIC (i.e., $\delta = \infty$). Figure 5a shows the BER versus the number of terms at various SNR values in Eq. (37) for the case of the 2×2 MIMO-OFDM configuration. Figure 5b shows BER versus the number of covariance matrix terms (i. e., $\delta$). expressed in Eq. (37), which indicates the elevation view of Fig. 5a. Figure 5c shows BER versus the SNR at different numbers of covariance matrix terms expressed in Eq. (37), which indicates the side view of Fig. 5a. Figure 5 ensures that according to the number of terms, the BER performance is close to that of the MMSE-SIC. Hence, according to your needs, you can select the value of $\delta$. Through the computational complexity section, the number of flops for the proposed equalizer will be estimated under the two conditions. The first condition is at $\delta = 3$, and the second case is at $\delta = \infty$.

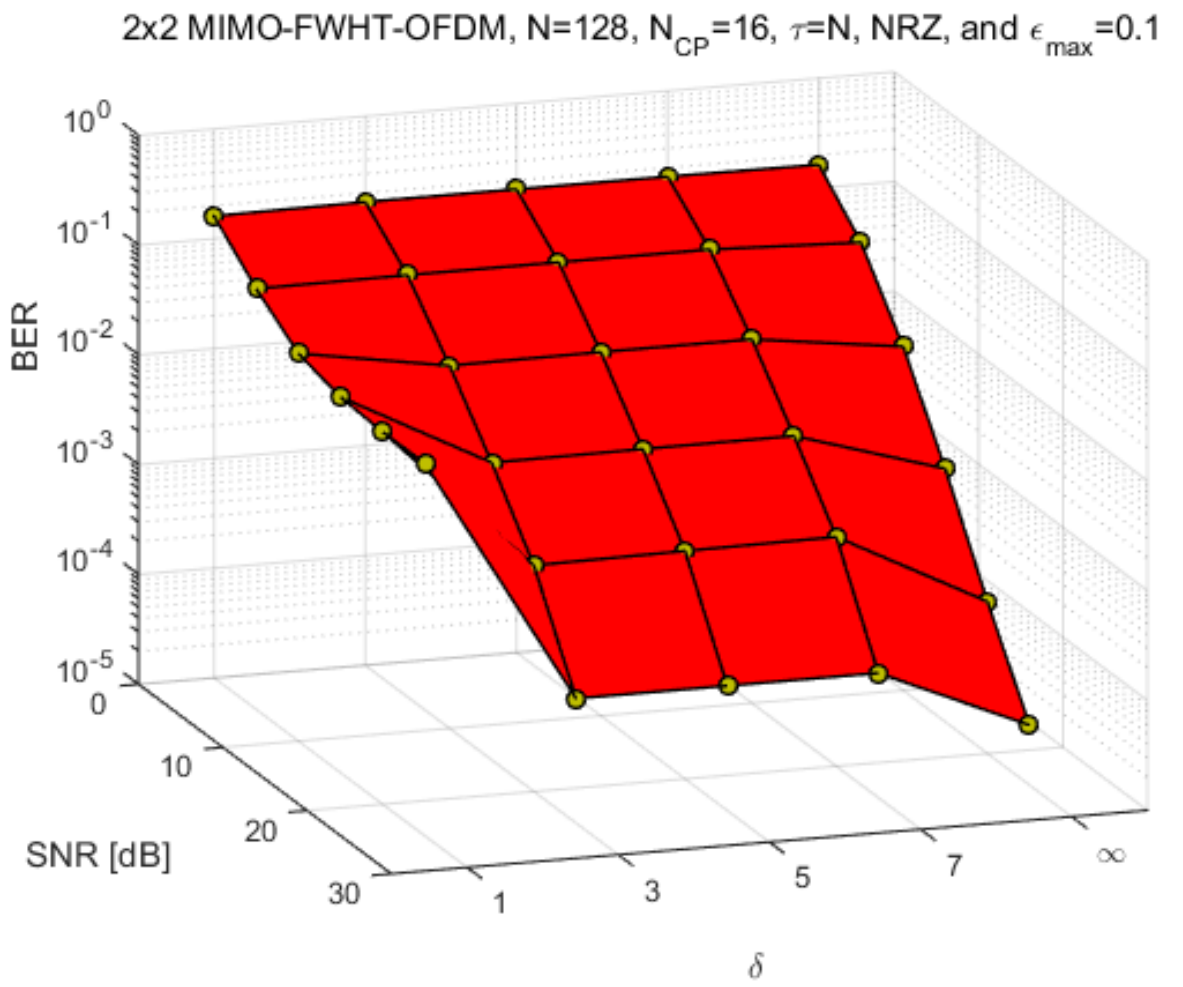

**Figure 5**a. The BER vs. number of terms of approximated covariance matrix (i. e., $\rho$) in Eq. (37)

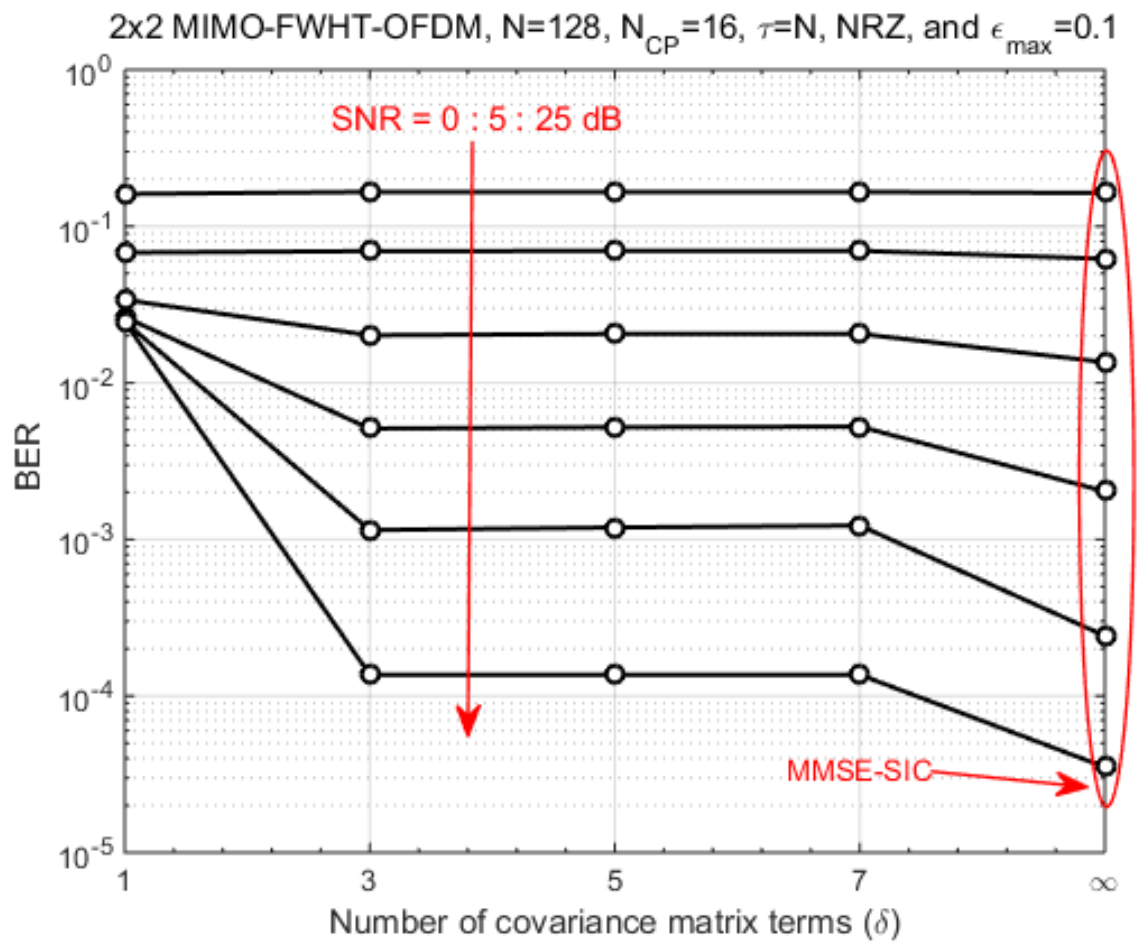

**Figure 5b**. The elevation view of Fig. 5a

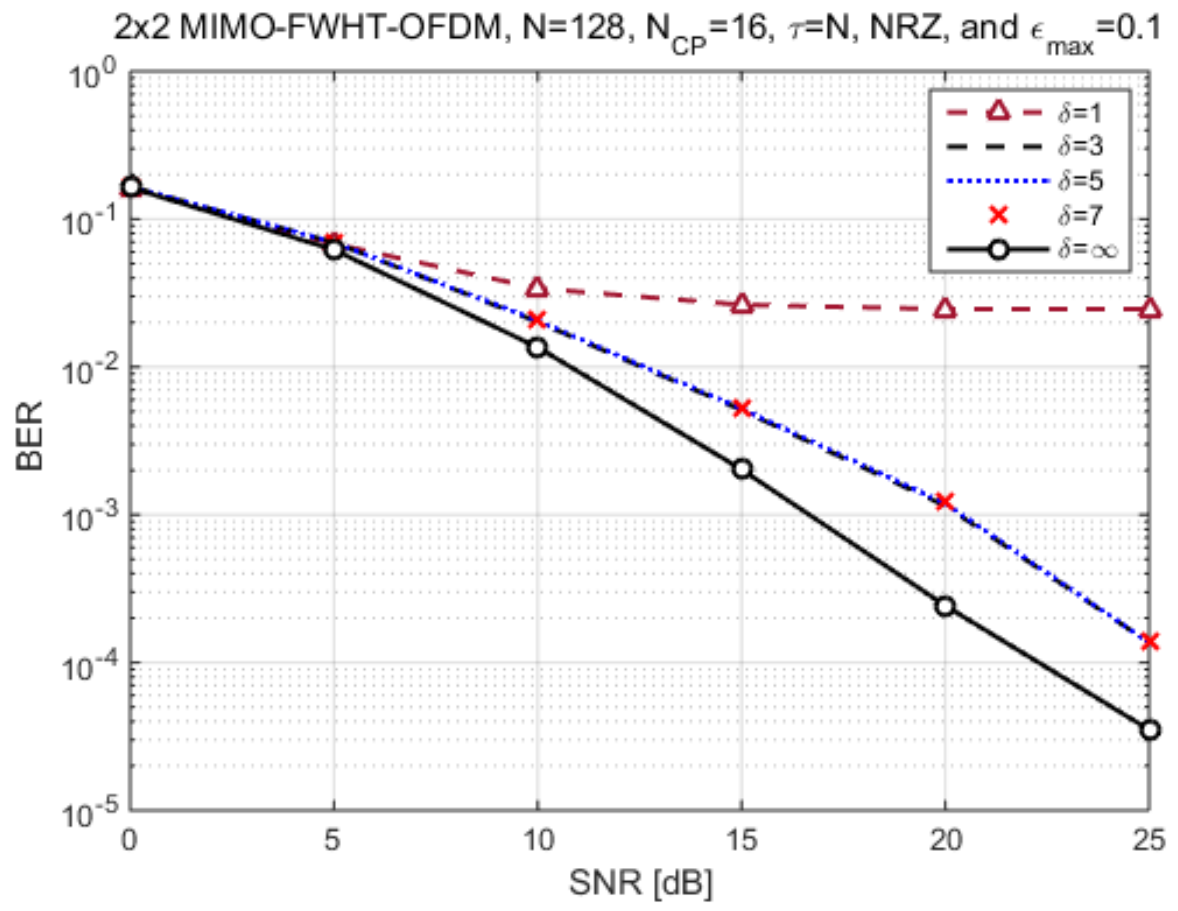

**Figure 5c**. The side view of Fig. 5a

## IV. The Computational Evaluation

This section is dedicated to calculating the computational complexity of various equalizers. The flop count is an important efficiency measure for algorithmic complexity [36]. Table1 lists the number of processes and flops for different arithmetic procedures. Table 2 tabulates the number of operations of different full matrices operations, while Table 3 tabulates the number of operations of different banded-matrices operations.

- *The LZF and LMMSE Equalizers*

The general solution matrix of the LZF equalizer for an $2^{\sigma} \times 2^{\sigma}$ MIMO-OFDM configuration can be expressed as [37]:

$$\boldsymbol{\mathcal{C}}_{\mathrm{LZF}} = (\boldsymbol{\pi}^{H}.\ \boldsymbol{\pi})^{-1}\,.\boldsymbol{\pi}^{H} \tag{40}$$

where $\boldsymbol{\pi} \in \mathbb{C}^{jN\times iN}$, referring to Eq. (40), the built of the LZF solution matrix needs matrix inversion plus two times complex matrix-by-matrix multiplication. This is followed by the multiplication of the constructed LZF solution matrix expressed in Eq. (40) is then multiplied by the received vector expressed in Eq. (16). This corresponds to $12(2^{\sigma}N)^{3} + 5(2^{\sigma}N)^{2} + 0.5(2^{\sigma}N)$ flops.

The general solution matrix of the LMMSE equalizer for an $2^{\sigma} \times 2^{\sigma}$ MIMO-OFDM configuration can be expressed as [37]:

$$\boldsymbol{\mathcal{C}}_{\mathrm{LMMSE}} = \left(\boldsymbol{\pi}^{H}.\ \boldsymbol{\pi} + \frac{\mathbf{R}_{\mathrm{z}}}{\sigma_{\mathbf{X}}^{2}}\mathbf{I}_{iN\times iN}\right)^{-1}.\boldsymbol{\pi}^{H} \tag{41}$$

where $\sigma_{\mathbf{X}}^{2}$ is the power of the transmitted signal, $\mathbf{R}_{\mathrm{z}} = \mathbb{E}\{\mathrm{z}\,.\,\mathrm{z}^{H}\}$ is the AWGN covariance matrix, and the term $\left(\frac{\sigma_{\mathbf{X}}^{2}}{\mathbf{R}_{\mathrm{z}}}\right)$ is the estimated value of the SNR. Comparing Eqs. (40), and (41), the two LMMSE equalizer needs the same number of flops as that of the LZF equalizer plus the term that includes the estimation of the SNR. Thus, the LMMSE equalizer needs approximately $12(2^{\sigma}N)^{3} + 5(2^{\sigma}N)^{2} + 1.5(2^{\sigma}N)$ flops, in addition to the estimation of the SNR.

- *The MMSE-SIC Equalizer*

According to the procedure steps listed in the previous section, the flops count for general $2^{\sigma} \times 2^{\sigma}$ MIMO-OFDM configuration in the case of MMSE-SIC equalizer can be summarized in the following pints:

1) The construction of the covariance matrix will be repeated $(2^{\sigma} - 1)$ times. According to Tables 1, and 2, the complex matrices multiplications are corresponding total flops number of:

$$\sum_{i=0}^{2^{\sigma}-2} \left((2^{\sigma} - i - 1) \times 4(2^{\sigma}N)^3 + 0.5(2^{\sigma}N)\right) \tag{42}$$

2) In the same manner, and according to Tables 1, and 2, the real/complex matrices additions are corresponding total flops number of:

$$\sum_{i=0}^{2^{\sigma}-1} \left((2^{\sigma} - i - 1) \times (2^{\sigma}N)^2 + 0.5(2^{\sigma}N)\right) \tag{43}$$

3) Each constructed matrix in the previous steps is powered to negative half. According to Tables 1, and 2, this is corresponding to flops amount of:

$$(2^{\sigma} - 1) \times \left(10(2^{\sigma}N)^3 + 3(2^{\sigma}N)^2 + 2.5(2^{\sigma}N)\right) \tag{44}$$

4) In the case of polar NRZ, the SIC process requires a total flop of:

$$(2^{\sigma} - 1) \times 1.5(2^{\sigma}N)^2 \tag{45}$$

5) Each covariance matrix is multiplied by the received vector/SIC output. The number of flops required for this step is:

$$(2^{\sigma} - 1) \times 4(2^{\sigma}N)^2 \tag{46}$$

6) Finally, the LMMSE equalizer will be built $2^{\sigma}$ times, this is corresponding to flops number of:

$$2^{\sigma} \times (12N^3 + 5N^2 + 1.5N) \tag{47}$$

- *The Proposed JLCOZF-SIC Equalizer*

According to the procedure steps listed in the previous section, the flops count for general $2^{\sigma} \times 2^{\sigma}$ MIMO-OFDM configuration in the case of JLCOZF-SIC equalizer can be summarized in the following pints:

1) The construction of the covariance matrix will be repeated $(2^{\sigma}-1)$ times. According to Tables 1, and 3, the complex matrices multiplications are corresponding total flops number of:

$$\sum_{i=0}^{2^{\sigma}-2}(2^{\sigma}-i-1)\times 2^{\sigma}N(16\tau^{2}+8\tau+4) \tag{48}$$

2) In the same manner, and according to Tables 1, and 3, the real/complex matrices additions are corresponding total flops number of:

$$\sum_{i=0}^{2^{\sigma}-1}(2^{\sigma}-i-1)\times 2^{\sigma}N(8\tau+4) \tag{49}$$

3) Each constructed matrix in the previous steps is powered to negative half. Two situations for this step are presented. For the case of $\delta=\infty$, this step will require the same flops of that of the MMSE-SIC equalizer as:

$$(2^{\sigma}-1)\times\left(10(2^{\sigma}N)^{3}+3(2^{\sigma}N)^{2}+2.5(2^{\sigma}N)\right) \tag{50}$$

In this case, let's check the shape of the covariance matrix for the case of $2\times 2$ MIMO-OFDM based on Walsh transform with $\delta=3$. Figure 6 shows the normalized amplitude for Eq. (37), with $\rho=0.5$, and $\delta=3$. This figure indicates the concept of the BMA can be also taken into account. But at least half of the sub-carriers must be taken into account. Thus, according to Tables 1, and 3 the construction of Eq. (37) requires approximately flops number of:

$$2^{\sigma}\times 2^{\sigma}N(32\tau^{2}+32\tau+16)+2^{\sigma}N \tag{51}$$

Let's now check the number of flops for $\delta=3$. Referring to Fig. 6, the constructed matrix $\left(\text{i.e.,}\ \left(\mathbf{K}_{\boldsymbol{N}_{\mathrm{j}}}\right)^{\frac{1}{2}}\right)$ is inverted, which consists of $2^{\sigma}\times 2^{\sigma}$ sub-matrices, that can be treated as a banded-matrices. In general, the inversion of $2^{\sigma}\times 2^{\sigma}$ matrix can be calculated as [38]:

$$
\boldsymbol{\mathcal{A}}^{-1}=\left[\begin{array}{cc}
\overbrace{\begin{bmatrix}
\boldsymbol{\mathcal{A}}_{1,1} & \boldsymbol{\mathcal{A}}_{1,2} & \cdots & \boldsymbol{\mathcal{A}}_{1,2^{\sigma-1}}\\
\boldsymbol{\mathcal{A}}_{2,1} & \boldsymbol{\mathcal{A}}_{2,2} & \cdots & \boldsymbol{\mathcal{A}}_{2,2^{\sigma-1}}\\
\vdots & \vdots & \ddots & \vdots\\
\boldsymbol{\mathcal{A}}_{2^{\sigma-1},1} & \boldsymbol{\mathcal{A}}_{2^{\sigma-1},2} & \cdots & \boldsymbol{\mathcal{A}}_{2^{\sigma-1},2^{\sigma-1}}
\end{bmatrix}}^{\mathcal{A}_1} &
\overbrace{\begin{bmatrix}
\boldsymbol{\mathcal{A}}_{1,2^{\sigma-1}+1} & \boldsymbol{\mathcal{A}}_{1,2^{\sigma-1}+2} & \cdots & \boldsymbol{\mathcal{A}}_{1,2^{\sigma}}\\
\boldsymbol{\mathcal{A}}_{2,2^{\sigma-1}+1} & \boldsymbol{\mathcal{A}}_{2,2^{\sigma-1}+2} & \cdots & \boldsymbol{\mathcal{A}}_{2,2^{\sigma}}\\
\vdots & \vdots & \ddots & \vdots\\
\boldsymbol{\mathcal{A}}_{2^{\sigma-1},2^{\sigma-1}+1} & \mathbf{H}_{2^{\sigma-1},2^{\sigma-1}+2} & \cdots & \mathbf{H}_{2^{\sigma-1},2^{\sigma}}
\end{bmatrix}}^{\mathcal{A}_2}\\
\underbrace{\begin{bmatrix}
\boldsymbol{\mathcal{A}}_{2^{\sigma-1}+1,1} & \boldsymbol{\mathcal{A}}_{2^{\sigma-1}+1,2} & \cdots & \boldsymbol{\mathcal{A}}_{2^{\sigma-1}+1,2^{\sigma-1}}\\
\boldsymbol{\mathcal{A}}_{2^{\sigma-1}+2,1} & \boldsymbol{\mathcal{A}}_{2^{\sigma-1}+2,2} & \cdots & \boldsymbol{\mathcal{A}}_{2^{\sigma-1}+2,2^{\sigma-1}}\\
\vdots & \vdots & \ddots & \vdots\\
\boldsymbol{\mathcal{A}}_{2^{\sigma},1} & \boldsymbol{\mathcal{A}}_{2^{\sigma},2} & \cdots & \boldsymbol{\mathcal{A}}_{2^{\sigma},2^{\sigma-1}}
\end{bmatrix}}_{\mathcal{A}_3} &
\underbrace{\begin{bmatrix}
\boldsymbol{\mathcal{A}}_{2^{\sigma-1}+1,2^{\sigma-1}+1} & \boldsymbol{\mathcal{A}}_{2^{\sigma-1}+1,2^{\sigma-1}+2} & \cdots & \boldsymbol{\mathcal{A}}_{2^{\sigma-1}+1,2^{\sigma}}\\
\boldsymbol{\mathcal{A}}_{2^{\sigma-1}+2,2^{\sigma-1}+1} & \boldsymbol{\mathcal{A}}_{2^{\sigma-1}+2,2^{\sigma-1}+2} & \cdots & \boldsymbol{\mathcal{A}}_{2^{\sigma-1}+2,2^{\sigma}}\\
\vdots & \vdots & \ddots & \vdots\\
\boldsymbol{\mathcal{A}}_{2^{\sigma},2^{\sigma-1}+1} & \boldsymbol{\mathcal{A}}_{2^{\sigma},2^{\sigma-1}+2} & \cdots & \boldsymbol{\mathcal{A}}_{2^{\sigma},2^{\sigma}}
\end{bmatrix}}_{\mathcal{A}_4}
\end{array}\right]^{-1}
$$
$$
=\begin{bmatrix}\overline{\boldsymbol{\mathcal{A}}}_1 & \overline{\boldsymbol{\mathcal{A}}}_2\\ \overline{\boldsymbol{\mathcal{A}}}_3 & \overline{\boldsymbol{\mathcal{A}}}_4\end{bmatrix}=\begin{bmatrix}\boldsymbol{\Phi} & -\boldsymbol{\Phi}\boldsymbol{\mathcal{A}}_2{\boldsymbol{\mathcal{A}}_4}^{-1}\\ -{\boldsymbol{\mathcal{A}}_4}^{-1}\boldsymbol{\mathcal{A}}_3\boldsymbol{\Phi} & {\boldsymbol{\mathcal{A}}_4}^{-1}+{\boldsymbol{\mathcal{A}}_4}^{-1}\boldsymbol{\mathcal{A}}_3\boldsymbol{\Phi}\boldsymbol{\mathcal{A}}_2{\boldsymbol{\mathcal{A}}_4}^{-1}\end{bmatrix} \tag{52}
$$

where $\boldsymbol{\Phi}=\left(\boldsymbol{\mathcal{A}}_1-\boldsymbol{\mathcal{A}}_2{\boldsymbol{\mathcal{A}}_4}^{-1}\boldsymbol{\mathcal{A}}_3\right)^{-1}$. The $\overline{\boldsymbol{\mathcal{A}}}_1$ matrix construction needs two times of complex matrix-by-matrix multiplication, matrix-by-matrix subtraction, and finally, matrix inversion. According to Tables 1, and 3, this is corresponding to $2^{\sigma}N(32\tau^2+16\tau+8)+2^{\sigma}N(8\tau+4)$ flops plus two times of matrix inversion for complex matrix of dimension $2^{\sigma-1}\times 2^{\sigma-1}$. Both of $\overline{\boldsymbol{\mathcal{A}}}_2$ and $\overline{\boldsymbol{\mathcal{A}}}_3$ matrices construction needs approximately $2^{\sigma}N(32\tau^2+16\tau+8)$ flops. Finally, the construction of the $\overline{\boldsymbol{\mathcal{A}}}_4$ matrix requires approximately $2^{\sigma}N(48\tau^2+24\tau+12)+2^{\sigma}N(8\tau+4)$ flops. Thus, the matrix inversion in the form of Eq. (52) needs approximately flops number of:

$$2^{\sigma}N(144\tau^2+88\tau+44) \tag{53}$$

Besides Eq. (53) a two times of matrix inversion for complex matrix of dimension $2^{\sigma-1}\times 2^{\sigma-1}$ is required.

4) In the case of polar NRZ, the SIC process requires a total flop of:

$$(2^{\sigma}-1)\times\left(2^{\sigma}N(8\tau+4)\right) \tag{54}$$

5) Each covariance matrix is multiplied by the received vector/SIC output. The number of flops required for this step is:

$$(2^{\sigma}-1)\times 2^{\sigma}N(8\tau+4) \tag{55}$$

6) Finally, the JLCOLZF equalizer will be built $2^{\sigma}$ times, this is corresponding to flops number of:

$$2^{\sigma}\times(37\tau^2+26.5\tau+9.5) \tag{56}$$

Figure 7a shows the number of flops versus the configuration order of different equalizers with respect to the proposed JLCOZF-SIC equalizer at two situations. It is clear that the proposed equalizer with $\delta = 3$, and $\tau = N/2$ give the better performance in terms of number of flops than all other schemes, but a degradation in the BER performance will occur as depicted in Fig. 5c. While, the proposed equalizer with $\delta = \infty$, and $\tau = N/2$ still gives better performance in terms of number of flops than that of MMSE-SIC. This like a trade-off between the BER performance and number of flops. Note that any two multiplications of complex matrices will be limited to bandwidth of $\tau$ over all steps. Now let's define the computational complexity efficiency as:

$$\Gamma\% = \frac{\mathcal{N}_2 - \mathcal{N}_1}{\mathcal{N}_1} \times 100 \tag{57}$$

where $\mathcal{N}_1$ is the flops number of the proposed equalizer, and $\mathcal{N}_2$ is the flops number of the compared equalizer. It should note that negative values of Eq. (58) means that the compared equalizer takes lower number of flops compared to that of the proposed equalizer. Also, the proposed equalizer is compared at two cases; $\delta = \infty$, and $\delta = 3$, with $\tau = 80$.

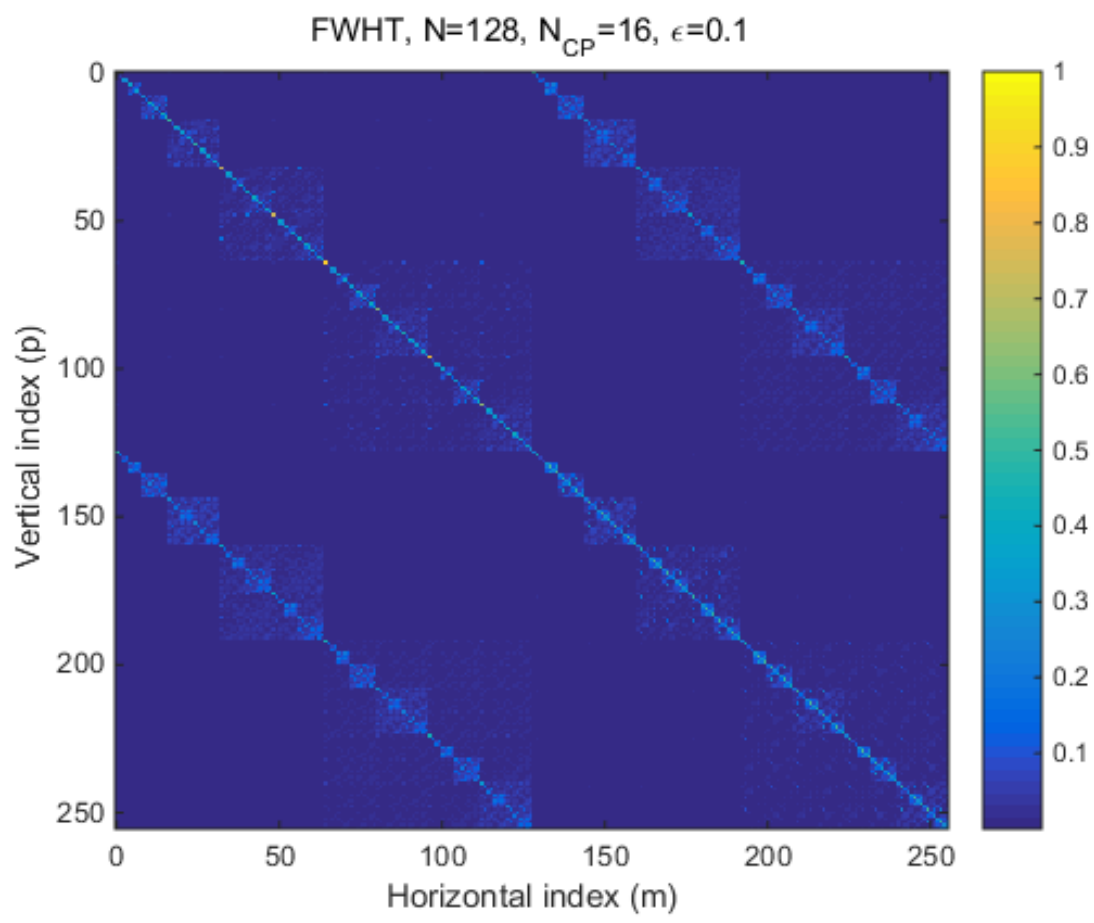

**Figure 6**. The normalized magnitude of the covariance matrix vs. sub-carrier indices using for $\rho = 0.5$, and $\delta = 3$.

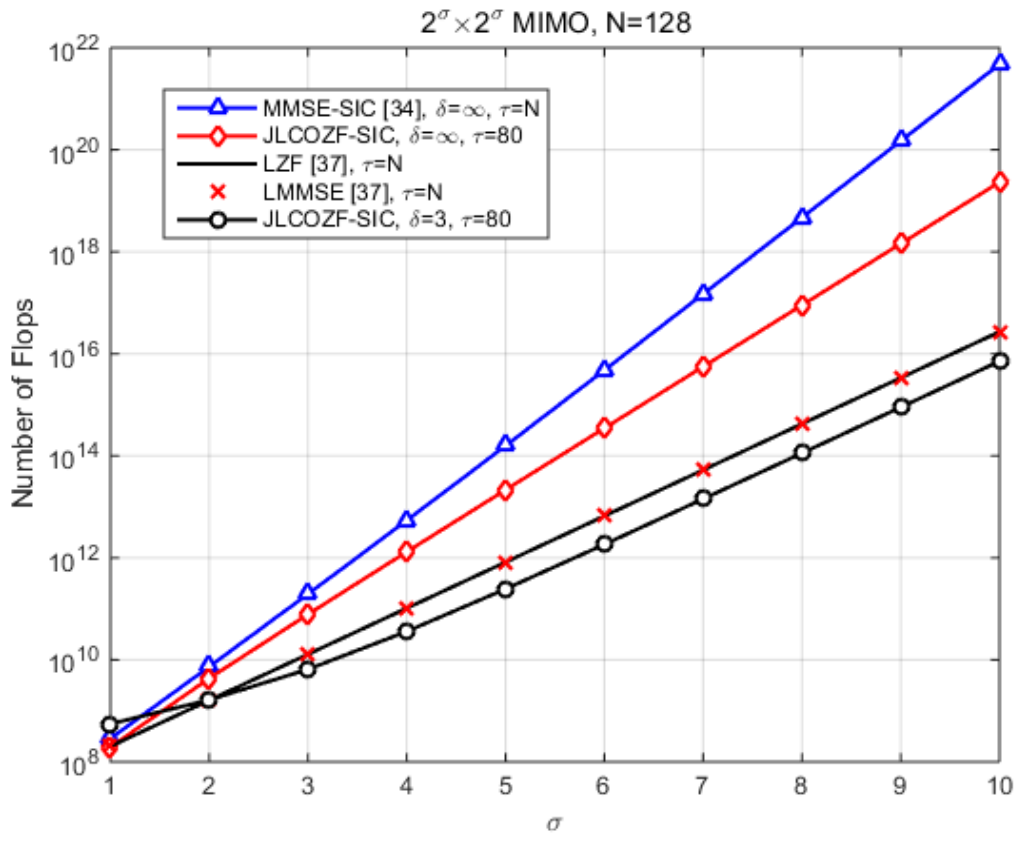

**Figure 7**a. The number of flops versus the configuration orders of different equalizers.

Figure 7b shows the computational complexity efficiency of the proposed equalizer with respect to LZF and LMMSE equalizers. It is clear that in the case of $\delta = 3$, the proposed equalizer offers a reduction in the computational complexity compared to that of the linear schemes as the MIMO configuration order increases. On the other hand, for the case of $\delta = \infty$, the proposed equalizer consumes more flops than that of the linear schemes as the MIMO configuration order increased. In the same manner, Figure 7c shows the computational complexity efficiency of the proposed equalizer with respect to MMSE-SIC equalizers for the two cases of $\delta = \infty$ and, 3. It is clear that the proposed equalizer offers a significant reduction in computational complexity compared to that of the MMSE-SIC equalizer as the MIMO configuration order increases.

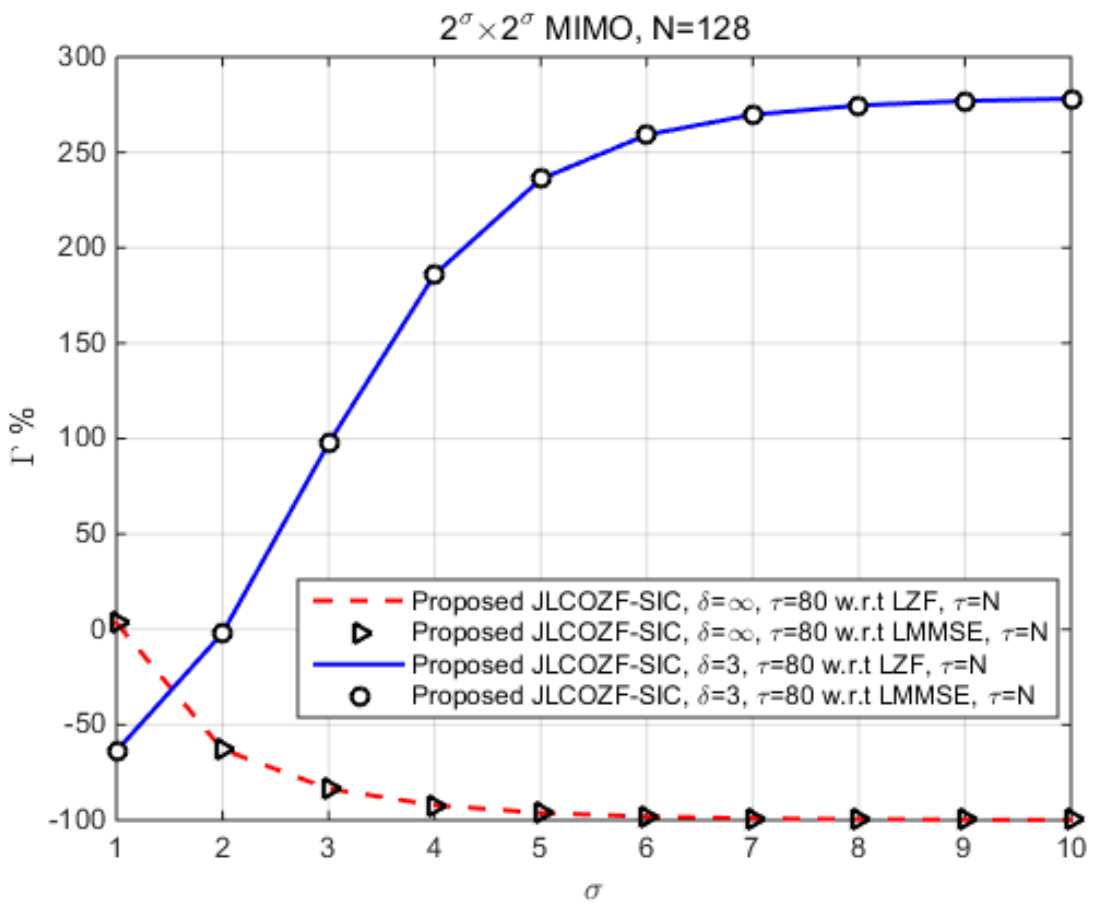

**Figure 7b**. The computational complexity efficiency of the proposed JLZOZF-SIC equalizer w.r.t linear equalizers

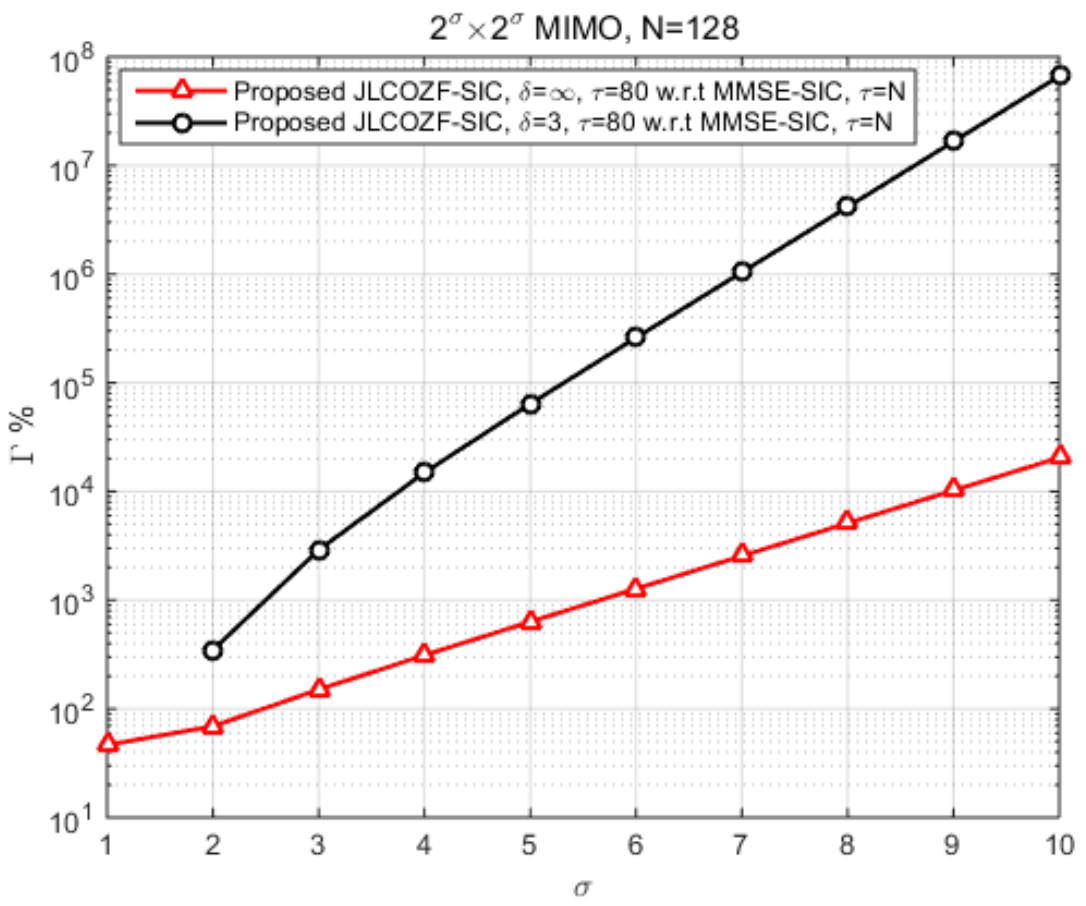

**Figure 7c**. The computational complexity efficiency of the proposed JLZOZF-SIC equalizer w.r.t MMSE-SIC equalizer

**Table 1**. The number of flops of different arithmetic operations [36].

| | Number of | | | | |
|---|---|---|---|---|---|
| Process | × | ± | ÷ | Operations | Flops |
| $a \pm b$ | 0 | 1 | 0 | 1 | 0.5 |
| $a \times b$ | 1 | 0 | 0 | 1 | 0.5 |
| $a \div b$ | 0 | 0 | 1 | 1 | 0.5 |
| $a + (b \pm jc)$ | 1 | 1 | 0 | 1 | 0.5 |
| $a \times (b \pm jc)$ | 2 | 0 | 0 | 2 | 1 |
| $(a \pm jb) \pm (c \pm jd)$ | 0 | 2 | 0 | 2 | 1 |
| $(a \pm jb) \times (c \pm jd)$ | 4 | 2 | 0 | 6 | 3 |
| $1 \div (a \pm jb)$ | 0 | 0 | 1 | 2 | 1 |

**Table 2**. The number of operations of different full matrices operations

| Process | Description | Full-matrix computation | | | | | | |
|---|---|---|---|---|---|---|---|---|
| | | No. of complex | | No. of real-complex | | | | Operations |
| | | × | ∓ | × | ∓ | ÷ | √ | |
| $\mathcal{A} \mp \mathcal{B}$ | $\mathcal{A}, \mathcal{B} \in \mathbb{C}^{N \times N}$ | --- | $2N^2$ | --- | --- | --- | --- | $2N^2$ |
| $\mathcal{A}.\mathcal{B}$ | | $N^3$ | $N^3$ | --- | --- | --- | --- | $8N^3$ |
| $\mathcal{A}^{-1}$ | | $N^3$ | $N^3$ | $N^2$ | --- | $N$ | -- | $8N^3 + 2N^2 + N$ |
| $\mathcal{A}^{-1/2}$ | | $2N^3$ | $2N^3$ | $N^2$ | $2N^2$ | $2N$ | $N$ | $20N^3 + 6N^2 + 5N$ |
| $\mathcal{A}.\mathcal{U}$ | $\mathcal{U} \in \mathbb{C}^{N \times 1}$ | $N^2$ | $N^2$ | --- | --- | --- | --- | $8N^2$ |

**Table 3**. The number of operations of different banded-matrices operations

| Process | Description | Banded-matrix computation | | | | | | |
|---|---|---|---|---|---|---|---|---|
| | | Number of | | | | | | Operations |
| | | Real | Complex | | | Real-complex | | |
| | | × | × | ∓ | ÷ | × | ∓ | |
| $\mathcal{A} \mp \mathcal{B}$ | $\mathcal{A}, \mathcal{B} \in \mathbb{C}^{N \times N}$ | -- | $N(2\tau+1)$ | $N(2\tau+1)$ | -- | -- | -- | $N(16\tau+8)$ |
| $\mathcal{A}.\mathcal{B}$ | | -- | $N(2\tau+1)^2$ | $N(2\tau+1)^2$ | -- | -- | -- | $N(32\tau^2+16\tau+8)$ |
| $\mathcal{A}^{-1}$ | | -- | $N(\tau^2+2\tau)$ | $N(\tau^2+2\tau)$ | $N(\tau+1)$ | -- | -- | $N(10\tau^2+21\tau+1)$ |
| $\mathcal{A}.\mathcal{U}$ | $\mathcal{U} \in \mathbb{C}^{N \times 1}$ | -- | $N(12\tau+6)$ | $N(4\tau+2)$ | -- | -- | -- | $N(16\tau+8)$ |

Table 4 illustrates the number of flops needed for various equalizers. This is followed by an assessment of the efficacy of the flops number expressed in Eq. (57) of different equalizers with regard to the proposed JLCOZF-SIC equalizer for $\delta = \infty, \tau = 80$. Because the proposed equalizer in this instance employs all terms without approximation (i.e., $\delta = \infty$), it outperforms the MMSE-SIC equalizer in terms of complexity reduction. On the other hand, the linear equalizers have a negative efficiency with regard to the proposed equalizer, which means they have fewer failures than the proposed equalizer.

**Table 4**. The number of flops and efficiency of different equalizer schemes for $2^\sigma \times 2^\sigma$MIMO configuration with respect to proposed JLCOZF-SIC, $\delta = \infty, \tau = 80$.

| Equalizer type | $\sigma = 2$ | Γ% | $\sigma = 4$ | Γ% | $\sigma = 6$ | Γ% |
|---|---|---|---|---|---|---|
| LZF [37], $\tau = 128$ | $1.6 \times 10^9$ | -62.9 | $1.0 \times 10^{11}$ | -92.2 | $3.5 \times 10^{14}$ | -98.1 |
| LMMSE [37], $\tau = 128$ | $1.6 \times 10^9$ | -62.9 | $1.0 \times 10^{11}$ | -92.2 | $6.6 \times 10^{12}$ | -98.1 |
| MMSE-SIC [34], $\tau = 128$ | $7.4 \times 10^9$ | 69.2 | $5.4 \times 10^{12}$ | 311.9 | $3.8 \times 10^{15}$ | 1273 |
| Proposed JLCOZF-SIC, $\delta = \infty, \tau = 80$ | $4.3 \times 10^9$ | -- | $1.3 \times 10^{12}$ | -- | $3.5 \times 10^{14}$ | -- |

In the same manner, Table 5 illustrates the number of flops needed for various equalizers. This is followed by an assessment of the efficacy of the flops number expressed in Eq. (57) of different equalizers with regard to the proposed JLCOZF-SIC equalizer for $\delta = 3$, $\tau = 80$. Because the proposed equalizer in this instance employs only three terms (i.e., $\delta = 3$), it outperforms all the equalizers in terms of complexity reduction. On the other hand, the linear equalizers have a negative efficiency with regard to the proposed equalizer only for a 2×2 MIMO channel configuration. As the configuration order increased, as the computational complexity and efficiency increased. Referring to Figs. 5c, 7a, Tables 5, and 6, there is a noticeable trade-off between the computational complexity and the BER performance.

**Table 5**. The number of flops and efficiency of different equalizer schemes for $2^{\sigma} \times 2^{\sigma}$MIMO configuration with respect to proposed JLCOZF-SIC, $\delta = 3, \tau = 80$.

| Equalizer type | $\sigma = 2$ | Γ% | $\sigma = 4$ | Γ% | $\sigma = 6$ | Γ% |
|---|---|---|---|---|---|---|
| LZF [37], $\tau = 128$ | $1.6 \times 10^{9}$ | -1.9 | $1.0 \times 10^{11}$ | 186 | $3.5 \times 10^{14}$ | 259.2 |
| LMMSE [37], $\tau = 128$ | $1.6 \times 10^{9}$ | -1.9 | $1.0 \times 10^{11}$ | 186 | $6.6 \times 10^{12}$ | 259.2 |
| MMSE-SIC [34], $\tau = 128$ | $7.4 \times 10^{9}$ | 347.7 | $5.4 \times 10^{12}$ | 14915 | $3.8 \times 10^{15}$ | 26014 |
| Proposed JLCOZF-SIC, $\delta = 3, \tau = 80$ | $1.6 \times 10^{9}$ | -- | $3.6 \times 10^{10}$ | -- | $1.8 \times 10^{12}$ | -- |

## V. The Simulation Results and Analysis

This section is devoted to a comparison study of the proposed equalizer with respect to various equalizers. The comparison is based on studies from different perspectives, as shown in the following paragraphs. According to the general section model presented in section II, the mathematical derivation studies in section III, and the simulation parameters tabulated in Table 6.

**Table 6**. The simulation parameters

| Parameter | Value | parameter | Value |
|---|---|---|---|
| The FFT/IFFT size | 128 | SNR range | 0:5:25 |
| The IFWHT/FWHT size | 128 | Power delay profile type | Vehicular A |
| User equipment velocity | 120 km/h | Average power [dB] | 0, -1, -9, -10, -15, and -20 |
| Number of channel paths | 6 | Type of co-CFO distribution | Uniform |
| Chip duration | 24.4 μ sec | Modulation type | BPSK |
| Carrier frequency | 2 GHz | Noise type | AWGN |
| CP length | 64 | Channel configuration type | 2×2, and $2^{\sigma} \times 2^{\sigma}$ |
| Data type | Polar NRZ | Data transmission type | Spatial Multiplexing |
| $\varepsilon_{max}$ | 0.1 | Number of channel realization | 104 |
| Channel model | Monto Carlo | Channel type | Jake's model [30] |
| Number of iterations | $10^{4}$ | | |

Firstly, we should specify the optimized values of the proposed equalizer. These optimized parameters are the real and imaginary parts of the optimized parameter (i.e., ξ), and the banded-matrix bandwidth (i.e., $\tau$). Figure 8a shows the variation of the real part of the optimization parameter (i.e., $\Re(\xi)$) as a function of the BER performance, while Fig. 8b shows the elevation view of Fig. 8a. The $\Re(\xi)$ parameter is assumed to change and takes positive and negative values. At each of these values, the BER performance is tested. Figure 8 shows that the minimum BER performance can be achieved in one of the situations as $\Re(\xi) = \pm 10^{-2}$. Let's select that $\Re(\xi) = 10^{-2}$. However, we should test the value of the $\Re(\xi)$ as variation of $\Re(\xi) = 10^{-2} \rightarrow 0.1$. Figure 9a shows the variation of the real part of the optimization parameter within the range of $\Re(\xi) = 10^{-2} \rightarrow 0.1$ as a function of the BER performance. It is clear that the minimum BER performance can be achieved as $\Re(\xi) = 0.01$, and $0.02$. So, let's check the BER performance versus the SNR at these values as depicted in Fig. 9b. These values are closely matched over the variation range of the SNR from 0 up to 25 dB with a small deviation. In the same manner as 0≤SNR≤23 dB, the case of $\Re(\xi) = 0.01$ superior the case of that $\Re(\xi) = 0.02$. At BER=$2\times10^{-4}$, the case of $\Re(\xi) = 0.01$ gives an SNR gain of 1.1 dB compared than that of $\Re(\xi) = 0.02$. Thus, select of $\Re(\xi) = 0.01$ for the rest of the simulation.

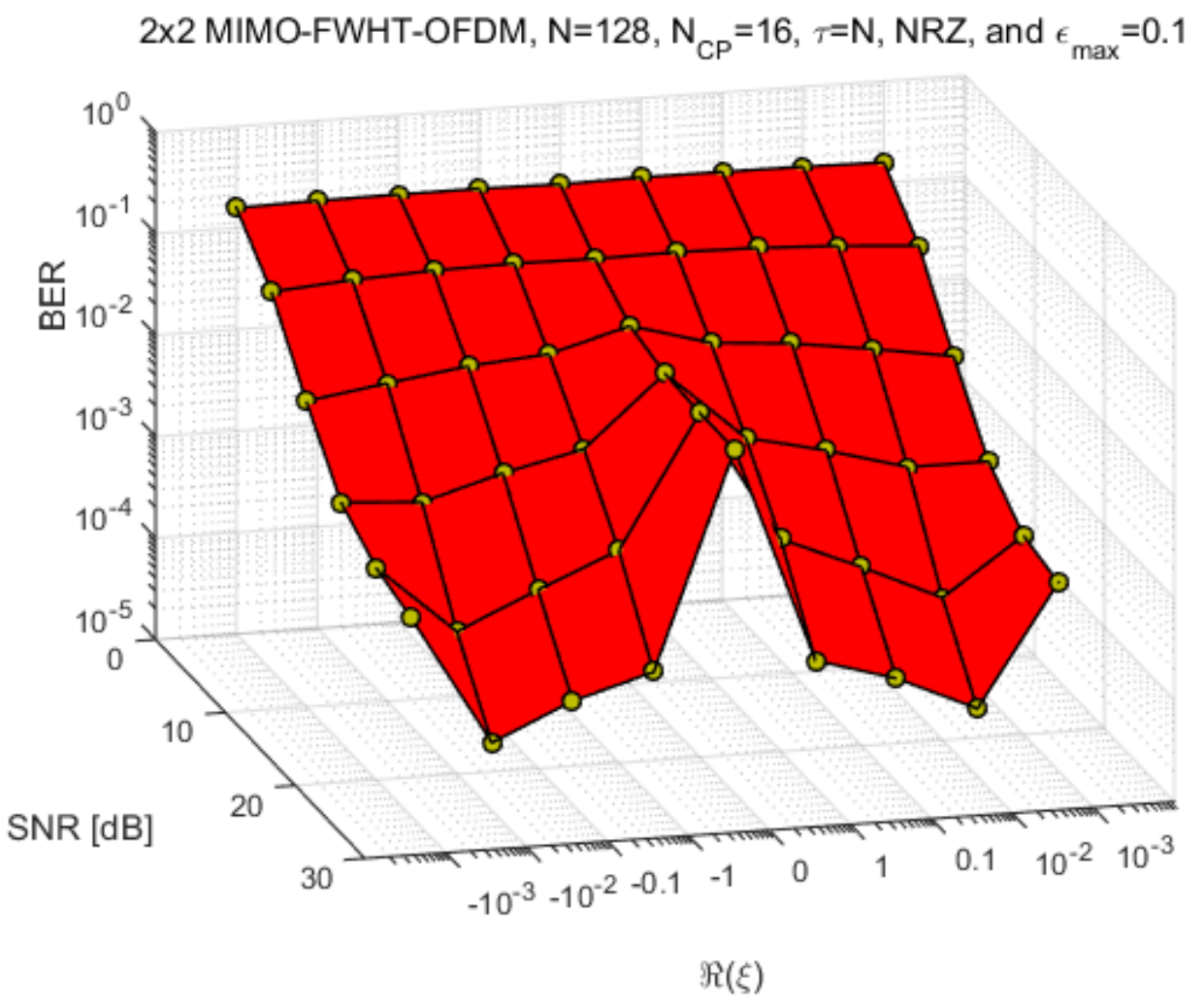

**Figure 8a**. The BER vs. the real part of the optimized parameter at different SNR values.

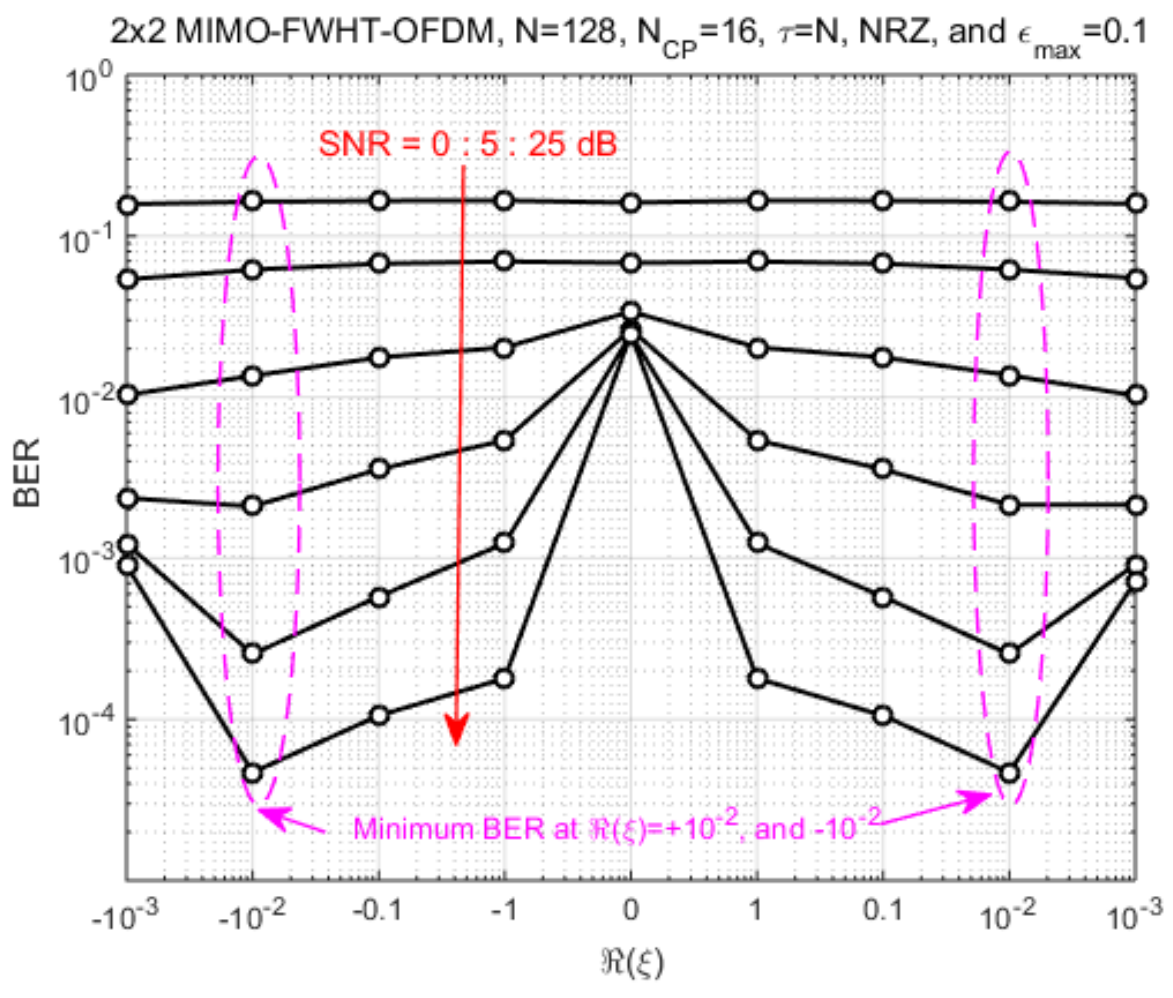

**Figure 8b**. The elevation view of Fig. 8a.

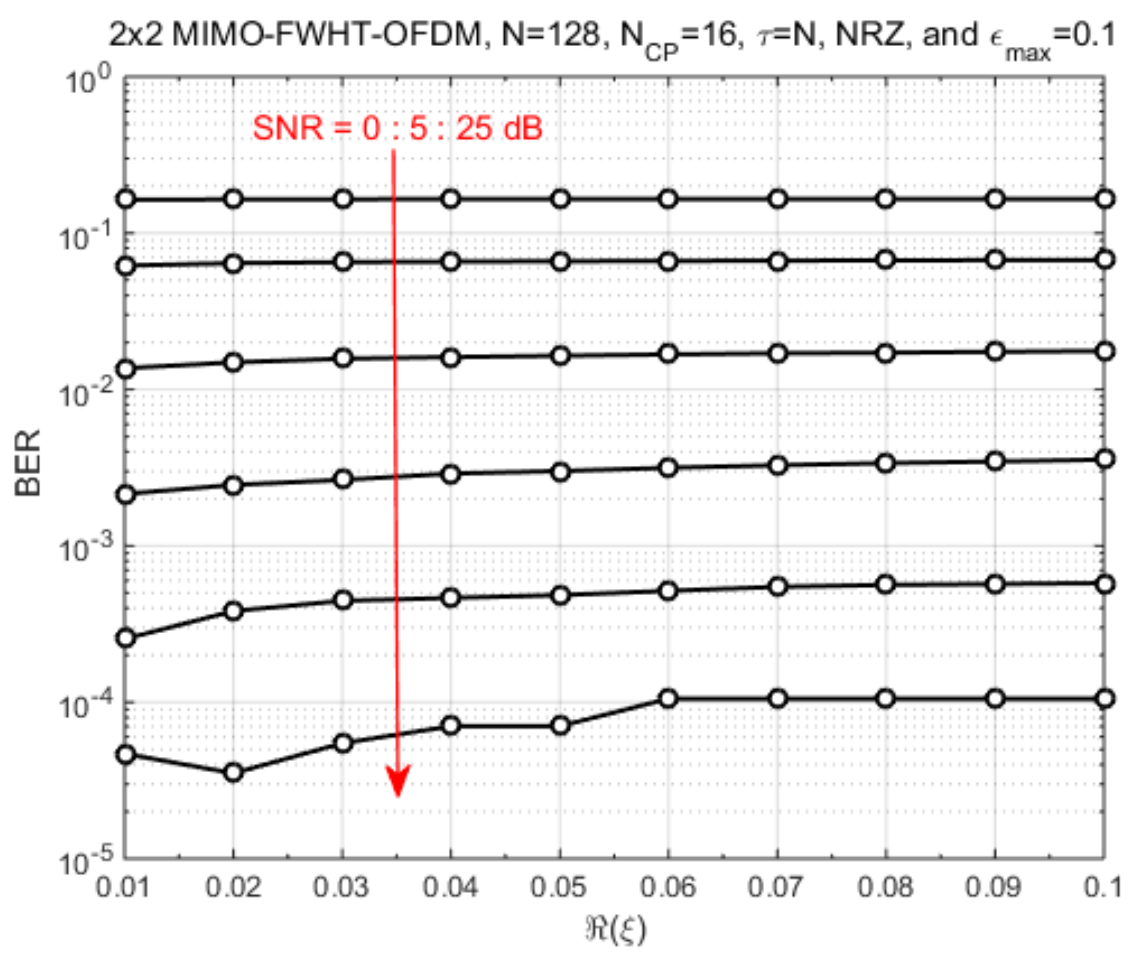

**Figure 9a**. The BER vs. the real part of the optimized parameter at different SNR values with the range of $\Re(\xi) = 10^{-2} \rightarrow 0.1$.

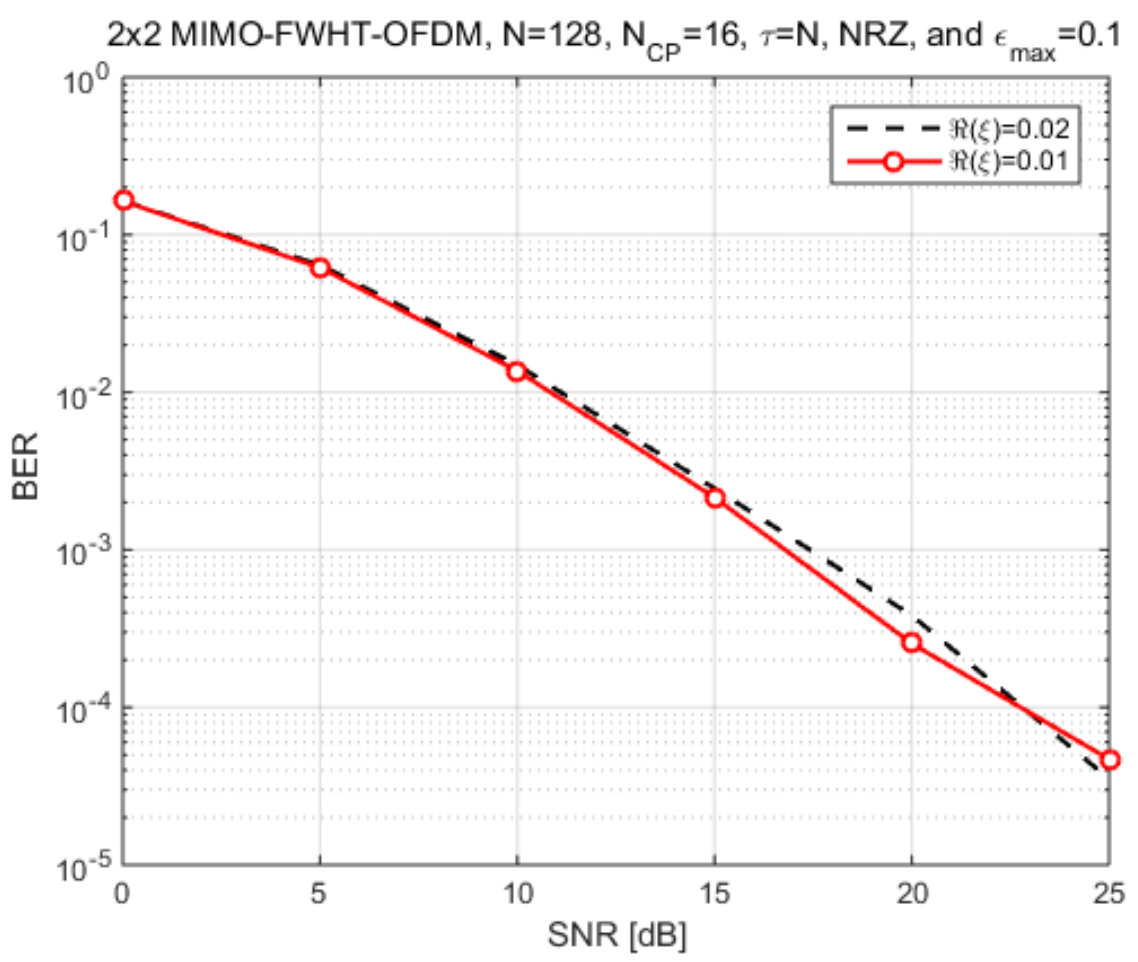

**Figure 9b**. The BER vs. the SNR at different values of $\Re(\xi)$.

Figure 10a shows the variation of the imaginary part of the optimization parameter $\left(\text{i.e.,}\ \Im(\xi)\right)$ as a function of the BER performance, while Fig. 10b shows the elevation view of Fig. 10a. The $\Im(\xi)$ parameter is assumed to change and takes positive and negative values. At each of these values, the BER performance is tested. Figure 10 shows that the minimum BER performance can be achieved at $\Im(\xi) = 1$. However, we should test the value of the $\Im(\xi)$ as variation of $\Im(\xi) = 0 \rightarrow 1$. Figure 11a shows the variation of the $\Im(\xi)$ within the range of $\Im(\xi) = 0 \rightarrow 1$ as a function of the BER performance. Figure 11b shows the BER performance versus the SNR $\Im(\xi) = 0.4$, and 1. Thus, select $\Im(\xi) = 1$ for the rest of the simulation. Hence, $\xi = \Re(\xi) + j\Im(\xi) = 0.01 + j$.

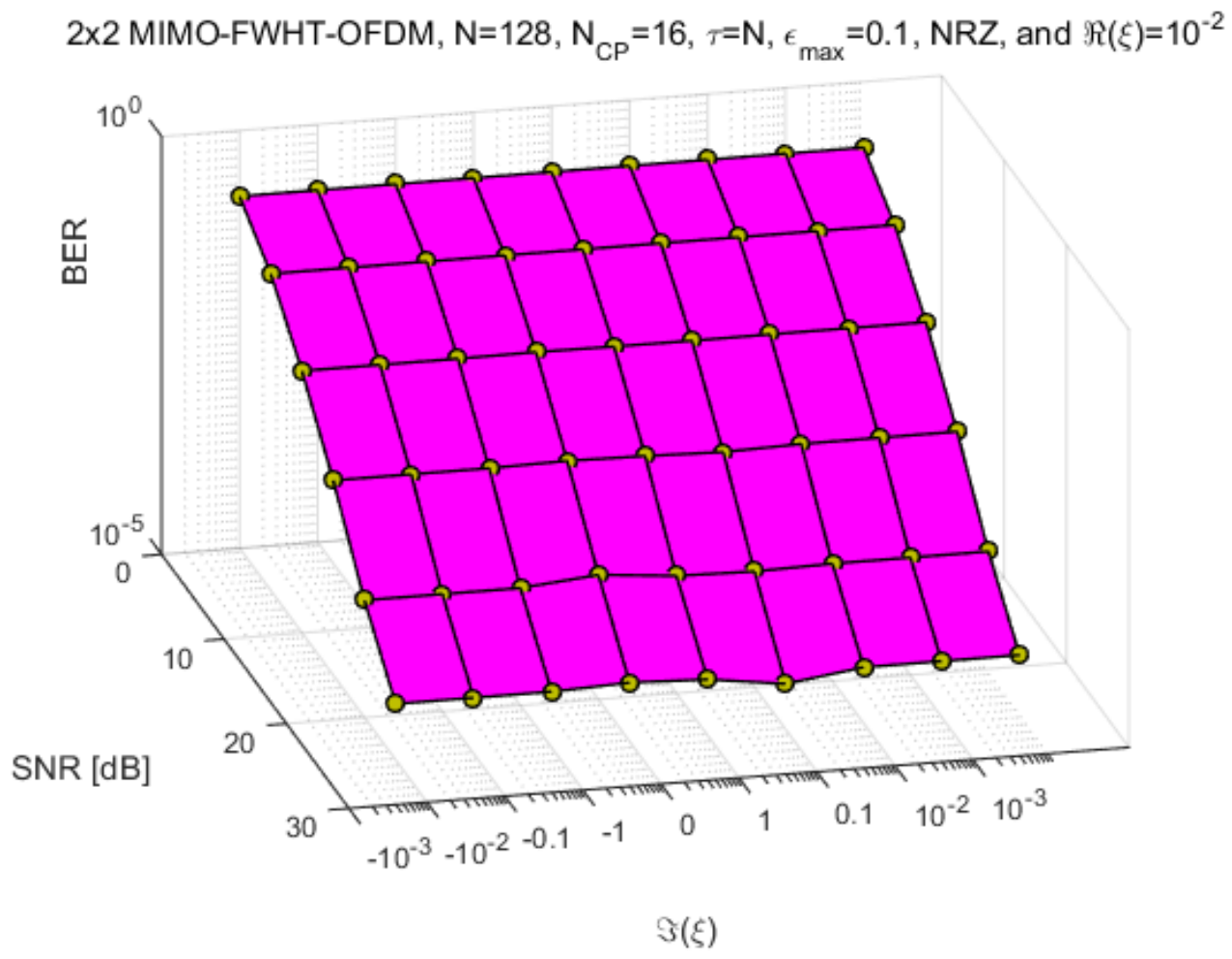

**Figure 10a**. The BER vs. the real part of the optimized parameter at different SNR values.

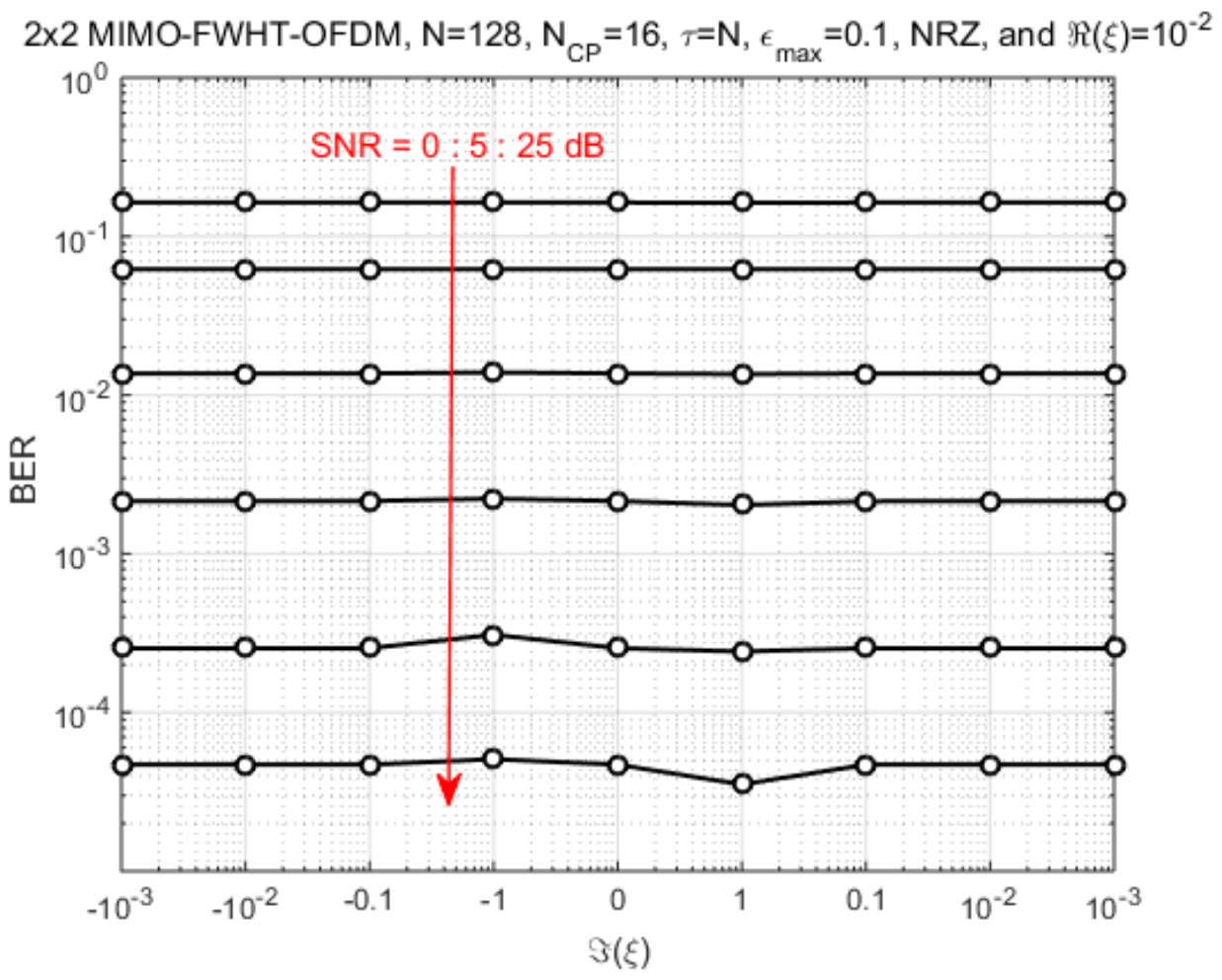

**Figure 10b**. The elevation view of Fig. 10a.

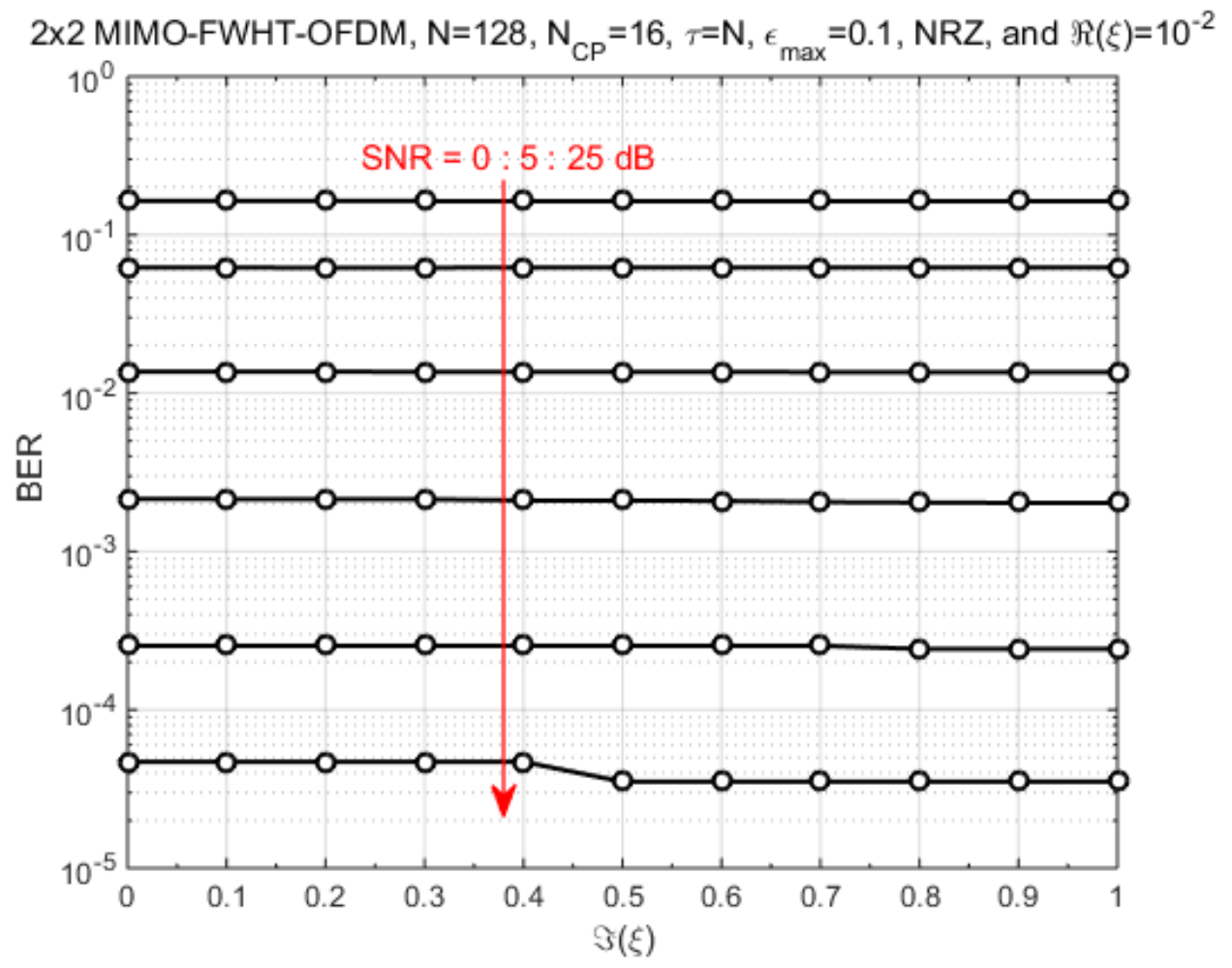

**Figure 11a**. The BER vs. the imaginary part of the optimized parameter at different SNR values with the range of $\Im(\xi) = 0.1 \rightarrow 1$.

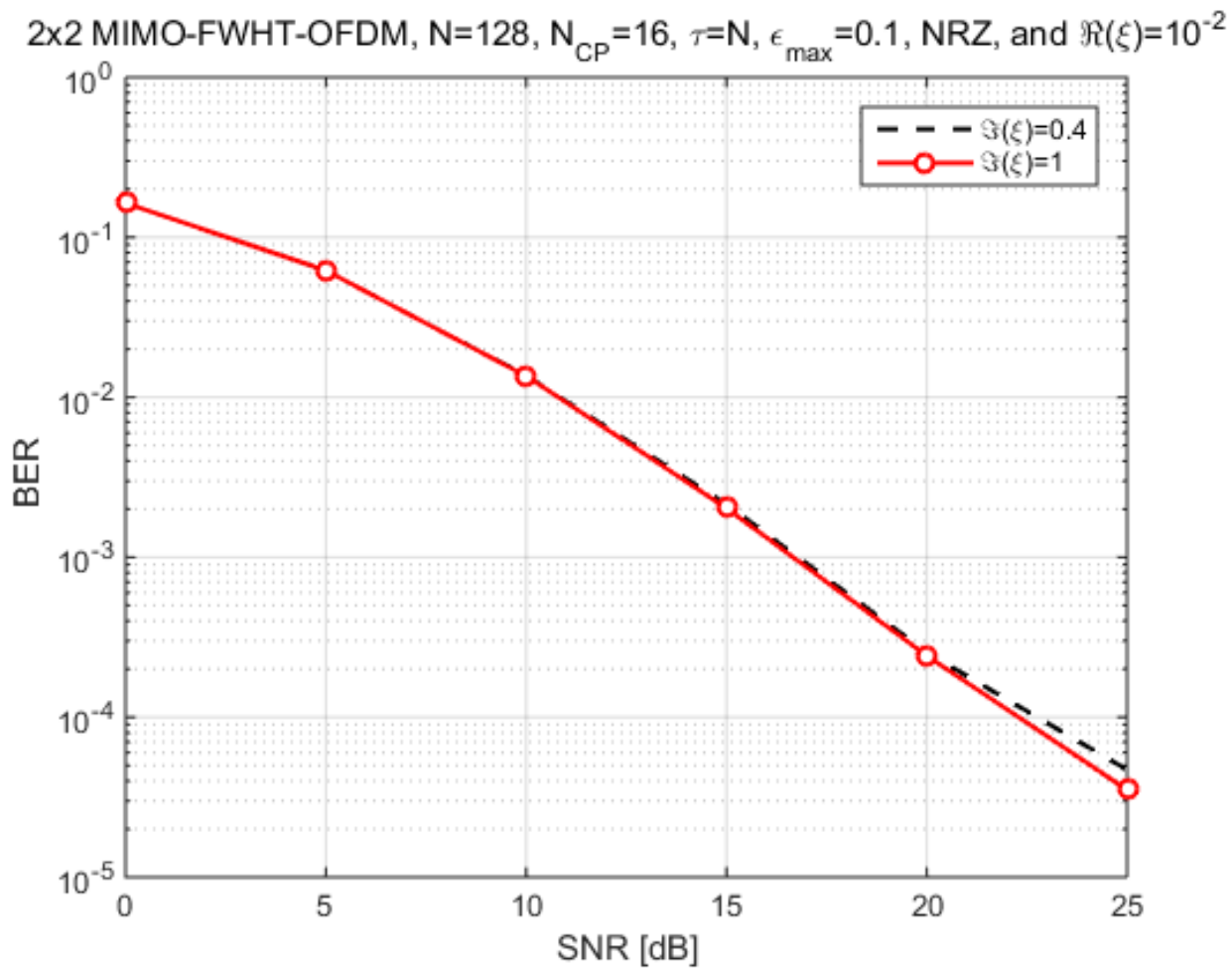

**Figure 11b**. The BER vs. the SNR at different values of $\Im(\xi)$.

The completing parameter of the proposed equalizer that must be stated is the number of sub-carriers considered to correspond to each diagonal element. Based on previous results regarding the intensity of different matrices in the case of the Walsh transform, we concluded that the banded-matrix can be considered with a minimum bandwidth of $N/2$ to guarantee that the BER performance is maintained. Figure 12a depicts the BER performance versus SNR under various correction situations. Furthermore, it is obvious that the BER performance suffers greatly as the compensation bandwidth below $N/2$. Figure 12b shows an elevated view of Fig. 12a, which shows the BER versus adjusted bandwidth at various SNR values. Figure 12c depicts a side view of Fig. 12a, which depicts BER performance versus SNR at various bandwidth-adjusted values.

Figure 12 shows that the OFDM based on the Walsh transform can use the idea of BMA and lowers processing complexity even when more than half of the adjacent elements corresponding to the diagonal are considered. However, at BER=$10^{-4}$ and with regard to complete matrix compensation (i.e., τ=128), the SNR degrades by about 0.3, 0.45, and 0.46 dB for the compensation instances of τ=80, 70, and 64, respectively (See Fig. 12c). As a result, for the remainder of the simulation, pick τ=80 and ensure that aboutness optimizes the system.

Figure 13a depicts the BER performance versus both compensation case variations and normalized CFO via various equalizers in comparison to the proposed JLCOZF-SIC equalizer. The proposed equalizer clearly performs well in terms of the variation of the normalized CFO. However, the adjusted bandwidth must be greater than $N/2$ in order to ensure the stability of the BER performance with respect to SNR variation, as mentioned in Fig. 12. According to Fig.12c, Table 7 tabulates the extra SNR of various compensation situations in comparison to that of the complete matrix computation at two BER levels (i.e., BER=$10^{-3}$ and $10^{-4}$). The BER can be conserved in the case of BMA compensation for $\tau > N/2$, as it requires an additional SNR of about 0.16 and 0.30 dB at BER=$10^{-3}$, and $10^{-4}$, respectively, to accomplish the same performance as the full compensation situation. On the other hand, as the compensation situation decreases, the required SNR increases for the same BER performance. The idea of a trade-off between BER efficiency and computational complexity is indicated.

**Table 7**. The additional SNR at different compensation scenarios with respect to full compensation case

| | BER=$10^{-3}$ | | BER=$10^{-4}$ | |
|---|---|---|---|---|
| Compensation scenario | SNR [dB] | Diff. [dB] | SNR [dB] | Diff. [dB] |
| $\tau$=0 | >25.00 | >8.34 | >25.00 | >2.70 |
| $\tau$=1 | >25.00 | >8.34 | >25.00 | >2.70 |
| $\tau$=60 | 18.50 | 1.84 | >25.00 | >2.70 |
| $\tau$=64 | 17.36 | 0.70 | 23.34 | 1.04 |
| $\tau$=70 | 16.82 | 0.16 | 22.75 | 0.45 |
| $\tau$=80 | 16.82 | 0.16 | 22.60 | 0.30 |
| $\tau$=128 | 16.66 | --- | 22.30 | --- |

Figure 13b is an elevated perspective of Fig. 13a, which depicts the BER versus adjusted bandwidth at different normalized CFO values. Figure 13c shows a side view of Fig. 13a, which shows BER performance versus normalized CFO under different situations of compensation. As a result, Fig. 13 guarantees that the proposed equalizer can perform very well when the normalized CFO grows to half the sub-carrier spacing and gives better performance than other equalizers at SNR=15 dB.

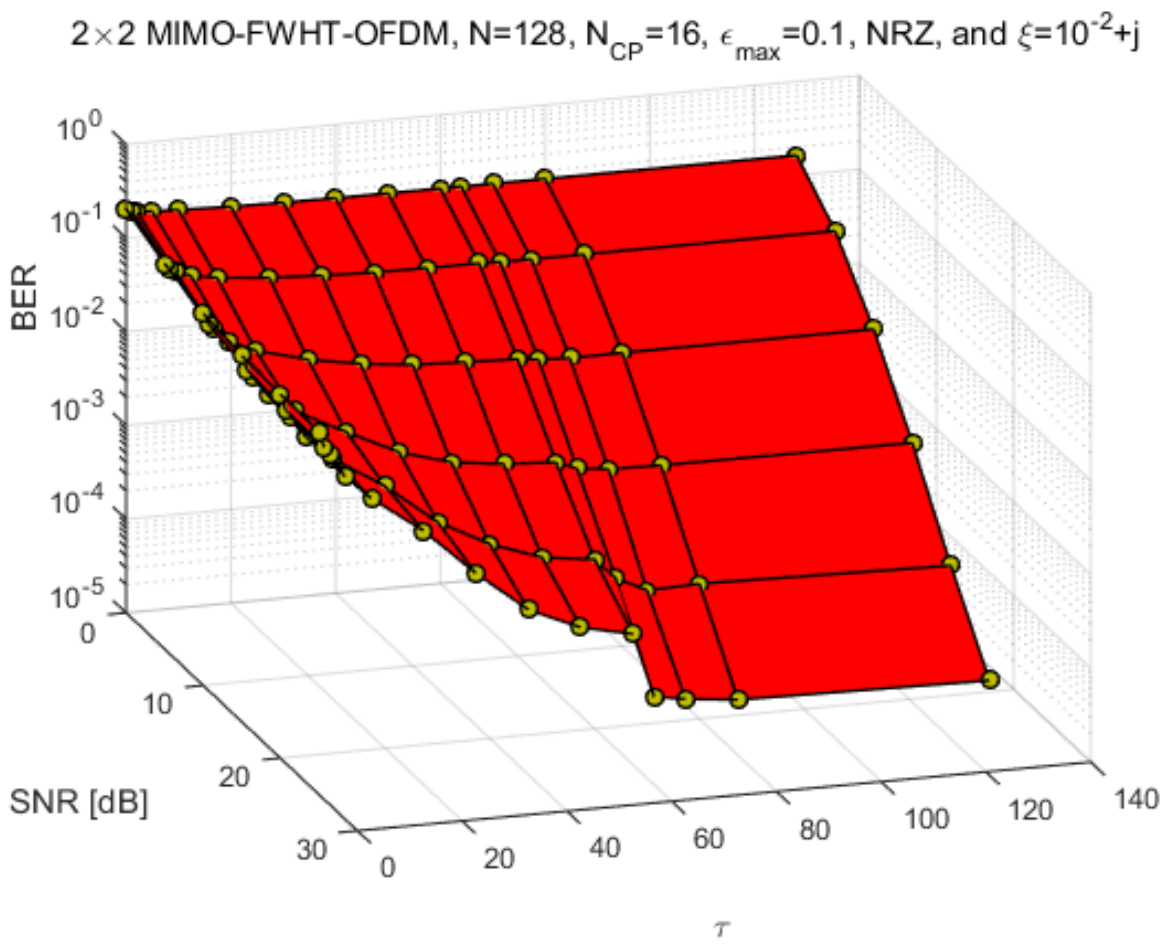

**Figure 12a**. The BER vs. SNR at various compensation scenarios (i.e., τ).

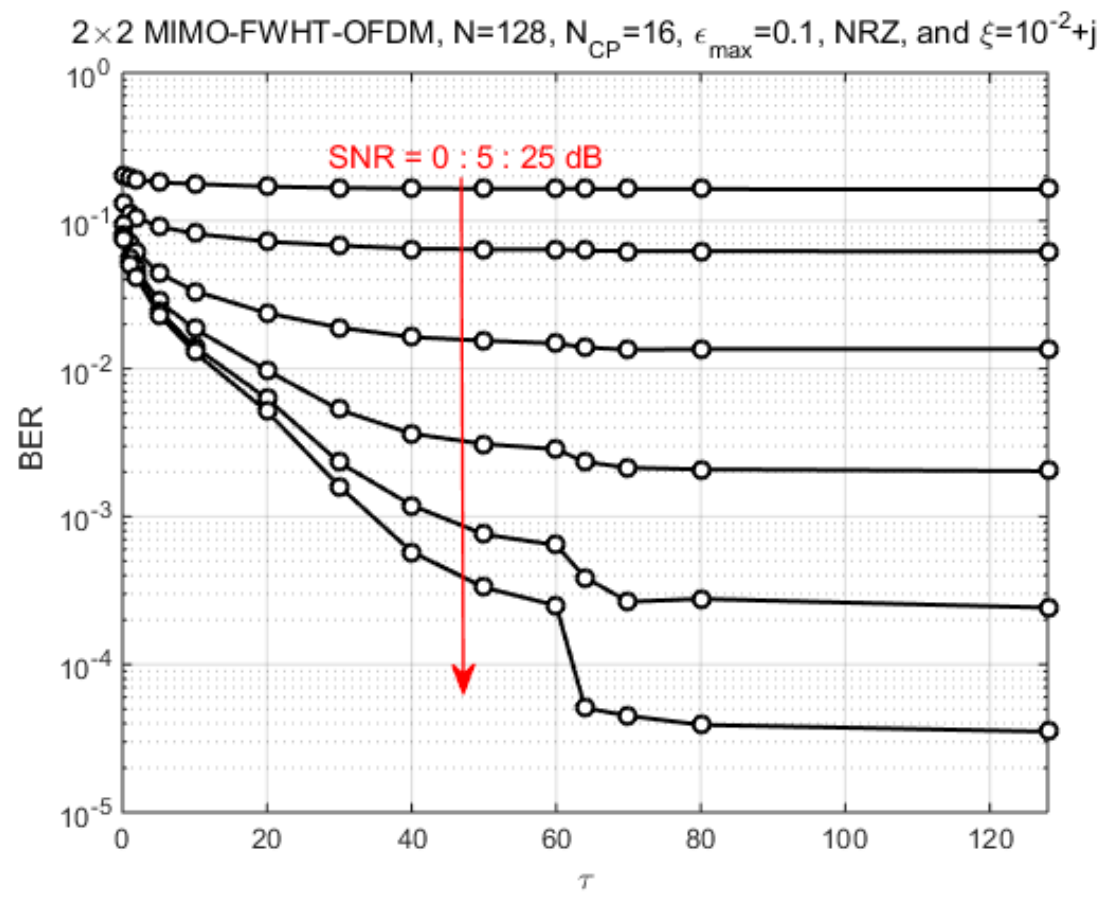

**Figure 12b**. The elevation view of Fig. 12a.

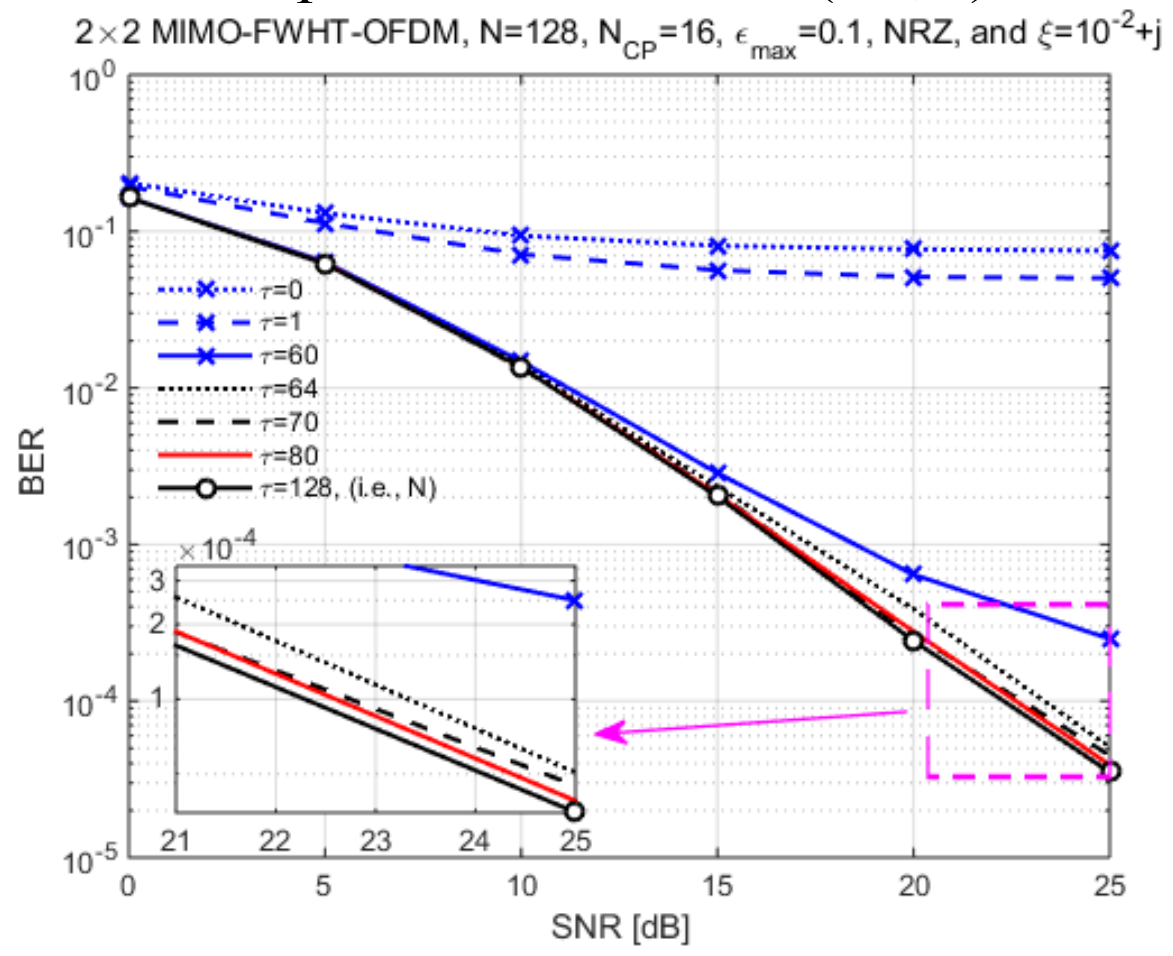

**Figure 12c**. The side view of Fig. 12a.

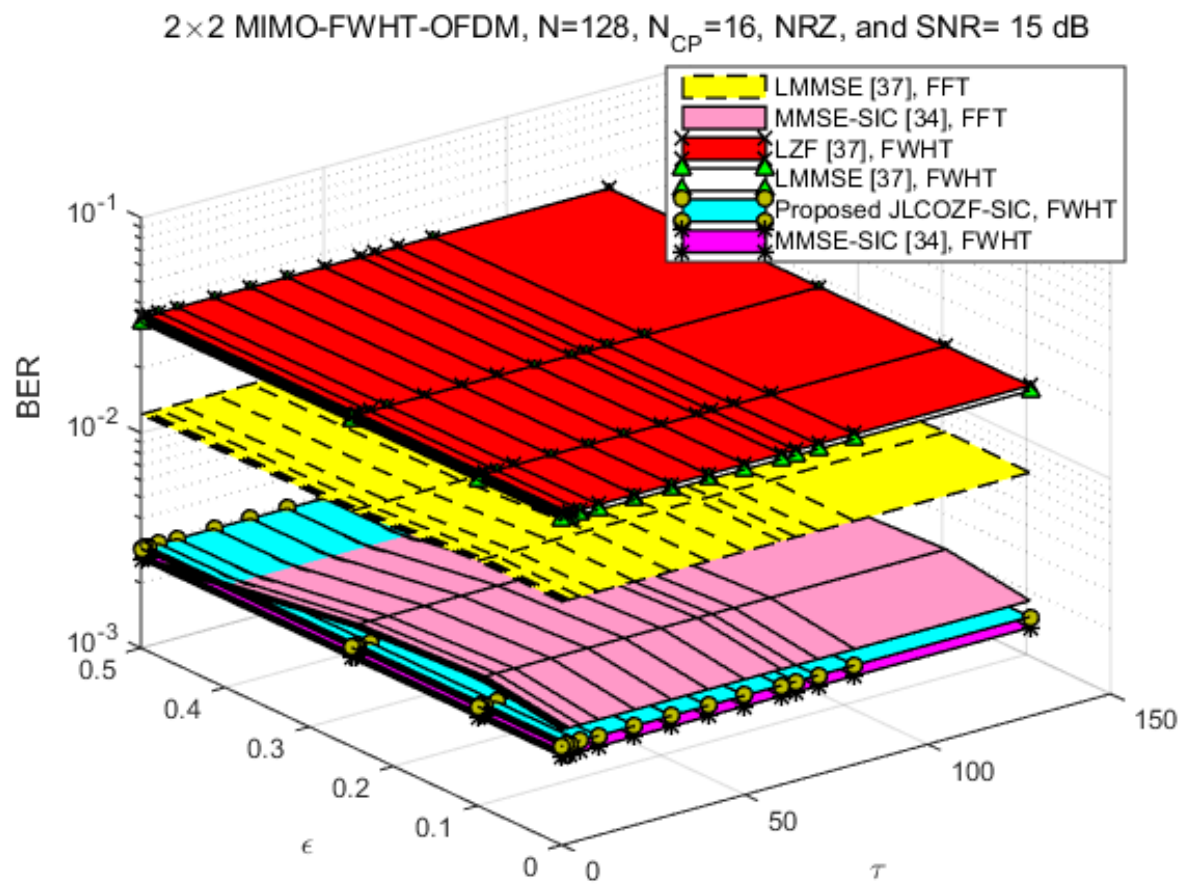

**Figure 13a**. The BER vs. normalized CFO at various compensation scenarios (i.e., τ).

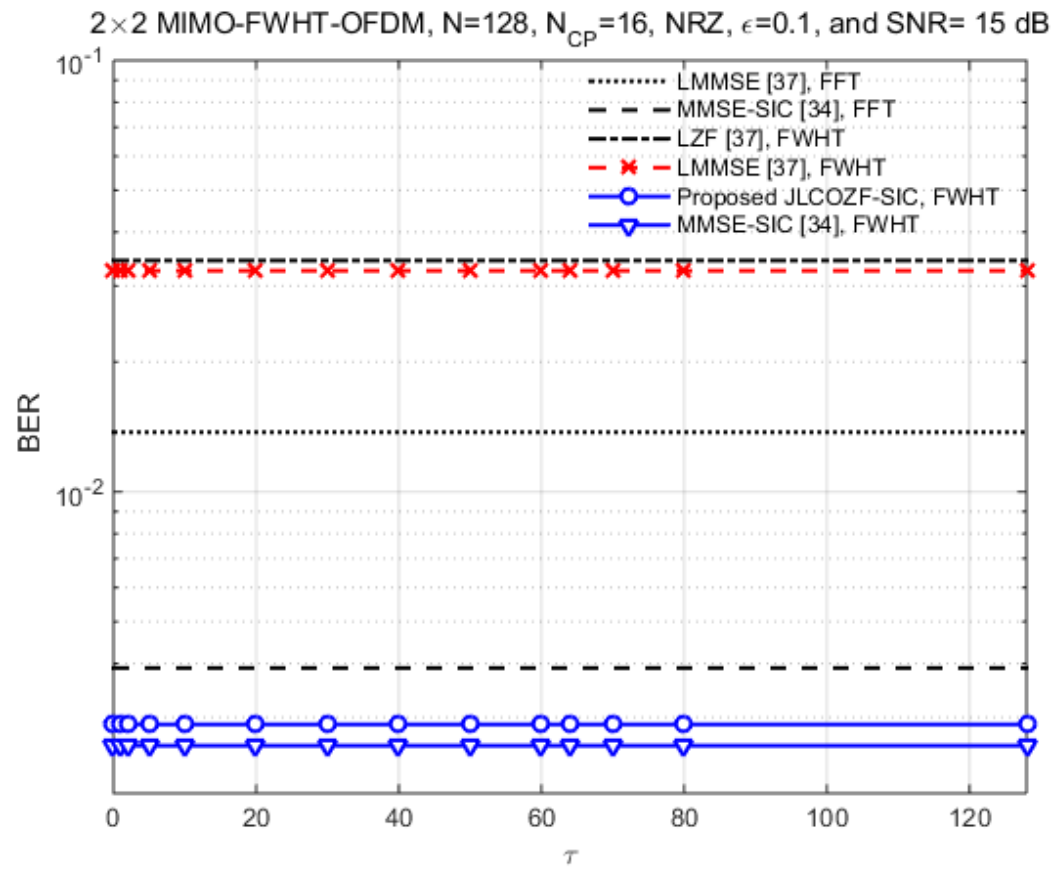

**Figure 13b**. The elevation view of Fig. 13a.

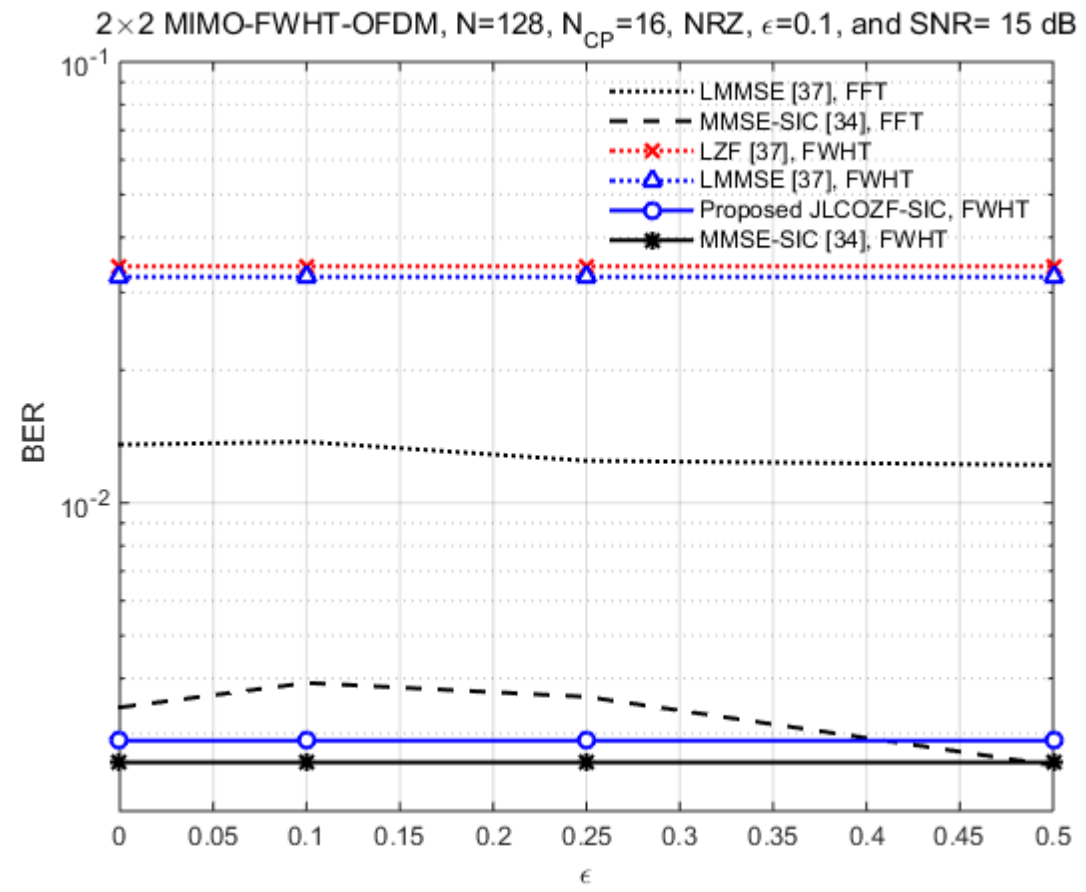

**Figure 13c**. The side view of Fig. 13a.

Figure 14 depicts the performance of the proposed JLCOZF-SIC equalizer and other linear and nonlinear equalizers in terms of BER versus SNR. It is obvious that the proposed equalizer closely fits the MMSE-SIC within $0 \leq \text{SNR} \leq 20$ dB. As previously stated, the suggested equalizer reduces computational complexity by employing the BMA idea, and its performance closely equals that of the MMSE-SIC equalizer (see Fig. 14). As a result, the proposed equalizer is the best option for coping with MIMO-OFDM impairments while maintaining a low complexity implementation. Figure 15 investigates BER effectiveness in the presence of estimation mistakes. This figure shows that the proposed equalizer fits the BER performance of the MMSE-SIC, which degrades as the estimation errors in both the channel and the CFO surpass the standard deviation limits of 0.01 ($\text{i.e.}, \ \sigma_{\varepsilon}, \ \sigma_{ch} > 0.01$).

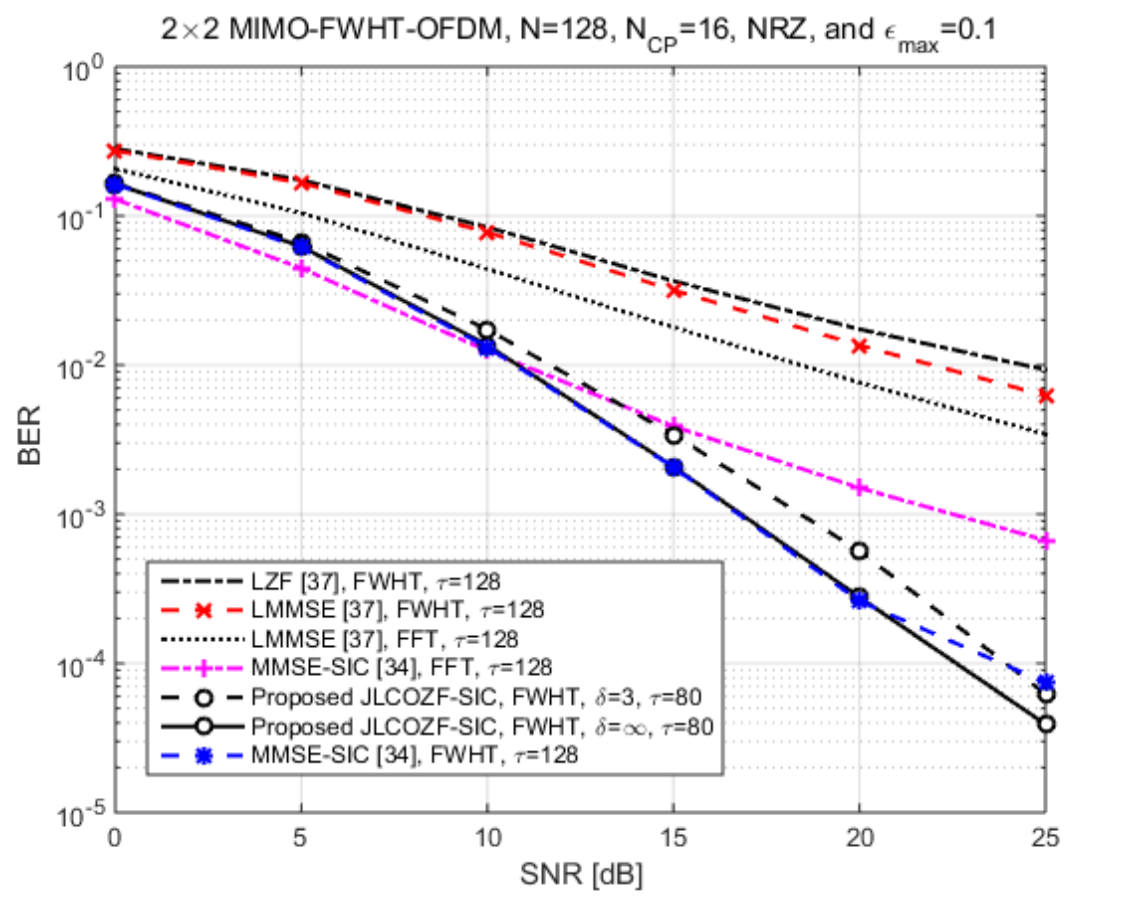

**Figure 14**. The BER vs. SNR of various equalizers

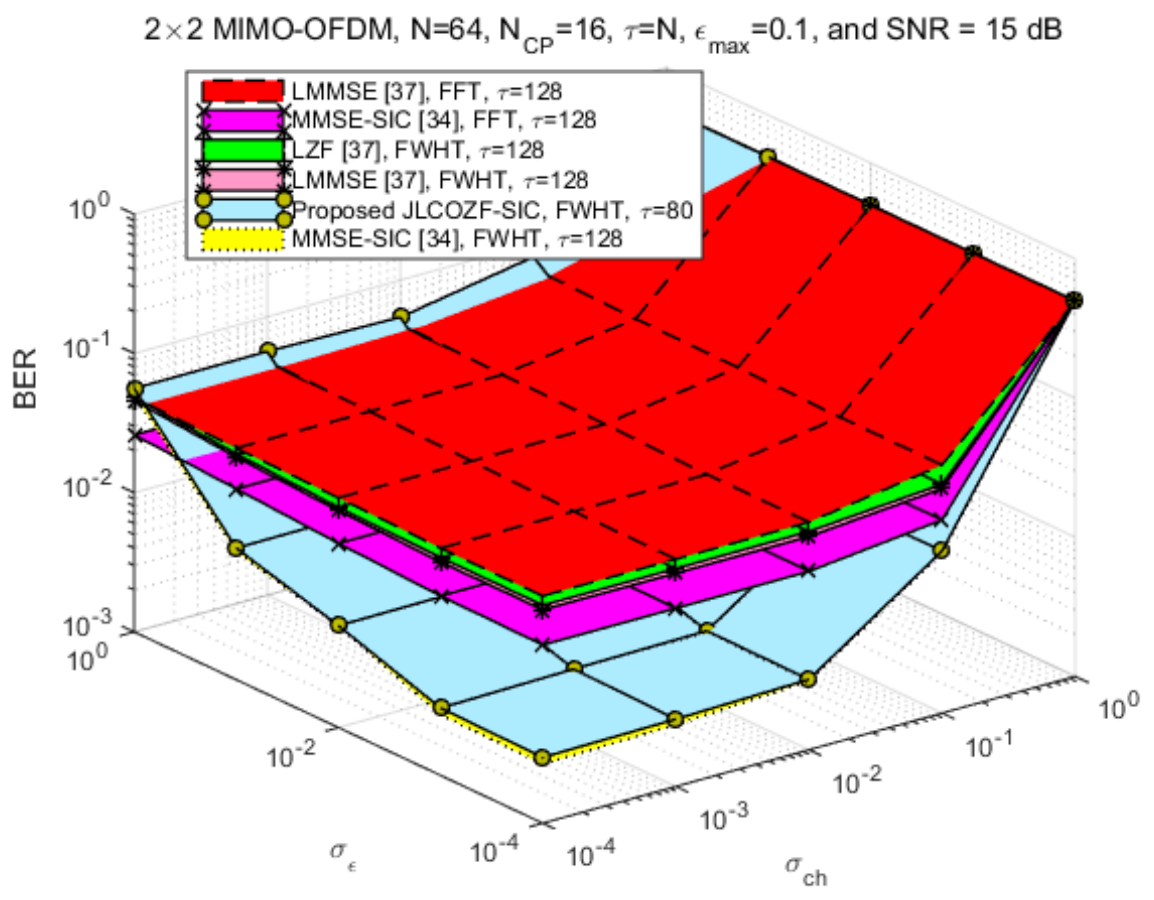

**Figure 15**. The BER performance in the case of estimation errors

According to Fig.14, Table 8 tabulates the extra SNR of various equalizers in comparison to that of the MMSE-SIC based on FWHT at two BER levels (i.e., BER=$10^{-3}$ and $10^{-4}$). It should be noted that the proposed JLCOZF-SIC equalizer employs the BMA concept of τ=80, in addition to the two cases of $\delta = 3$, and $\delta = \infty$. In the instance of $\delta = \infty$, the proposed JLCOZF-SIC equalizer requires an additional SNR of approximately 0.07 and 1.22 dB at BER=$10^{-3}$ and $10^{-4}$, respectively, to obtain the same BER performance as the MMSE-SIC. Similarly, in the instance of $\delta = 3$, the proposed JLCOZF-SIC equalizer requires an additional SNR of approximately 1.67 and 0.89 dB at BER=$10^{-3}$ and $10^{-4}$, respectively, to obtain the same BER performance as the MMSE-SIC. Thus, for the case of BER=$10^{-}$

$^{3}$, the proposed JLCOZF-SIC has an SNR gap of about 1.6 dB for the two instances of $\delta = 3$, and $\delta = \infty$. Similarly, for the case of BER=$10^{-4}$, there is an SNR gap of about 0.33 dB for the suggested JLCOZF-SIC for the two instances of $\delta = 3$, and $\delta = \infty$. Thus, depending on your desired robustness, BER performance or decrease in computational complexity. Note that, this proposed scheme can be applied across various applications, especially in satellite communication, to alleviate bandwidth limitation [39].

**Table 8**. The additional SNR at different equalizers with respect to MMSE-SIC

| | BER=$10^{-3}$ | | BER=$10^{-4}$ | |
|---|---|---|---|---|
| Equalizer type | SNR [dB] | Diff. [dB] | SNR [dB] | Diff. [dB] |
| LMMSE [37], FFT, τ=128 | >25.00 | >8.25 | >25.00 | >1.18 |
| MMSE-SIC [34], FFT, τ=128 | 22.51 | 5.76 | >25.00 | >1.18 |
| LZF [37], FWHT, τ=128 | >25.00 | >8.25 | >25.00 | >1.18 |
| LMMSE [37], FWHT, τ=128 | >25.00 | >8.25 | >25.00 | >1.18 |
| Proposed JLCOZF-SIC, FWHT, δ=∞, τ=128 | 16.82 | 0.07 | 22.60 | 1.22 |
| Proposed JLCOZF-SIC, FWHT, δ=3, τ=128 | 18.42 | 1.67 | 22.93 | 0.89 |
| MMSE-SIC [34], FWHT, τ=128 | 16.75 | --- | 23.82 | --- |

## VI. Conclusions

This article focuses on the construction of OFDM based on the Walsh transform for various compensation types. As is well known, using the BMA principle reduces complexity by at least half of the communicated data length $(\mathrm{i.e.,}\ N/2)$ in the case of OFM based on the Walsh transform. Otherwise, the BER degrades, and the effective retrieval of the transmitted data is threatened. In this paper, detailed expressions for the arithmetic operation, expressed in terms of the flop number of the proposed Walsh domain equalizer and other various equalizers, have been derived for the general MIMO configuration. These formulas were used to investigate the computational difficulty of each approach. Furthermore, computer models were performed, and several instances were found in which OFDM based on the Walsh transform outperformed OFDM based on the Fourier transform in terms of BER performance. In contrast to the MMSE-SIC equalizer, the proposed Walsh domain equalizer reduces complexity while maintaining BER performance. This advantage, combined with the excellent behavior of OFDM based on Walsh transform under the co-CFO described in earlier works, makes OFDM based on Walsh transform a fascinating alternative to OFDM based on Fourier transform.

## VII. Future work

The future work of this manuscript can be summarized in the following points:

1. The extension of the proposed scheme for Massive MIMO-OFDM by using Walsh Transform.
2. The extension of the proposed scheme for DRL principal-based resource allocation [25].
3. The extension of the proposed scheme for combining optimal power allocation and user association methods [26].
4. The extension of the proposed scheme for adaptive backhaul topology that can respond to changing traffic patterns [7].
5. The extension of the proposed scheme for a dynamic optimization approach to lower the overall energy consumption of fifth-generation (5G) heterogeneous networks [6].
6. The extension of the proposed scheme for Non-Orthogonal Multiple Access (NOMA) for future wireless communication systems.
7. The extension of the proposed scheme for Universal Filtered Multi-Carrier (UFMC) modulation, and Filter-Bank Muli-Carrier (FBMC) modulation with different modulation schemes.

## IX. Appendix

The general FWHT matrix can be expressed as [29]:

$$\mathbf{\Gamma}_N = \mathbf{W}_N(2^k) = \sqrt{N}\begin{bmatrix} \boldsymbol{\varphi}(2^{k-1}) & \boldsymbol{\varphi}(2^{k-1}) \\ \boldsymbol{\varphi}(2^{k-1}) & -\boldsymbol{\varphi}(2^{k-1}) \end{bmatrix} = \boldsymbol{\varphi}(2)\otimes\boldsymbol{\varphi}(2^{k-1})$$

where $\otimes$ denotes the Kronecker product, $2 \le k \in N$.

A fractional power of a matrix can be approximated by the partial sum of a similar power series, as demonstrated in [35]:

$$[\mathbf{I}_{N\times N} + \boldsymbol{\beta}_{N\times N}]^{\rho} = \mathbf{I}_{N\times N} + \rho\boldsymbol{\beta}_{N\times N} + \frac{\rho(\rho-1)}{2!}(\boldsymbol{\beta}_{N\times N})^2 + \frac{\rho(\rho-1)(\rho-2)}{3!}(\boldsymbol{\beta}_{N\times N})^3 + \cdots$$

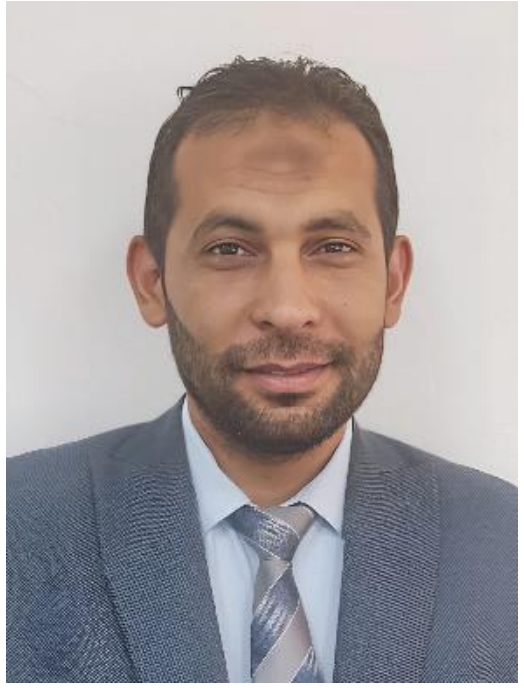

**Khaled Ramadan Mohamed** received his B.Sc. from the Higher Institute of Engineering (El-Shorouk Academy) in 2011, and his M.Sc. and Ph.D. from the Faculty of Electronic Engineering, Menoufia University, Menouf, Egypt, in 2018 and 2021, respectively. He joined the teaching staff of the Communication and Computer Engineering Department at the Higher Institute of Engineering in Al-Shorouk City, Egypt, in 2021. He has published several scientific papers in national and international conference proceedings and journals. He has received the most read paper from the International Journal of Communication Systems (IJCS) for 2018–2019. He has various online courses on the Udemy platform, and more than 1000 students from +48 countries around the world have enrolled in the courses with positive feedback; besides, one of the courses is the highest rated according to the Udemy platform. His research interests include multicarrier communication systems, Multiple-Input Multiple-Output (MIMO) systems, Digital Signal Processing (DSP), digital communications, channel equalization, Carrier Frequency Offset (CFO) estimation and compensation, Underwater Acoustic (UWA) wireless communication systems, and multicarrier Non-Orthogonal Multiple Access (NOMA) for 5G networks.

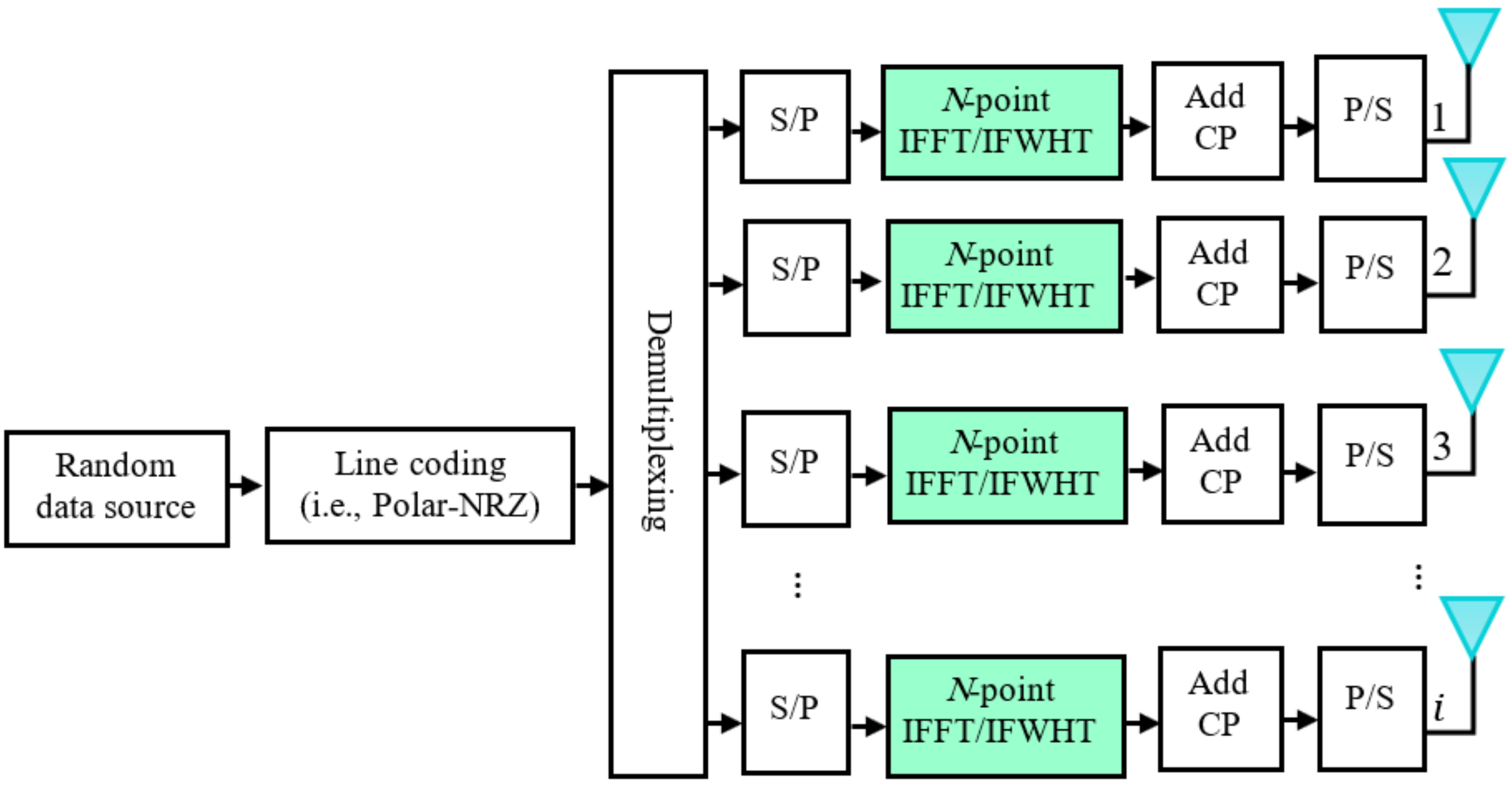
(a)

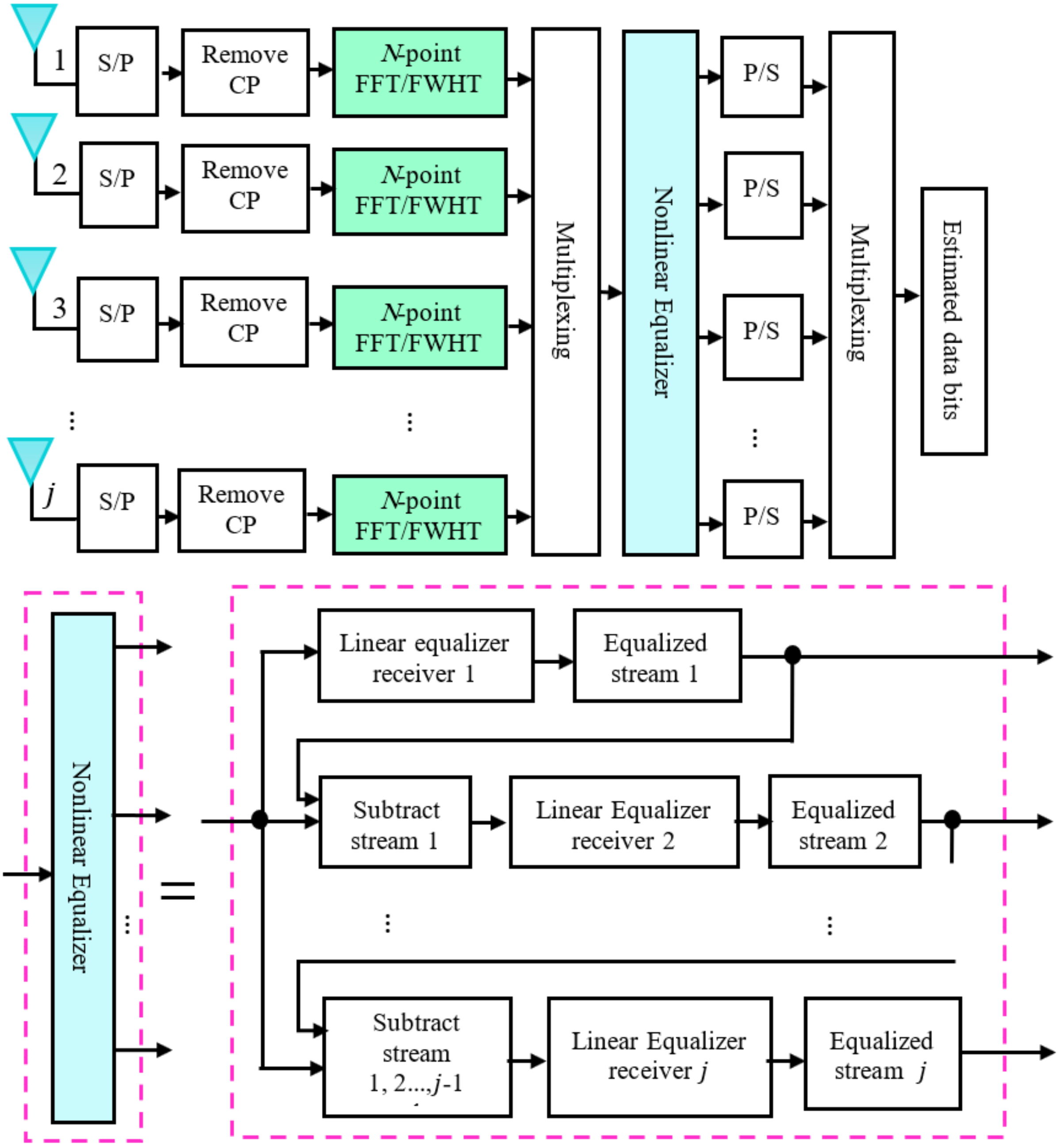

(b)

**Figure 1**. The transceiver block diagram of an $i \times j$ MIMO-OFDM communication system.

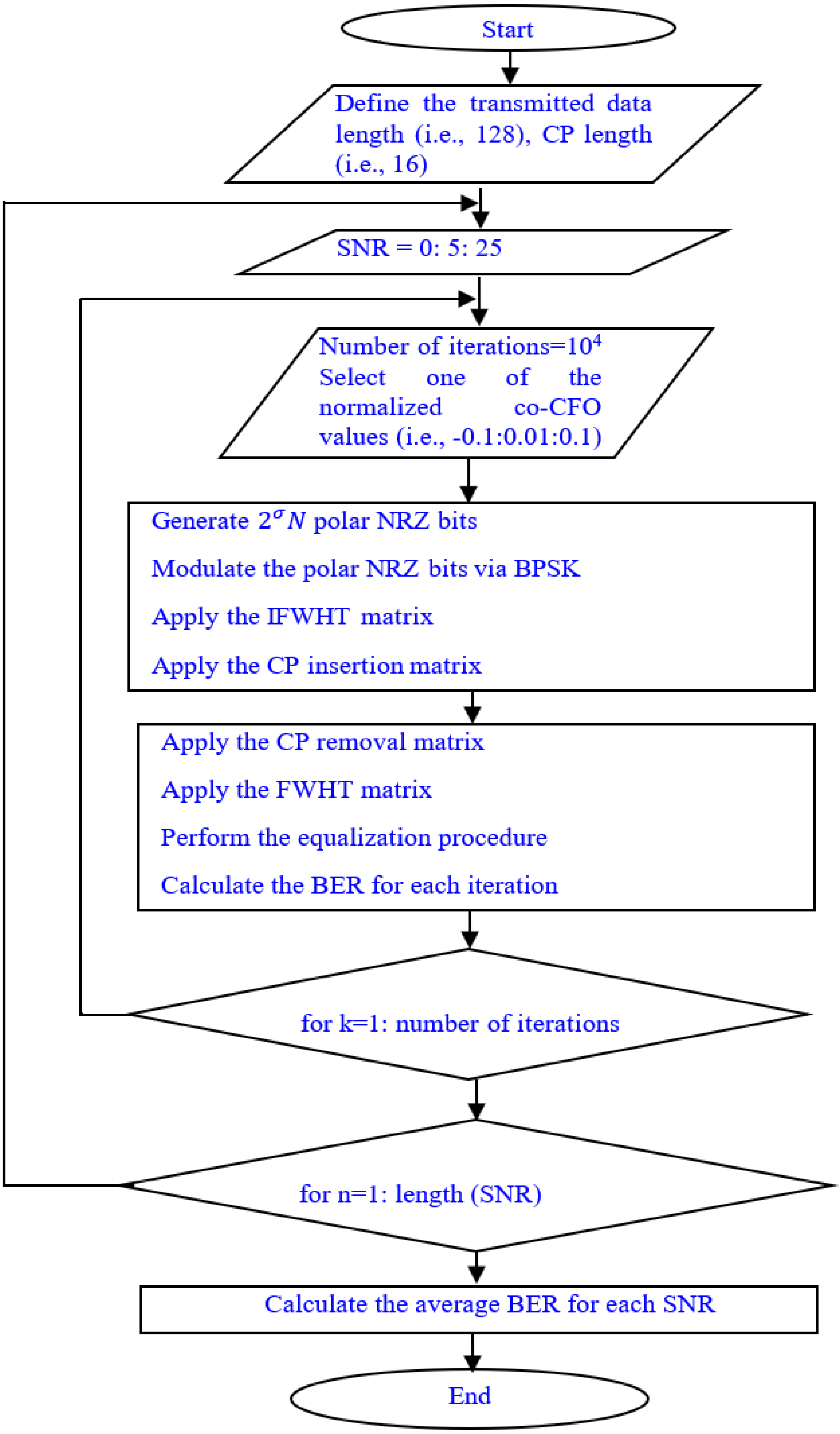

**Figure 1c**. Flowchart of an $i \times j$ MIMO-OFDM transceiver system.

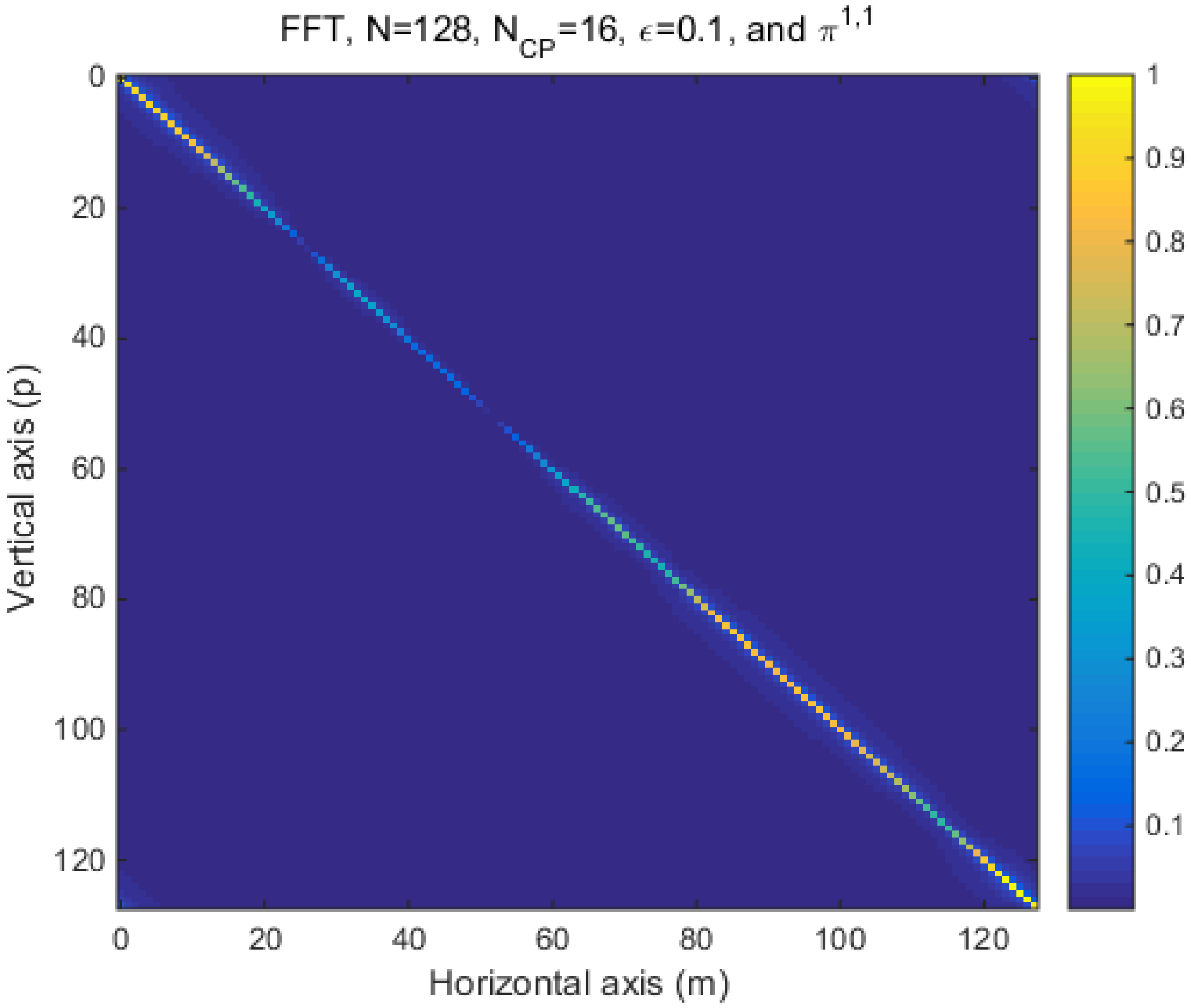

(a)

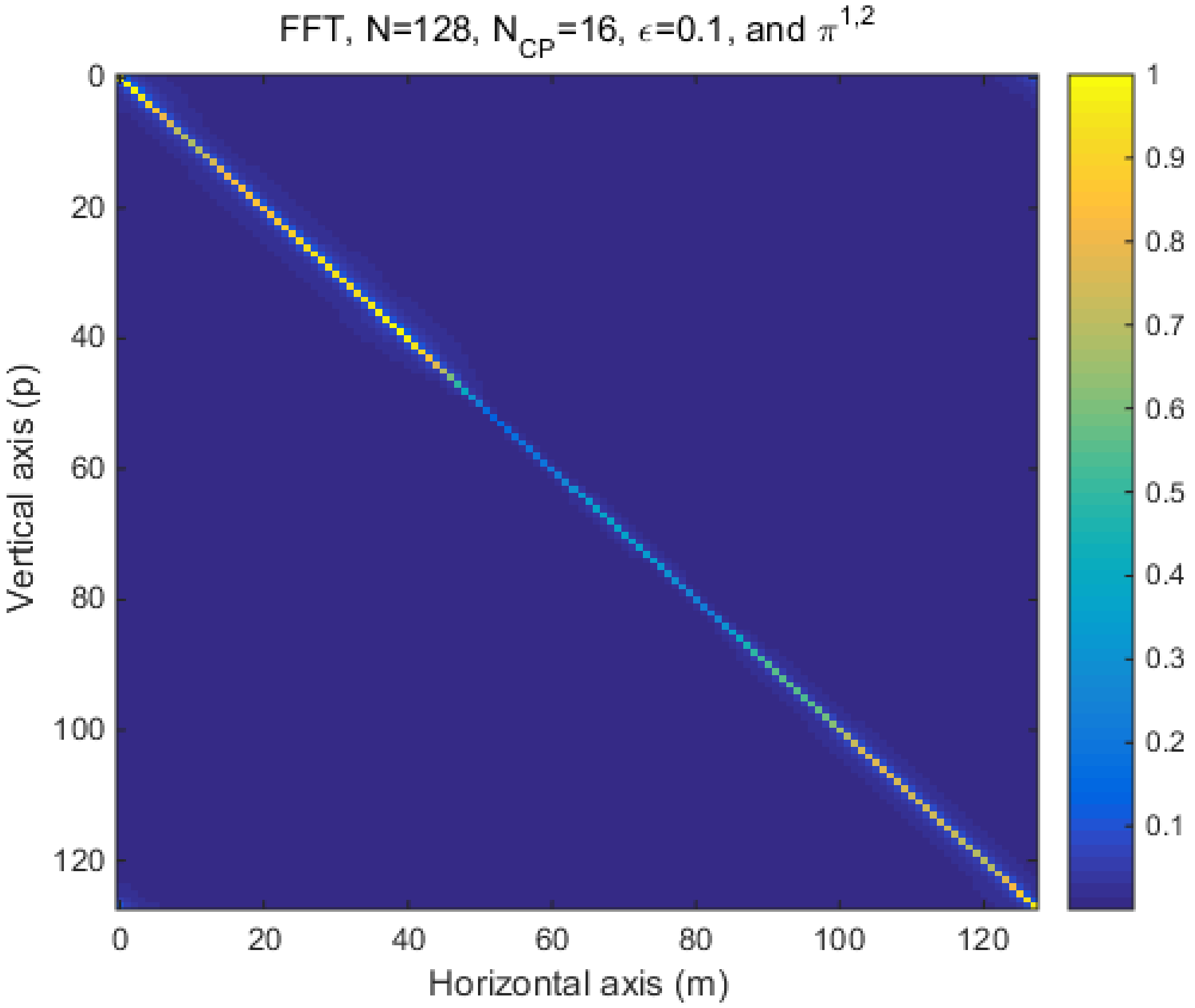

(b)

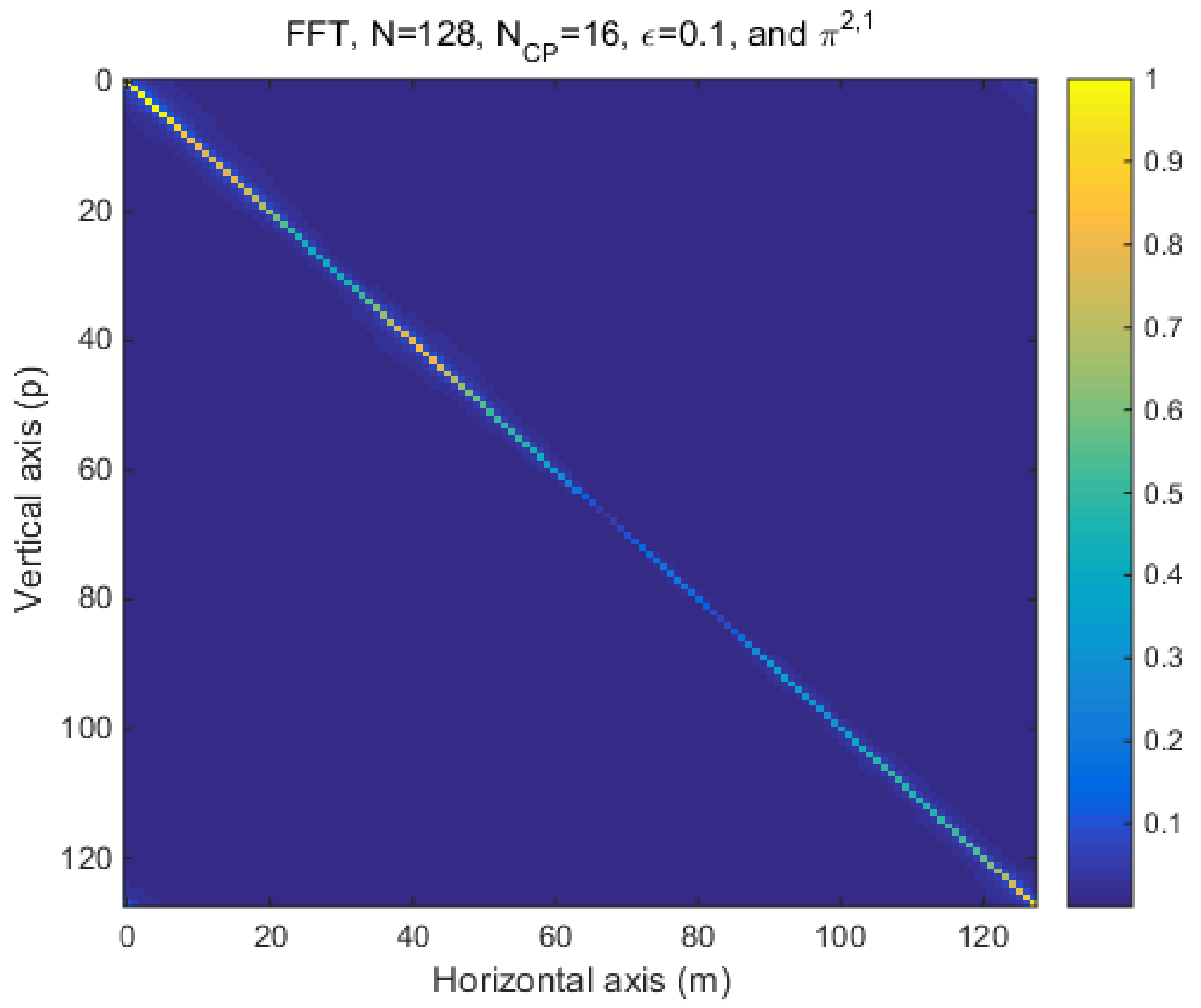

(c)

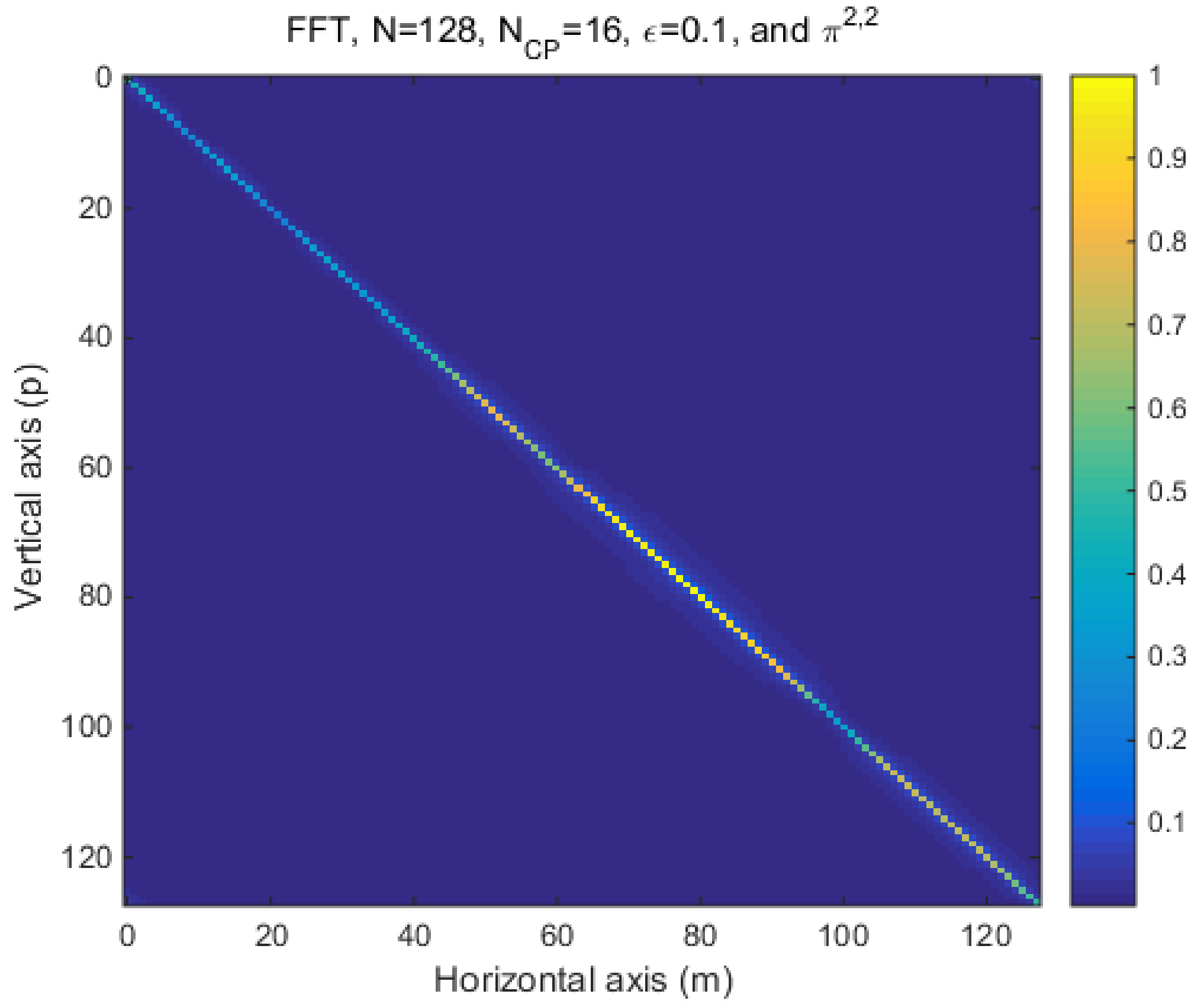

(d)

**Figure 2**. The normalized magnitude of the interference matrix vs. sub-carrier indices using Fourier transform.

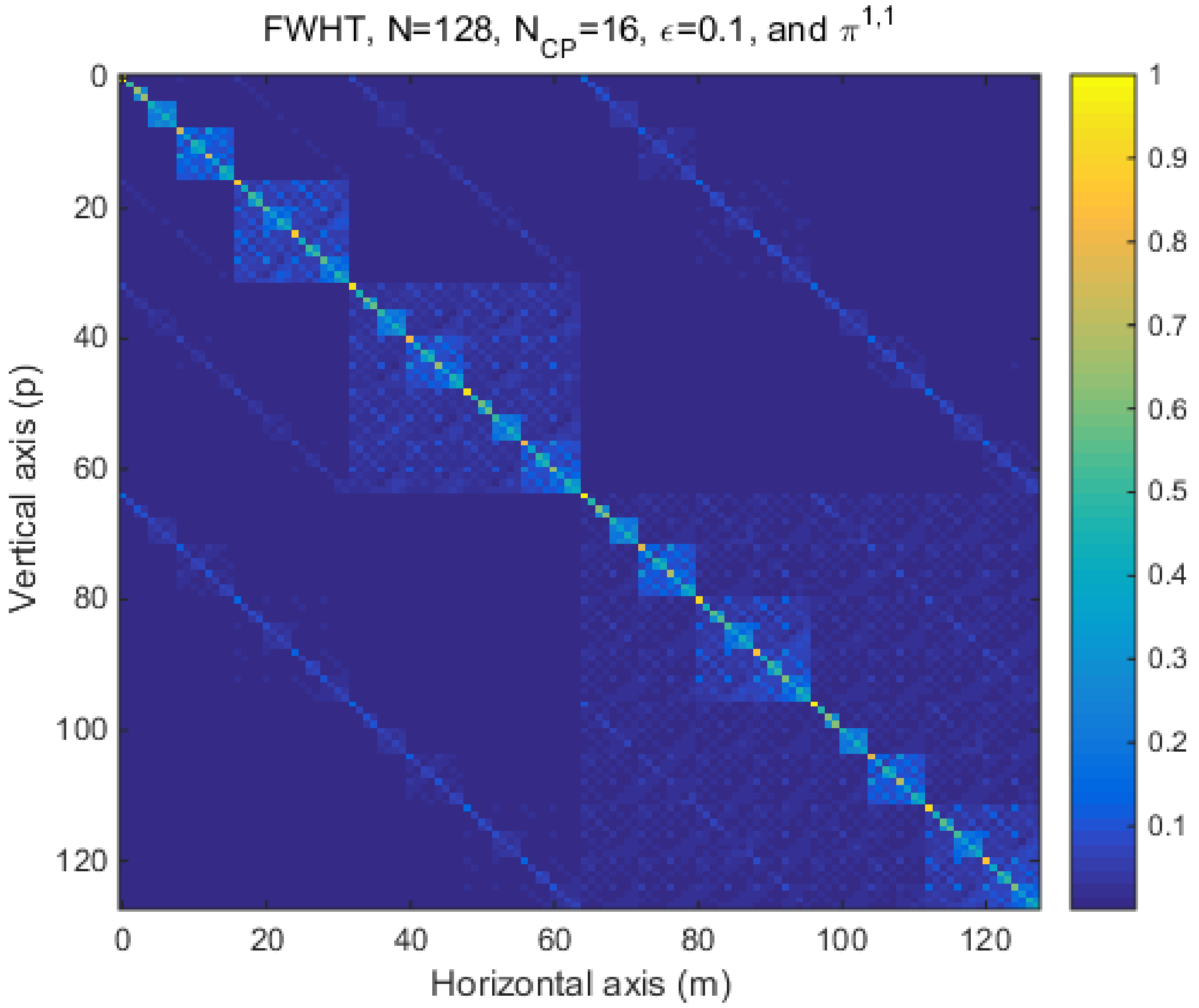

(a)

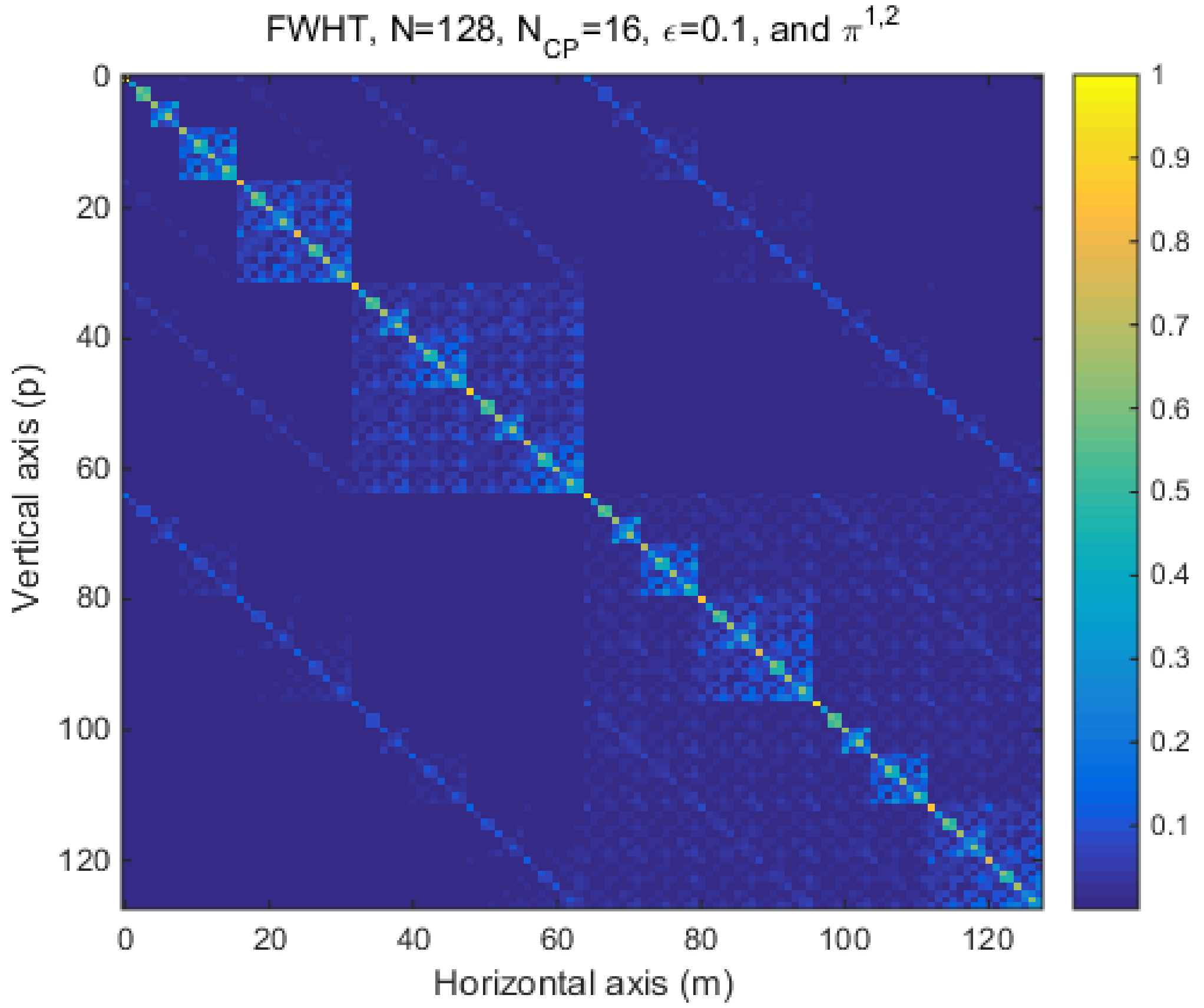

(b)

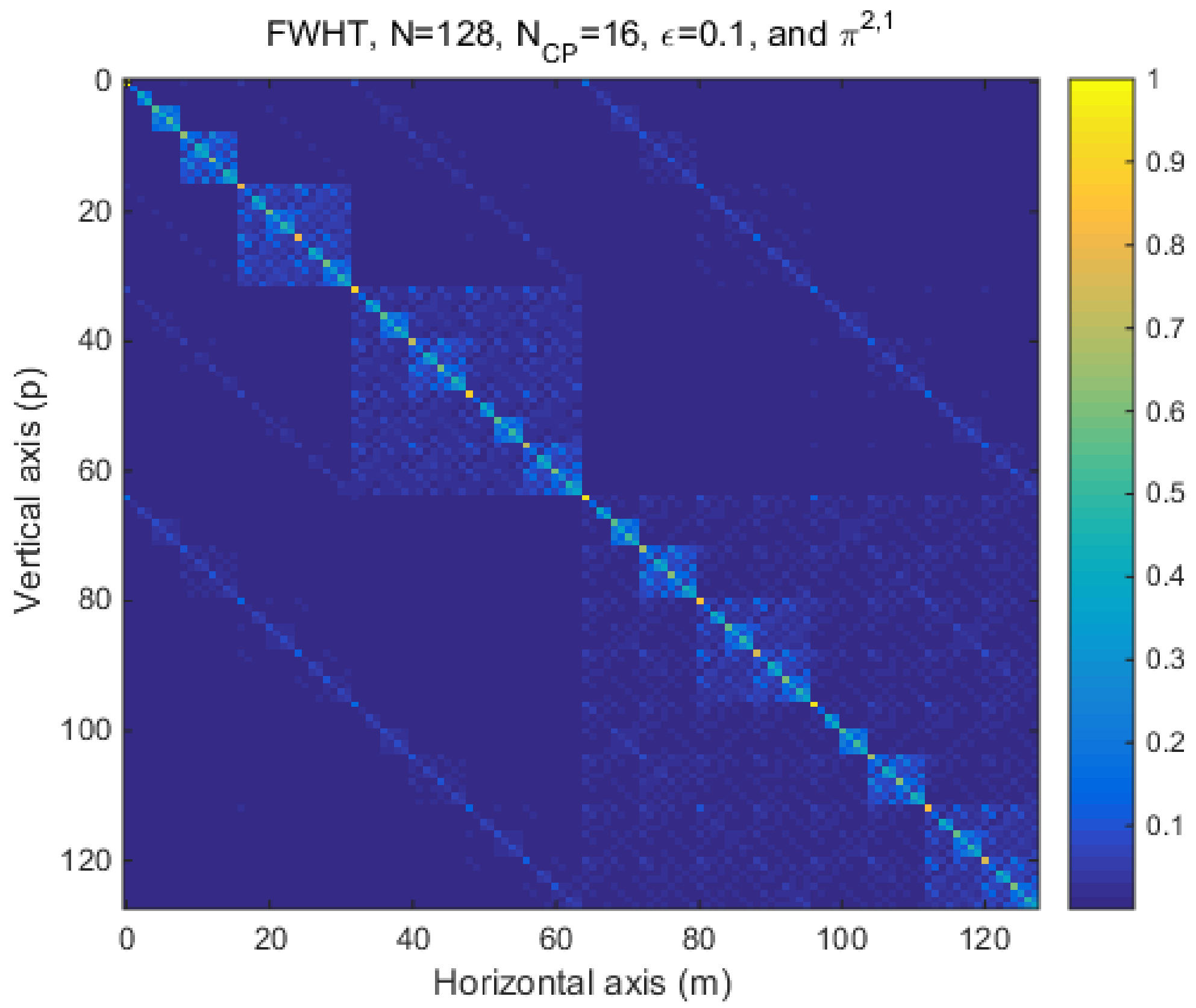

(c)

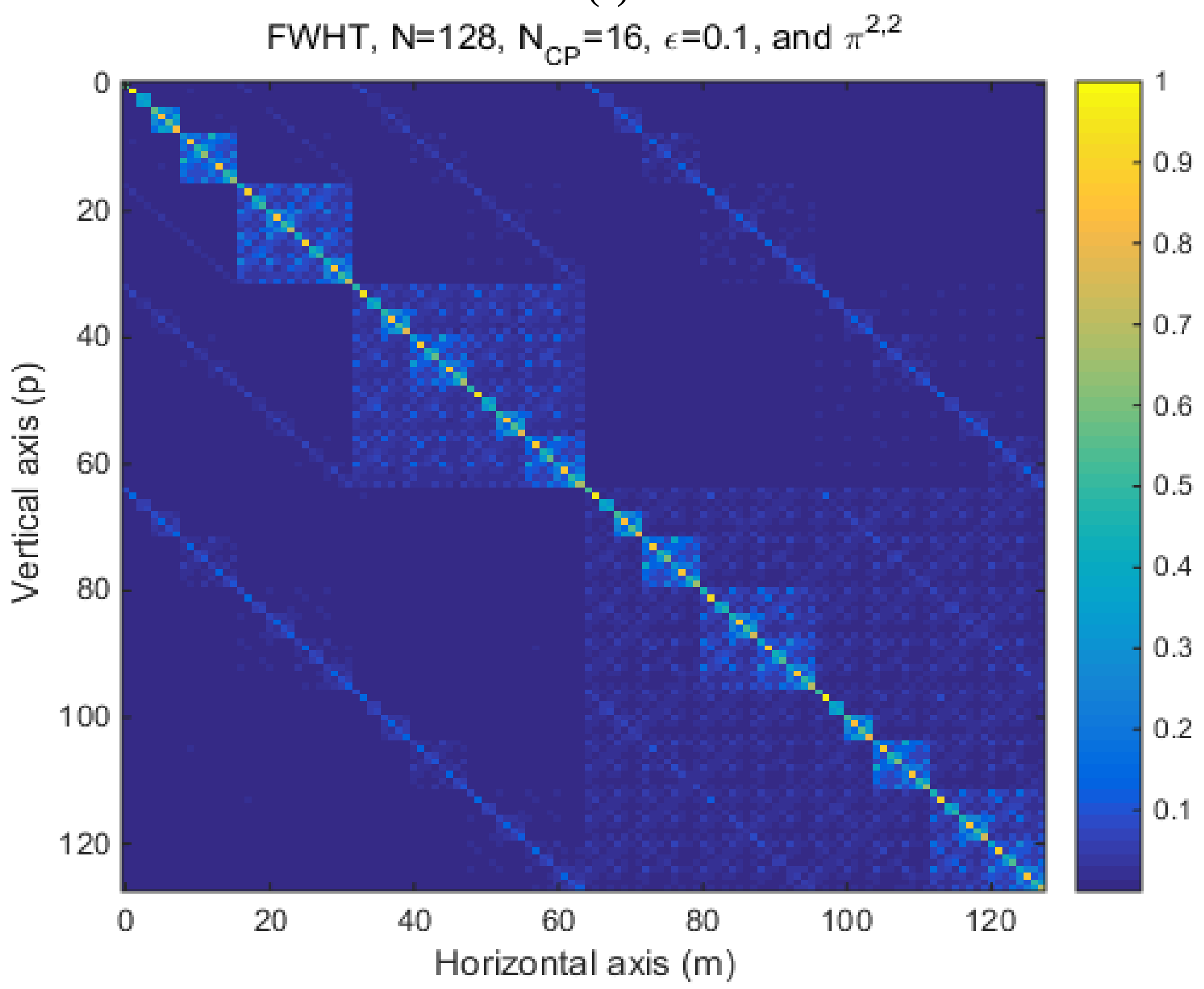

(d)

**Figure 3**. The normalized magnitude of the $\boldsymbol{\pi}^{j,i}$ , $j, i \in \{1,2\}$ matrix vs. sub-carrier indices using Walsh transform.

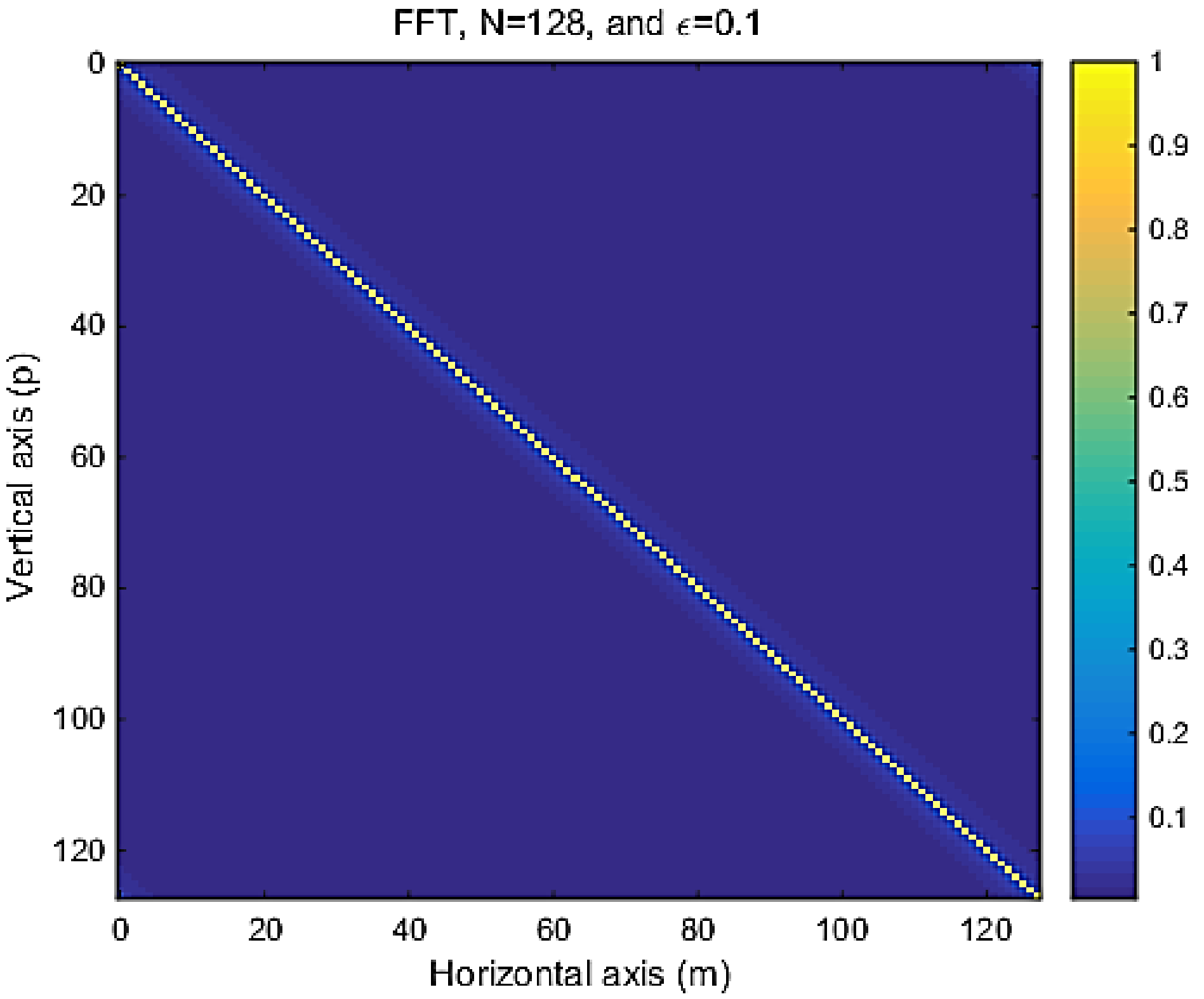

(a)

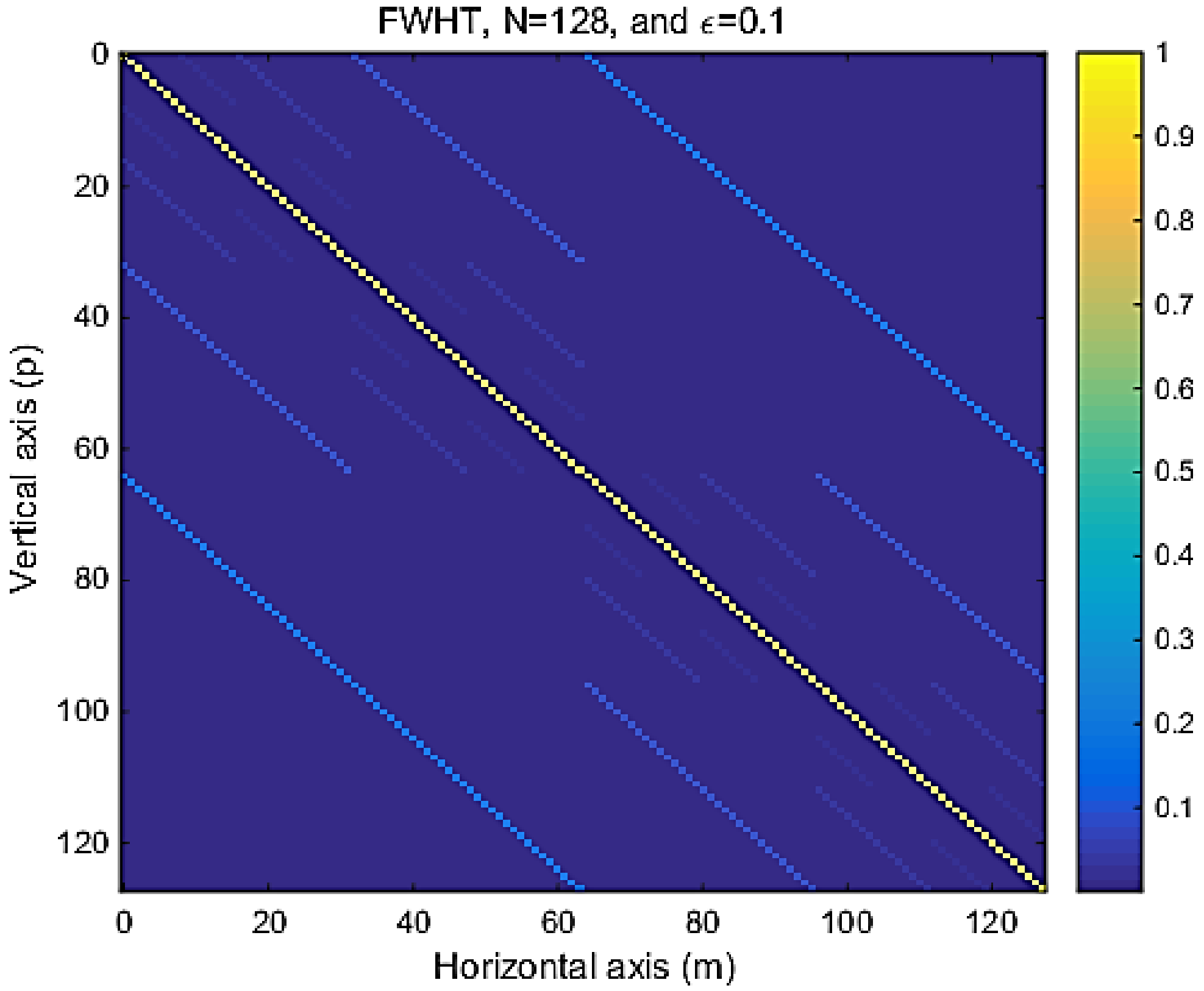

(b)

**Figure 4**. The normalized magnitude of the interference matrix vs. sub-carrier indices using different transforms.

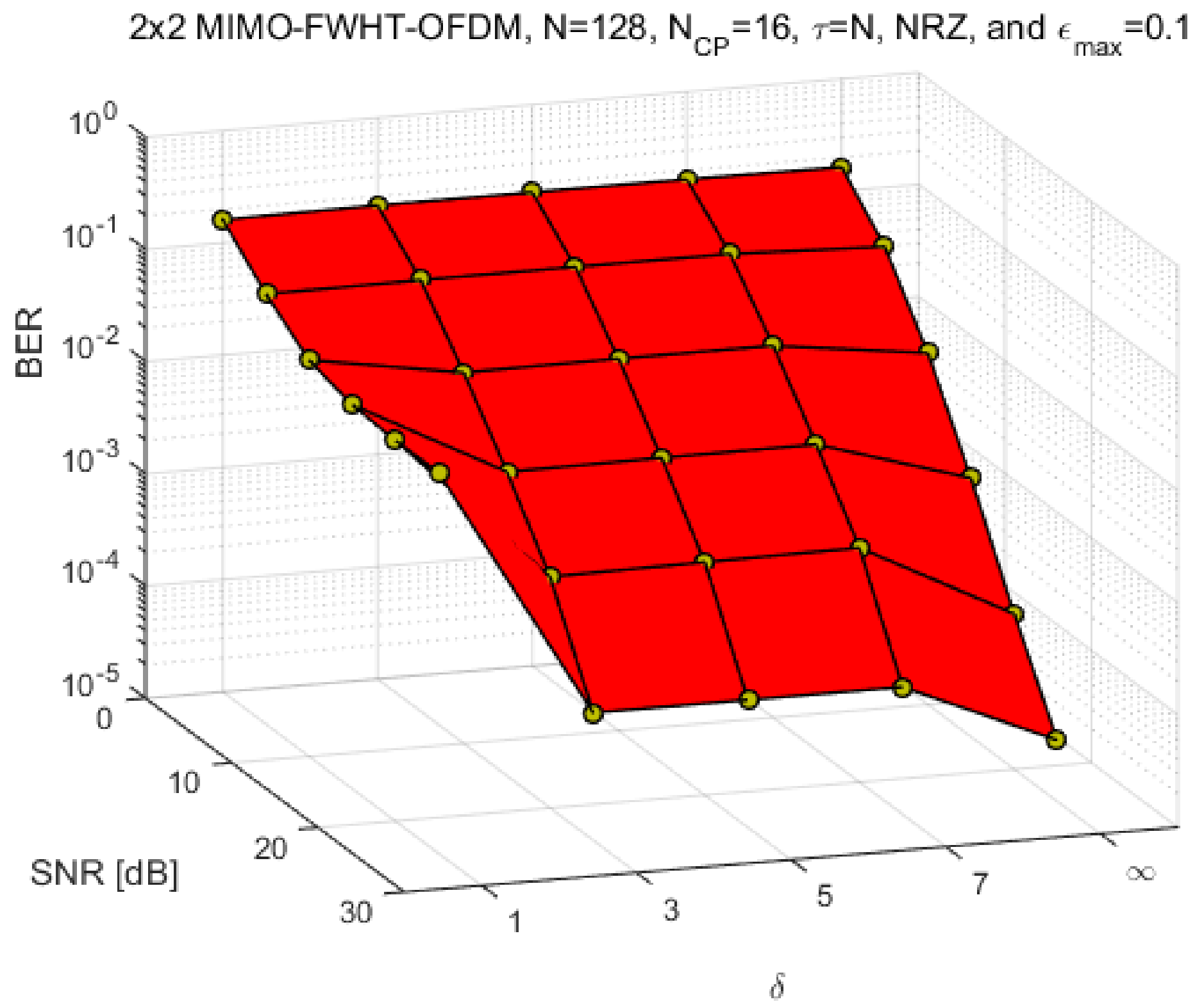

**Figure 5**a. The BER vs. number of terms of approximated covariance matrix (i. e., $\rho$) in Eq. (37)

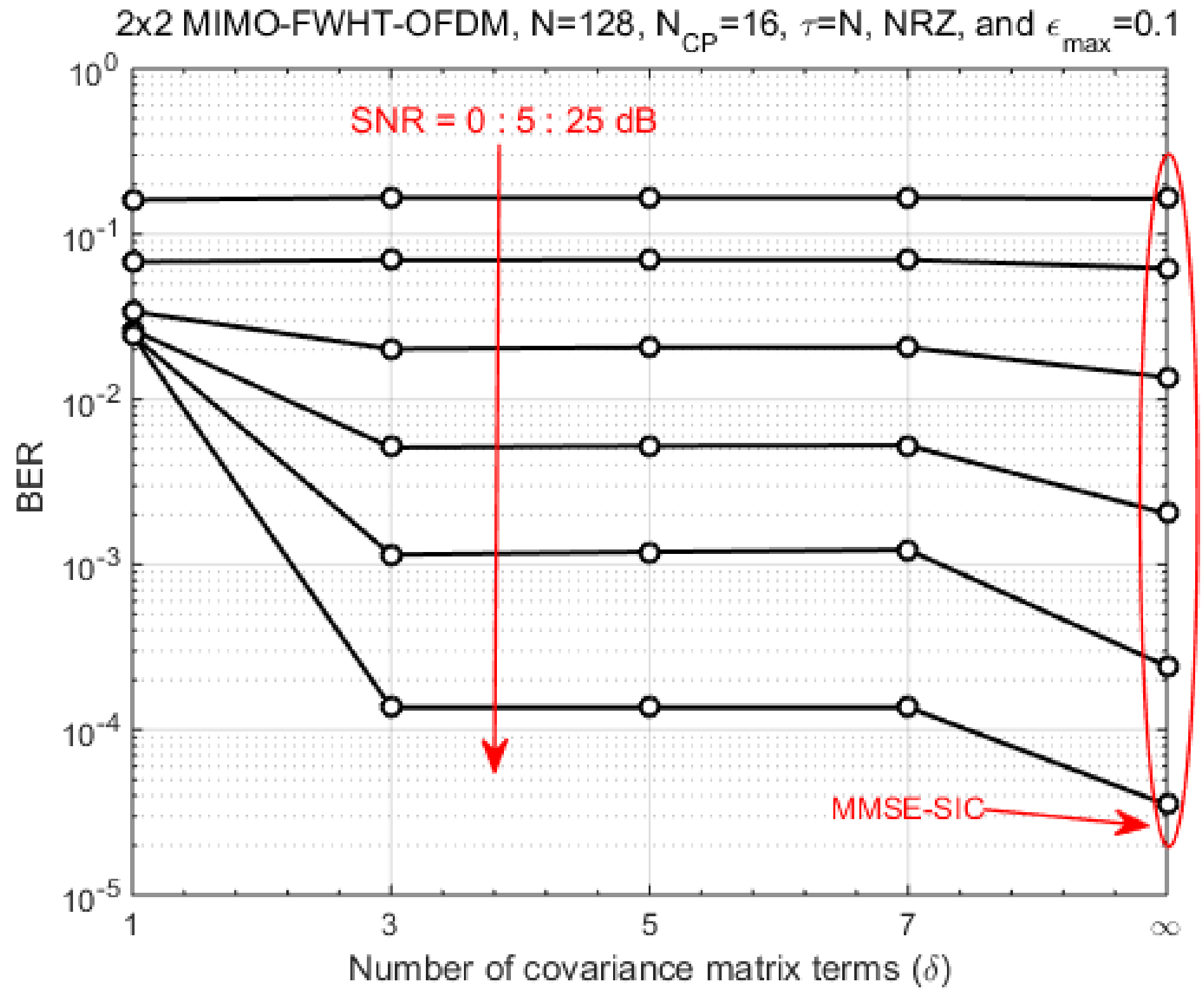

**Figure 5b**. The elevation view of Fig. 5a

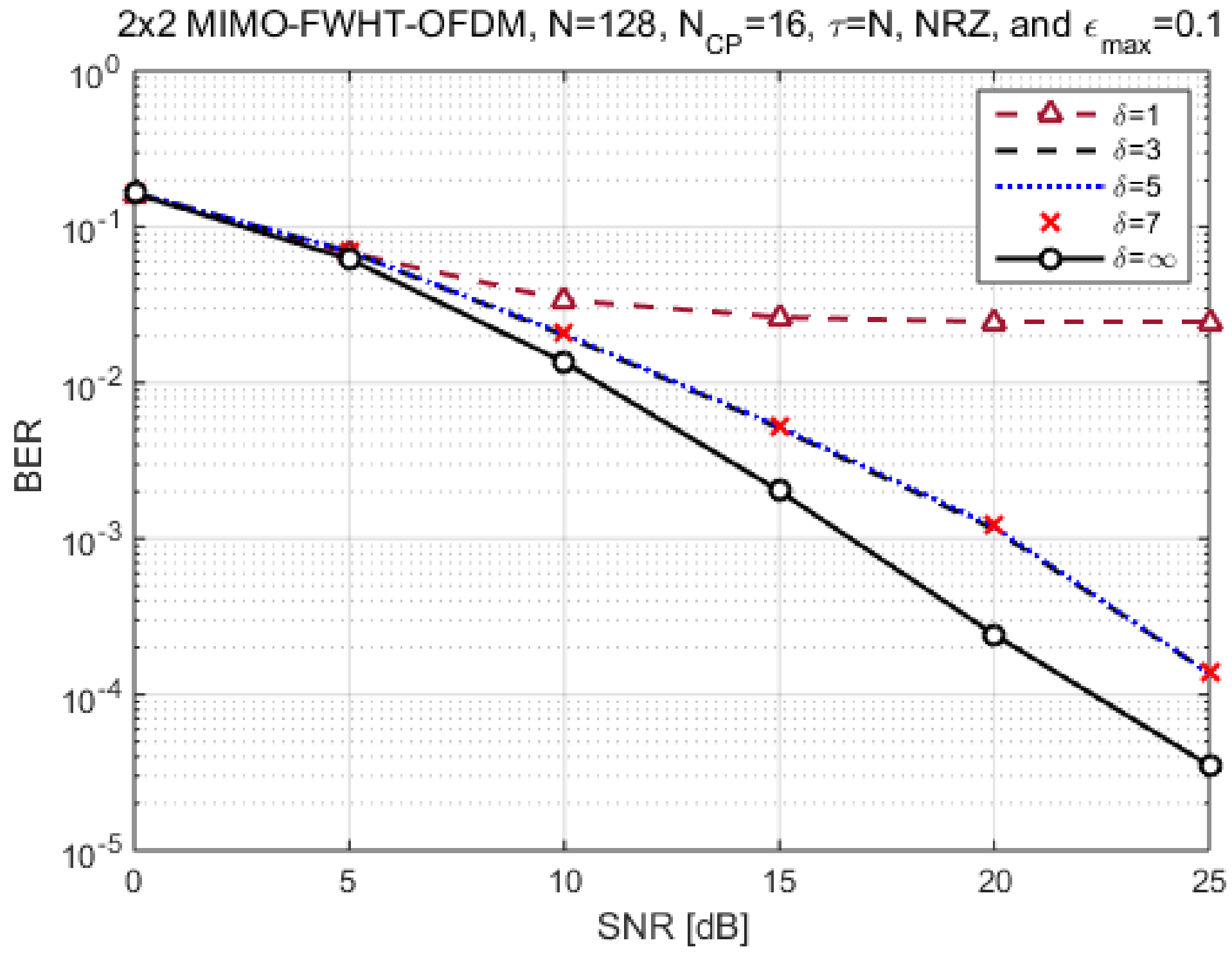

**Figure 5c**. The side view of Fig. 5a

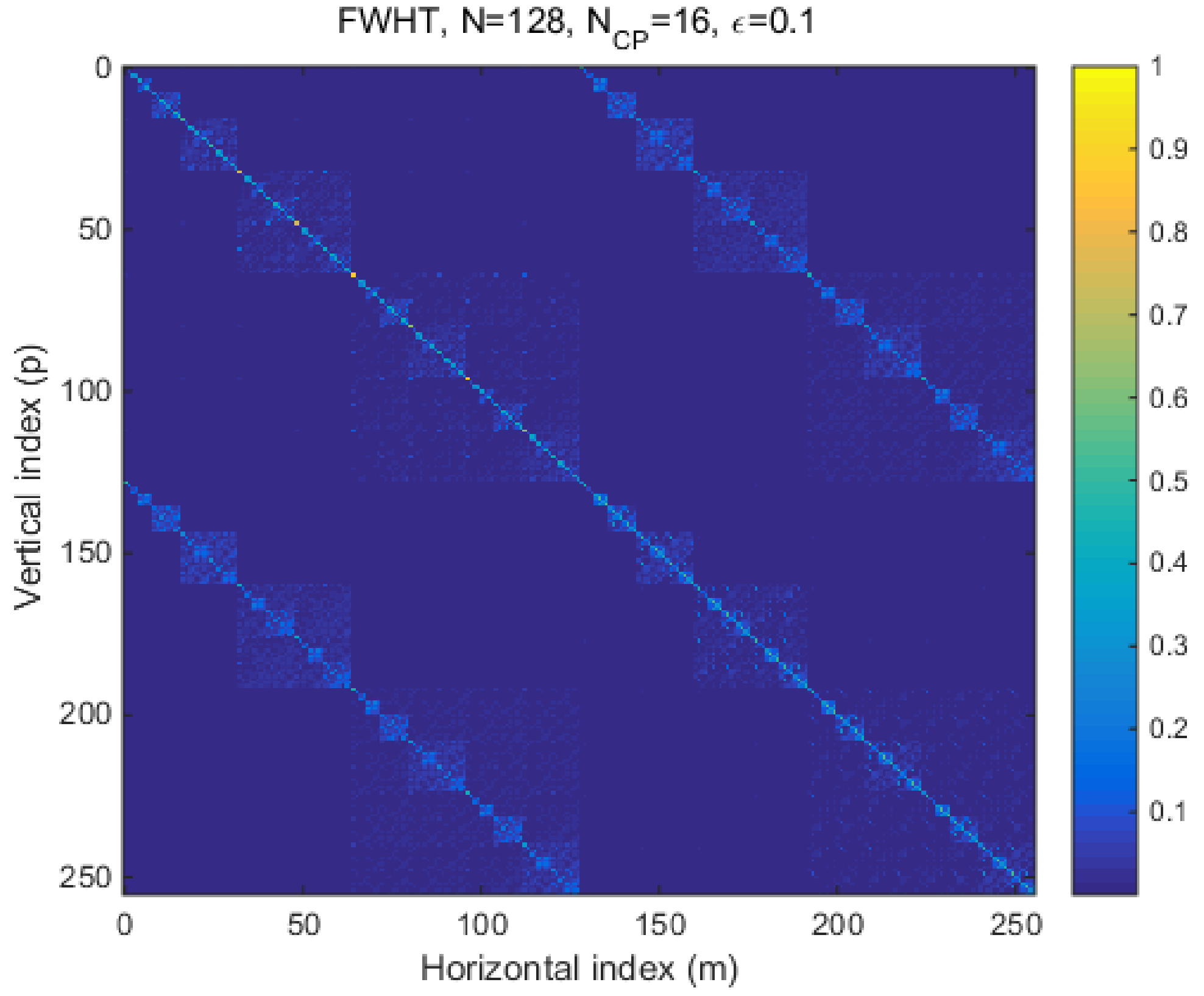

**Figure 6**. The normalized magnitude of the covariance matrix vs. sub-carrier indices using for $\rho = 0.5$, and $\delta = 3$.

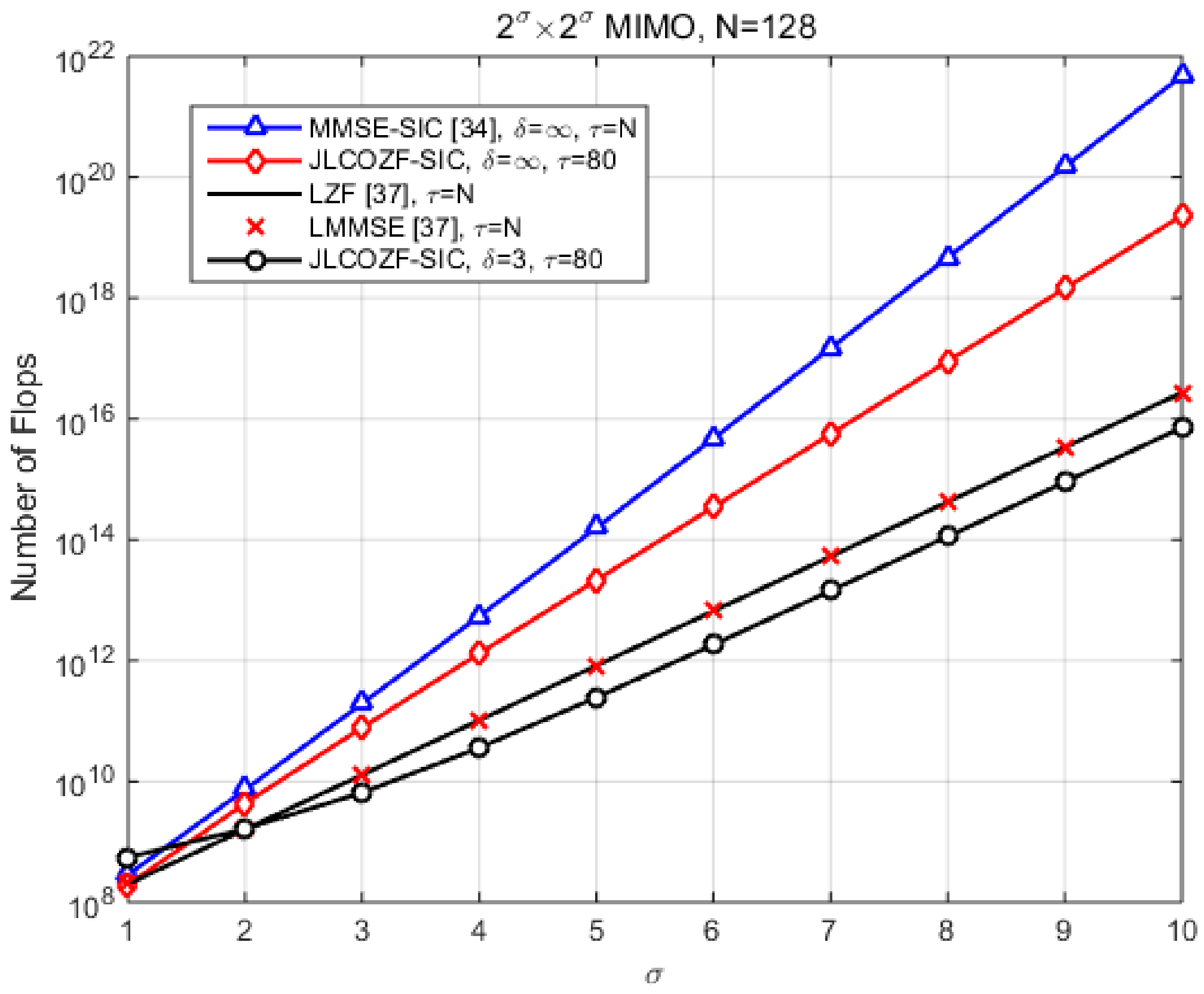

**Figure 7a**. The number of flops versus the configuration orders of different equalizers.

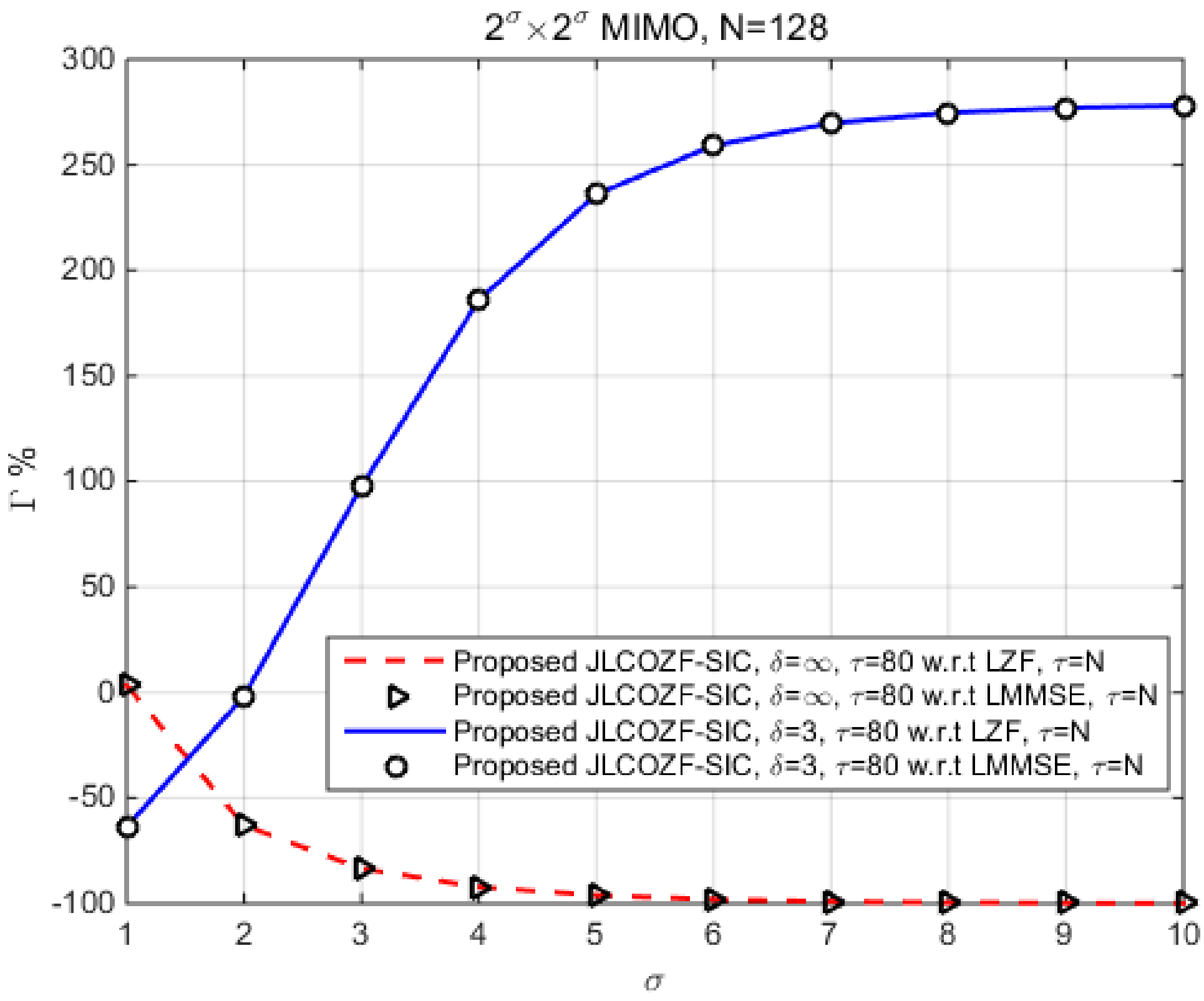

**Figure 7b**. The computational complexity efficiency of the proposed JLZOZF-SIC equalizer w.r.t linear equalizers

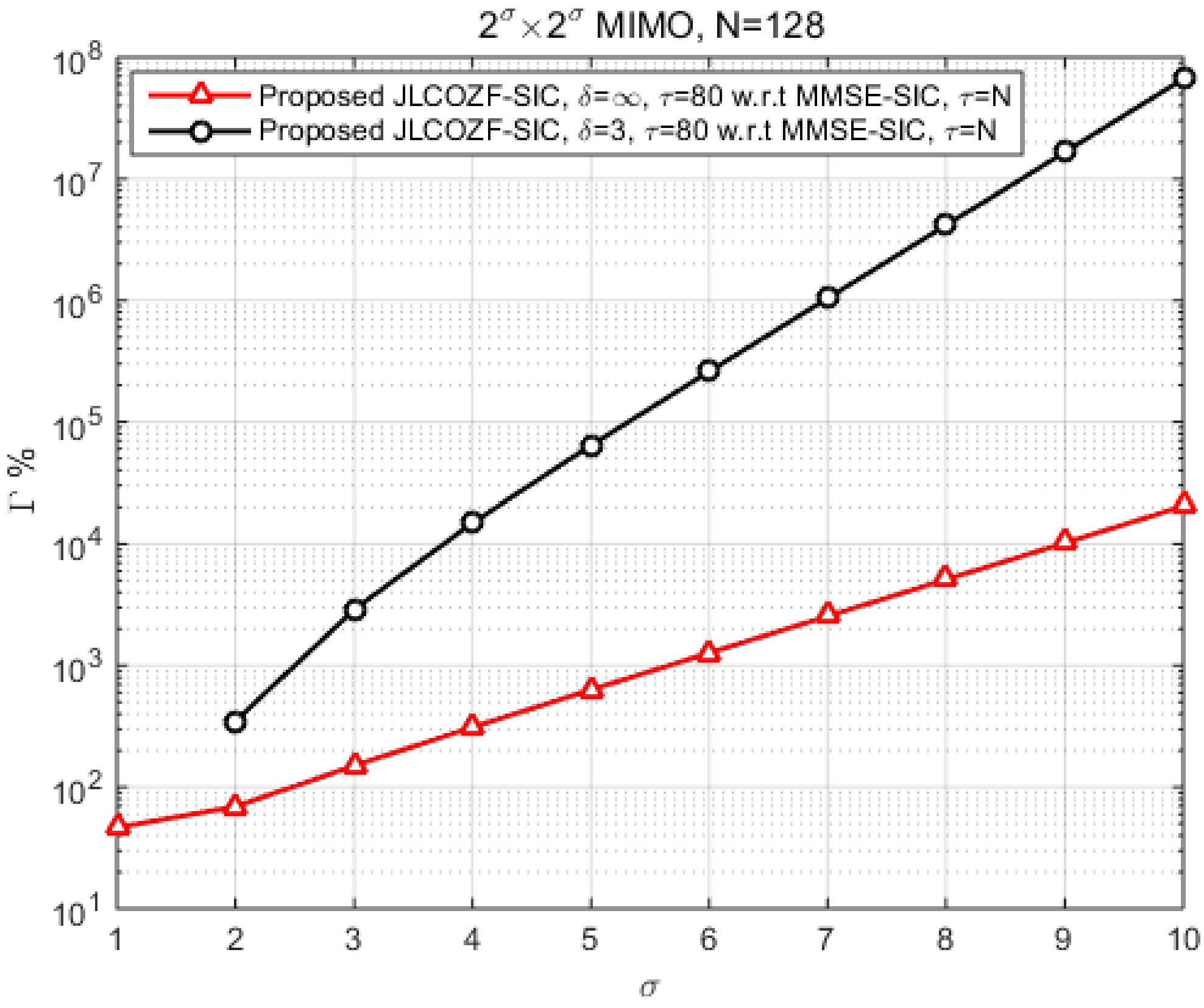

**Figure 7c**. The computational complexity efficiency of the proposed JLZOZF-SIC equalizer w.r.t MMSE-SIC equalizer

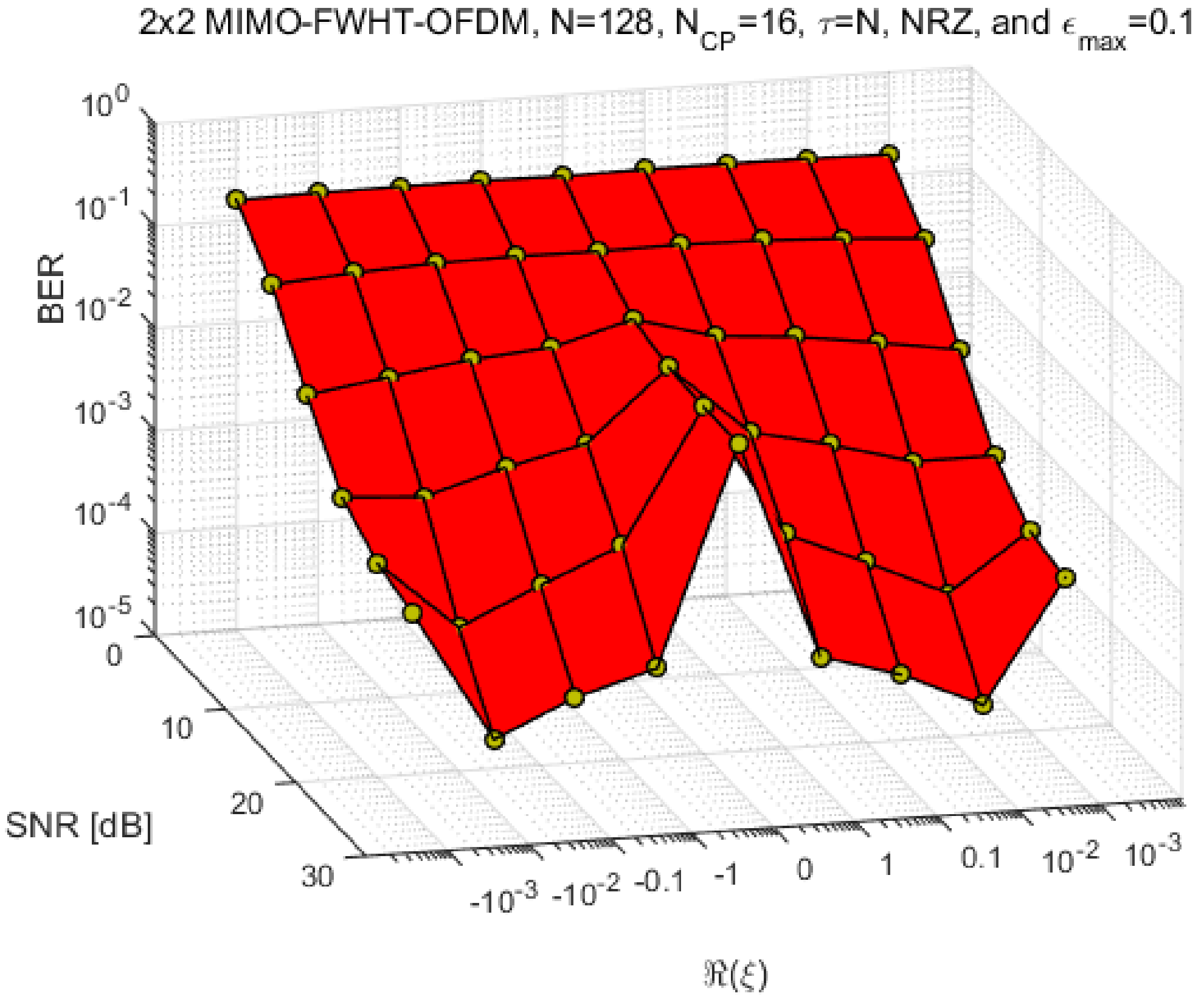

**Figure 8a**. The BER vs. the real part of the optimized parameter at different SNR values.

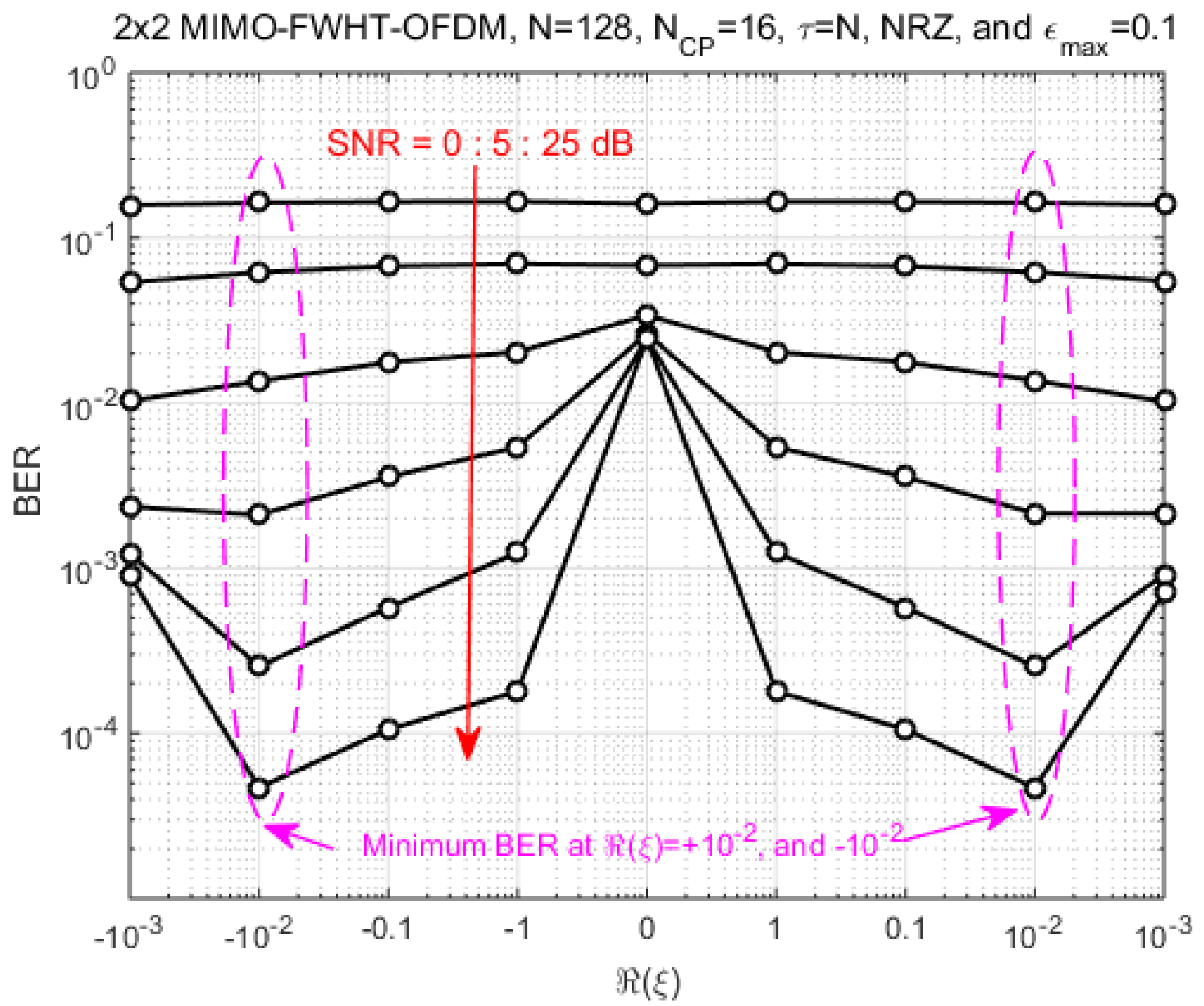

**Figure 8b**. The elevation view of Fig. 8a.

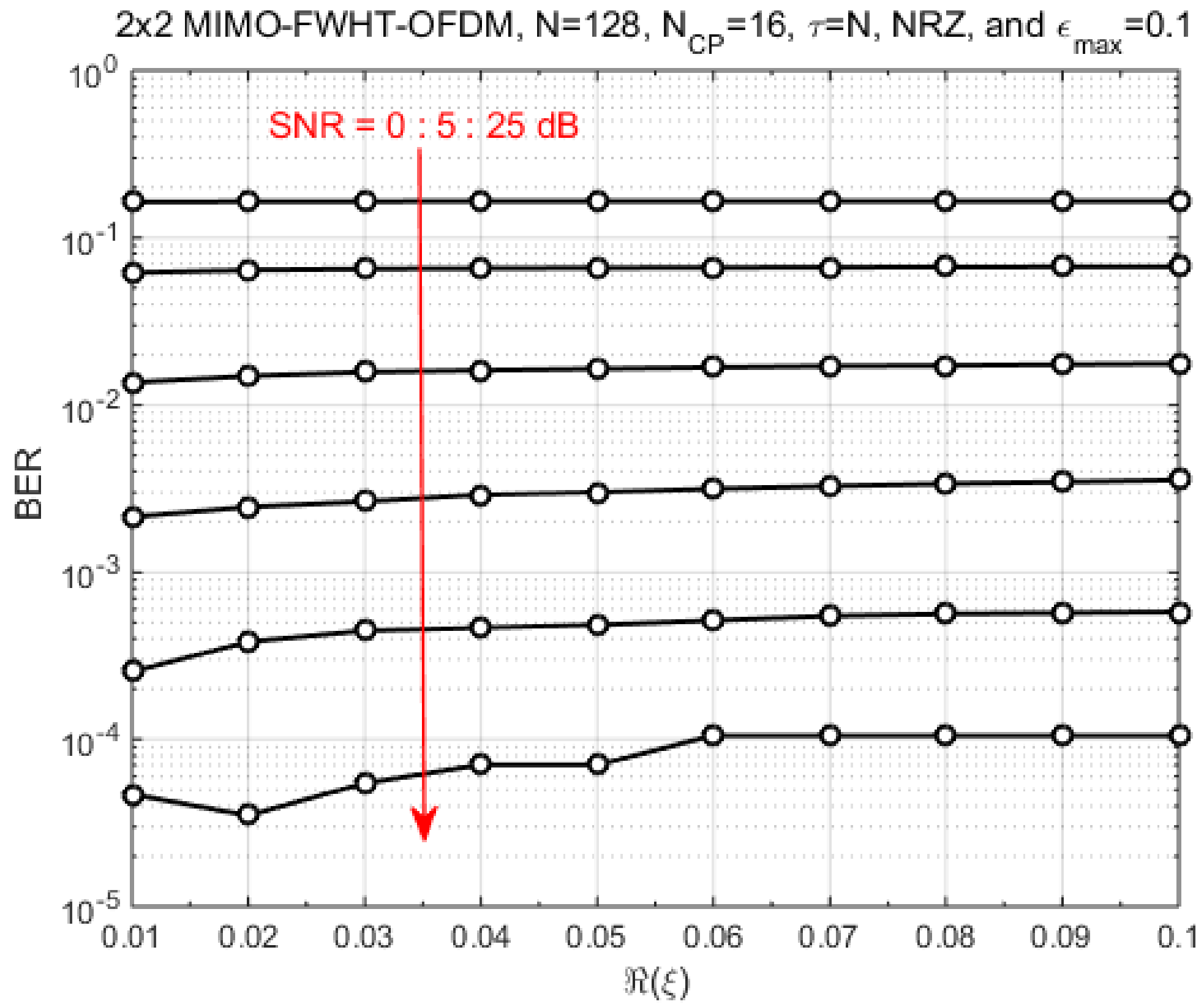

**Figure 9a**. The BER vs. the real part of the optimized parameter at different SNR values with the range of $\Re(\xi) = 10^{-2} \rightarrow 0.1$.

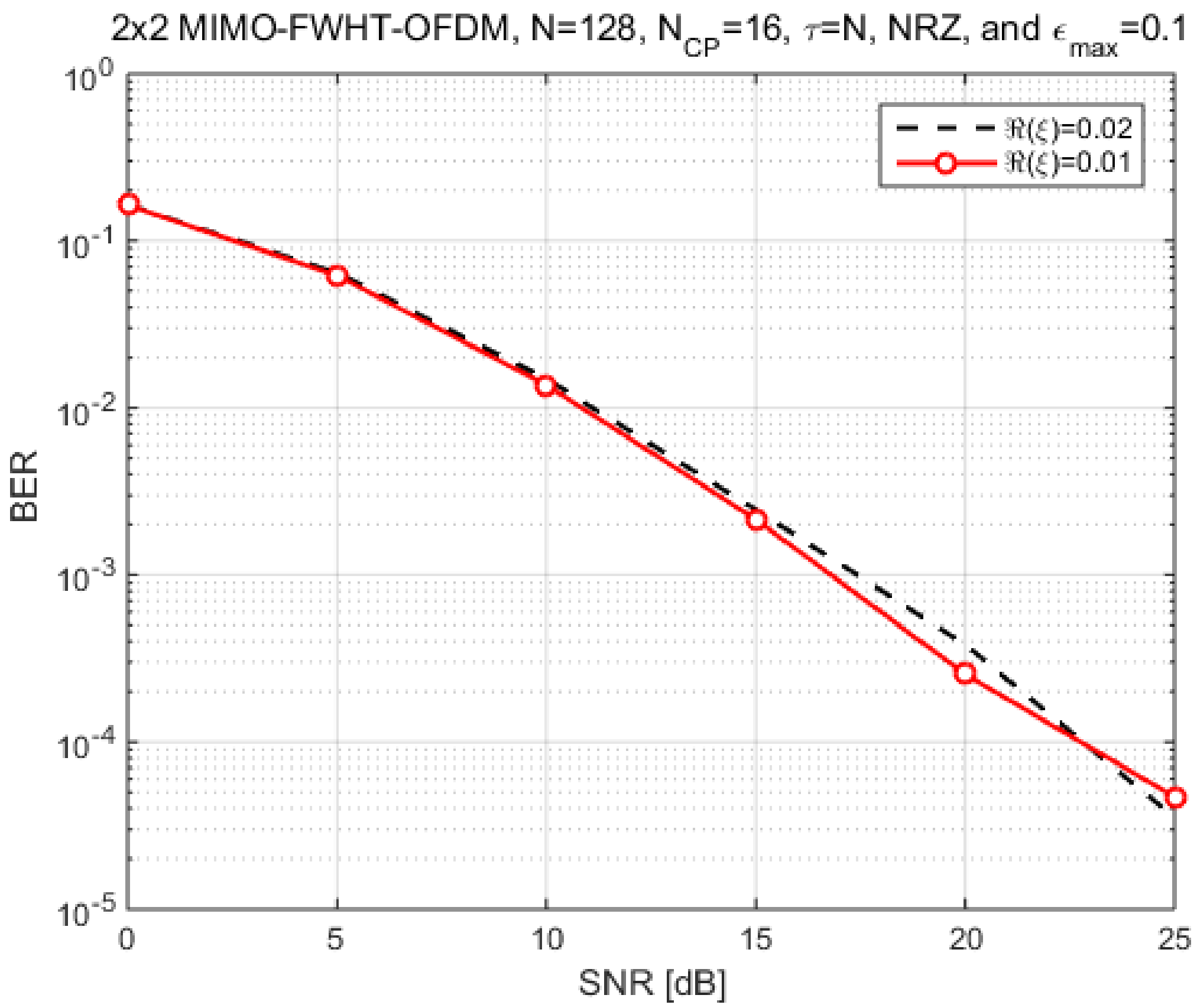

**Figure 9b**. The BER vs. the SNR at different values of $\Re(\xi)$.

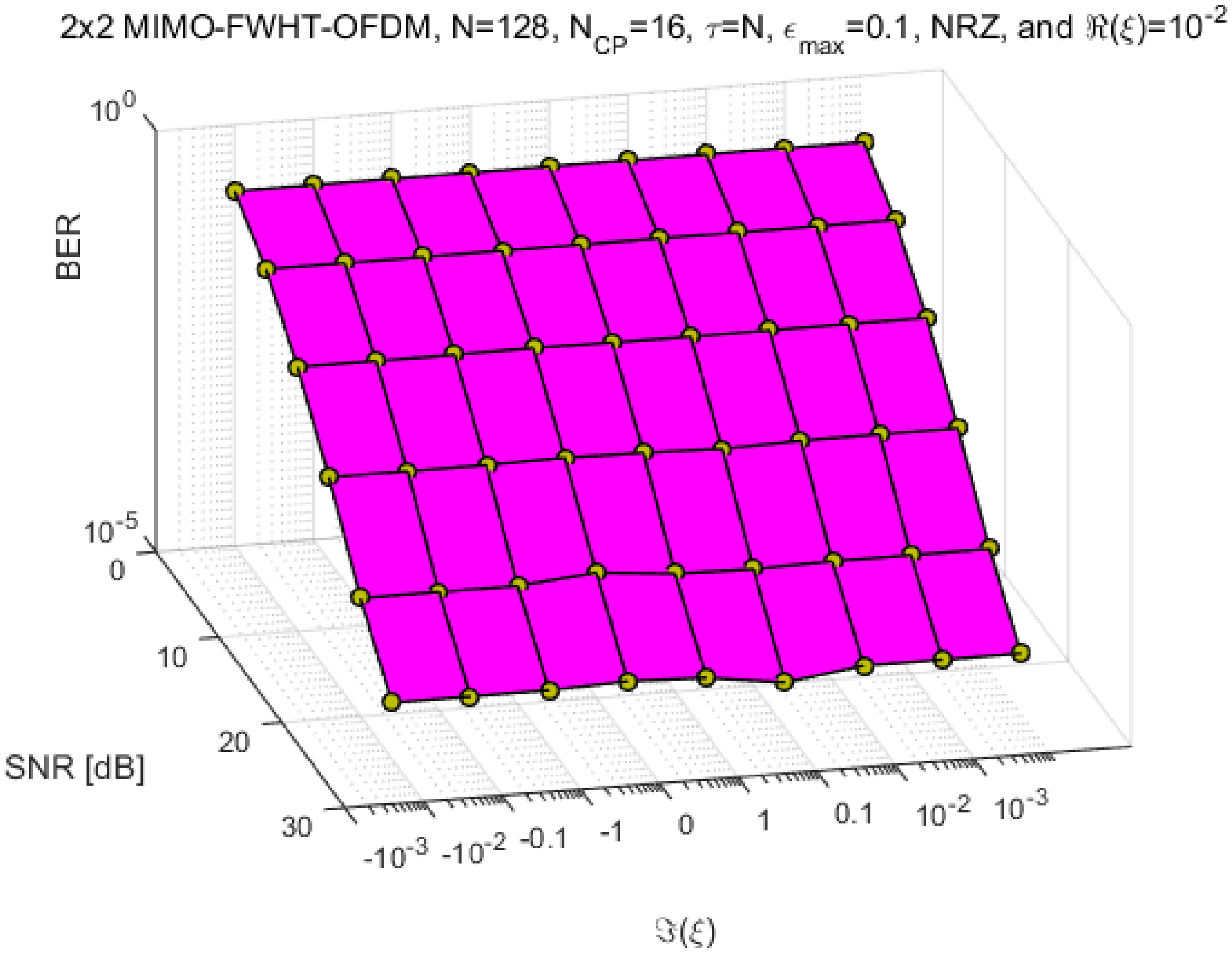

**Figure 10a**. The BER vs. the real part of the optimized parameter at different SNR values.

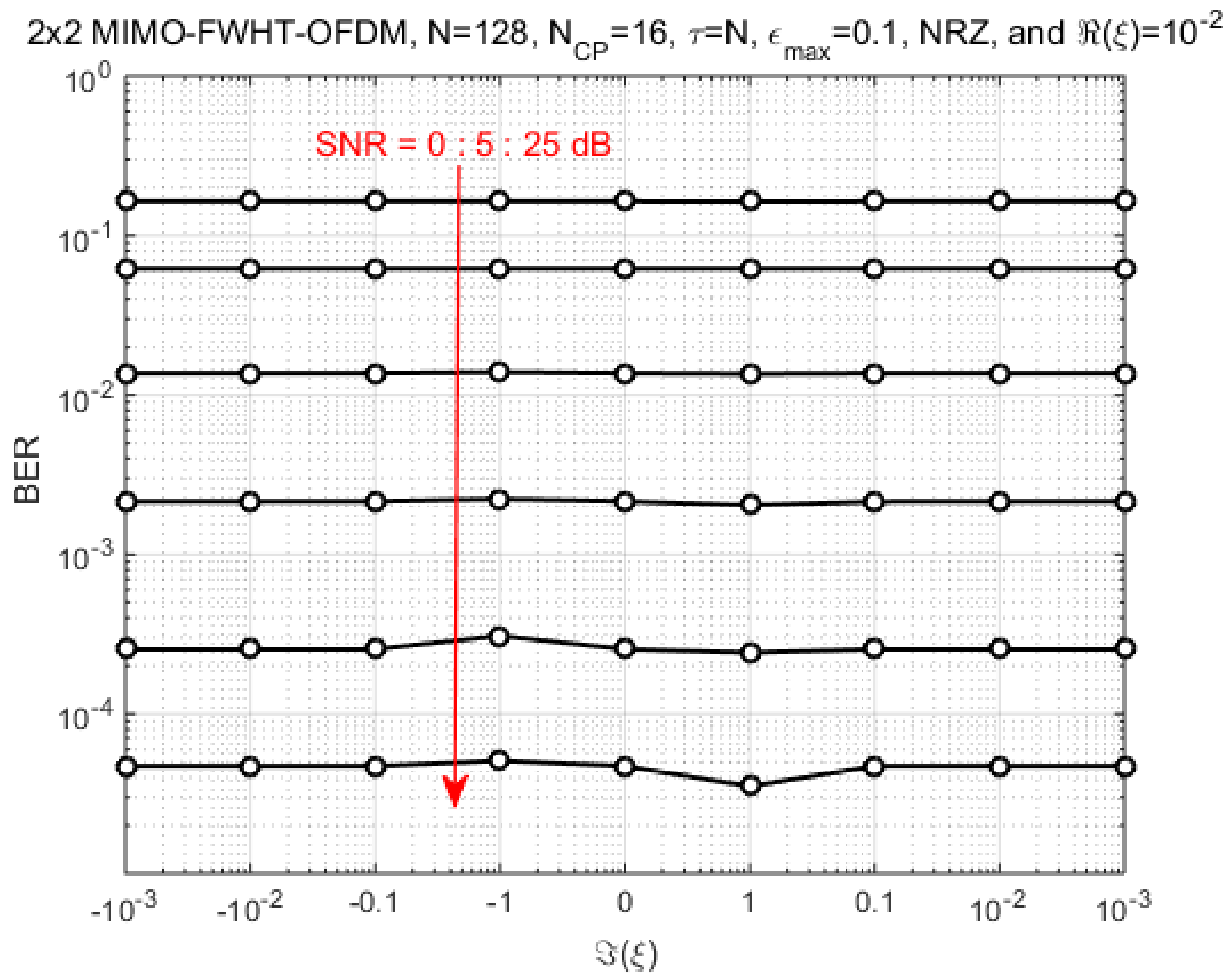

**Figure 10b**. The elevation view of Fig. 10a.

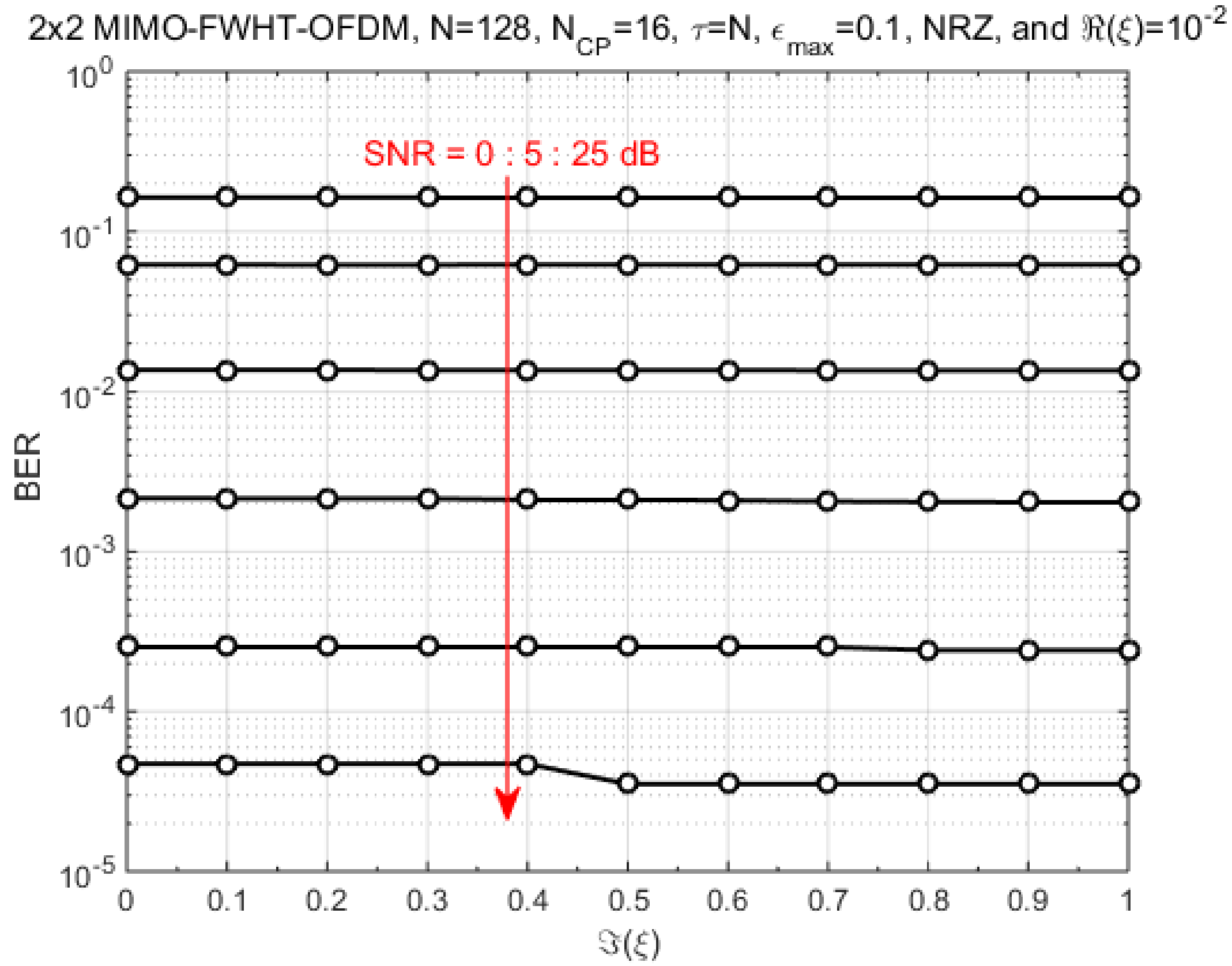

**Figure 11a**. The BER vs. the imaginary part of the optimized parameter at different SNR values with the range of $\Im(\xi) = 0.1 \rightarrow 1$.

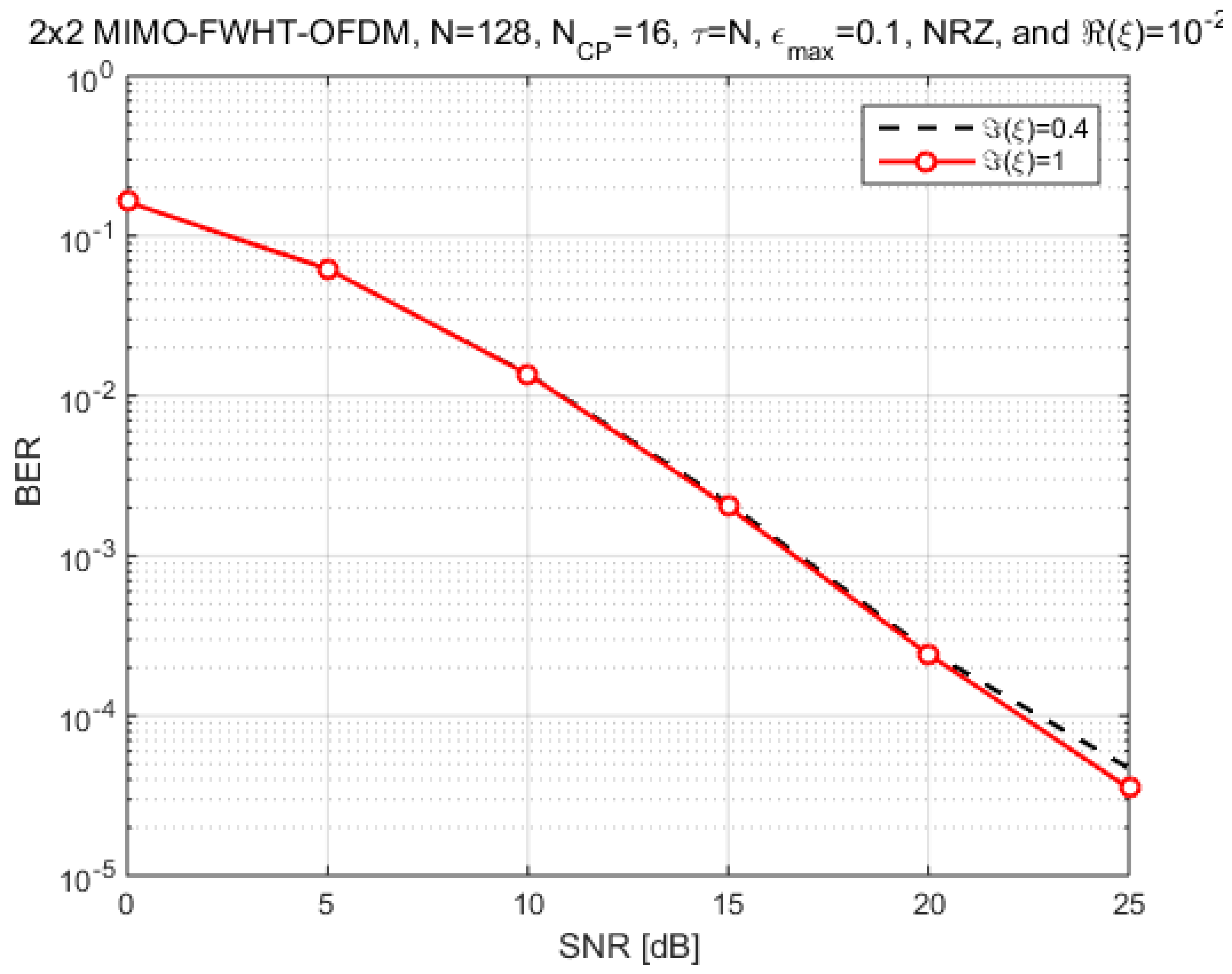

**Figure 11b**. The BER vs. the SNR at different values of $\Im(\xi)$.

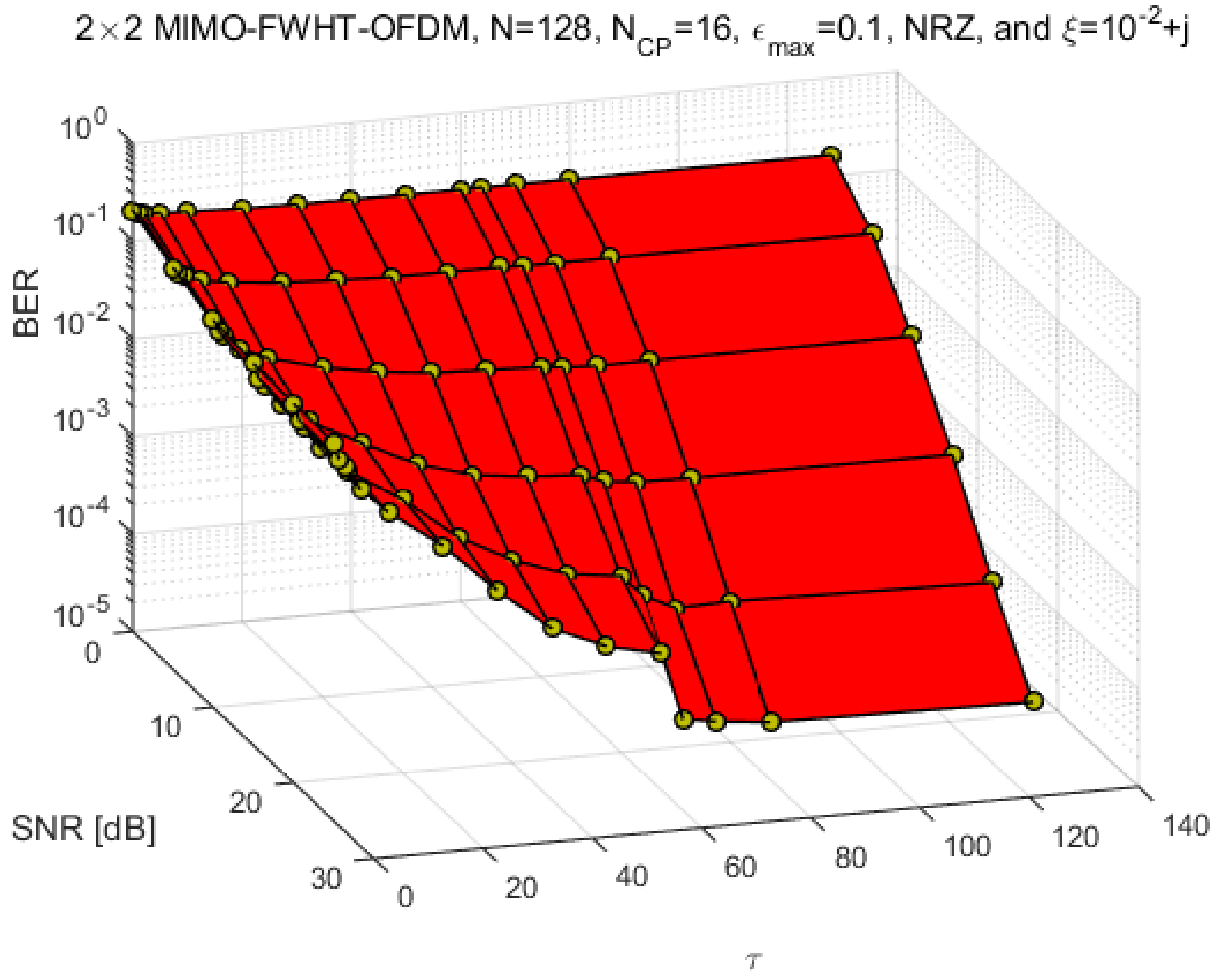

**Figure 12a**. The BER vs. SNR at various compensation scenarios (i.e., τ).

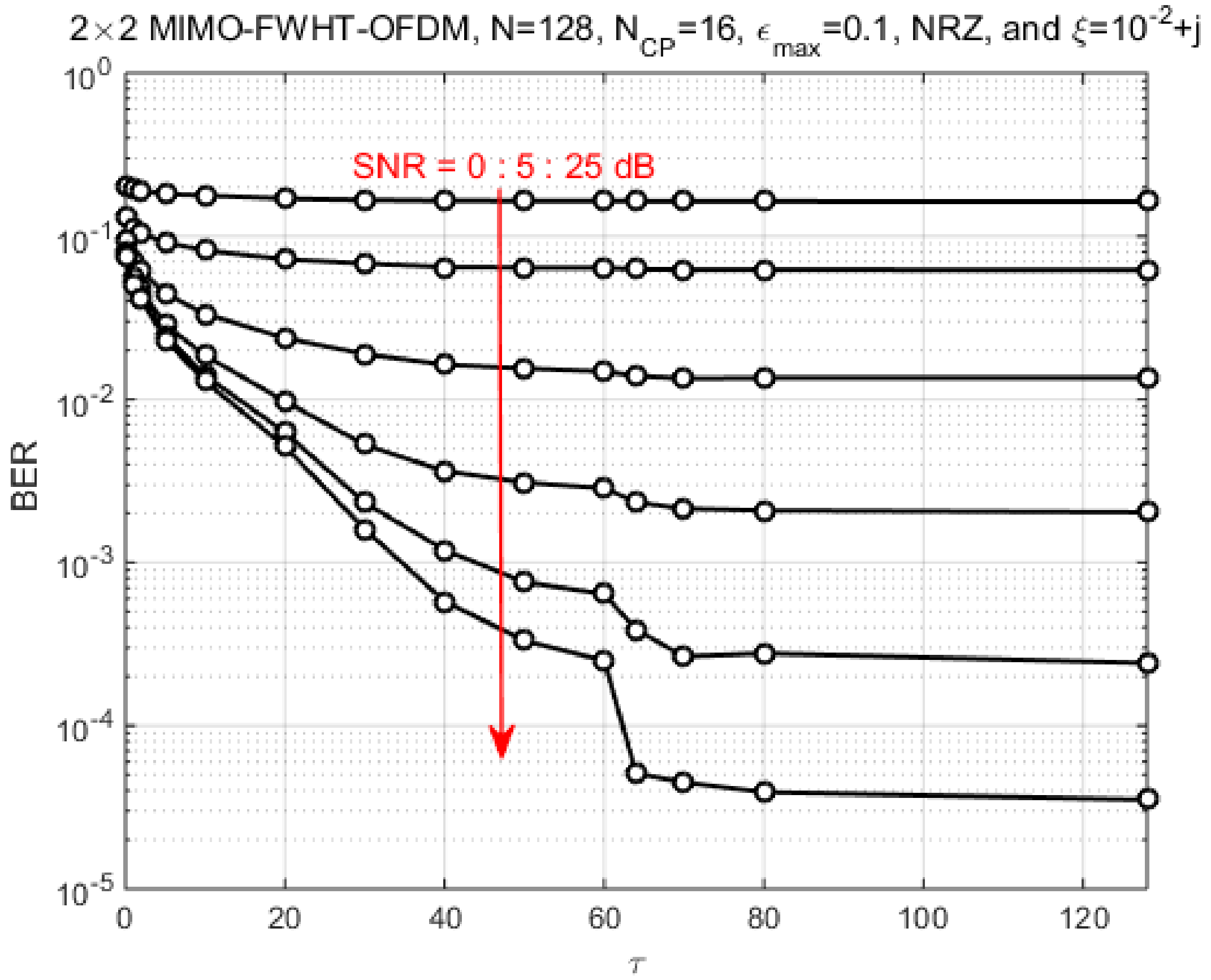

**Figure 12b**. The elevation view of Fig. 12a.

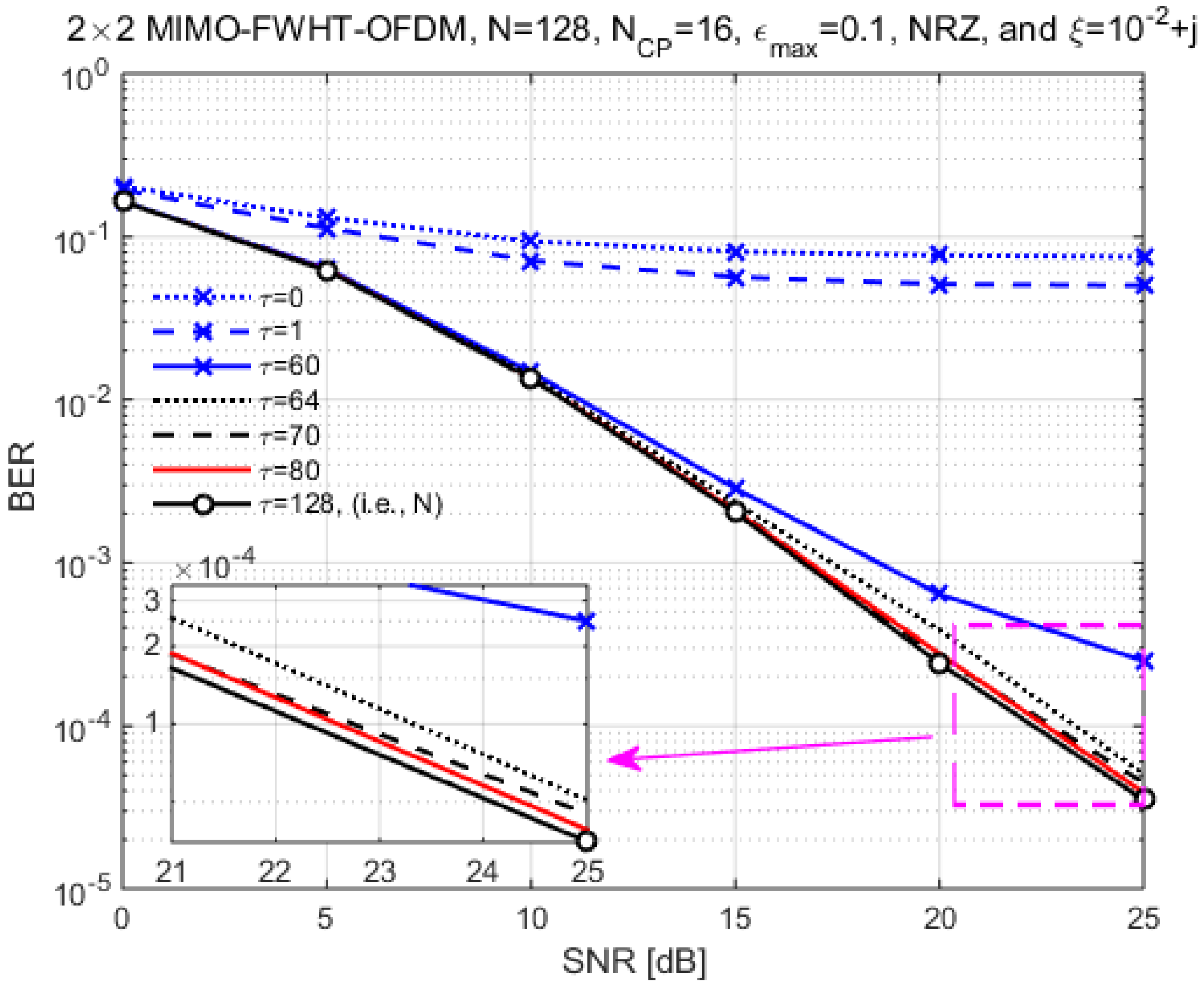

**Figure 12c**. The side view of Fig. 12a.

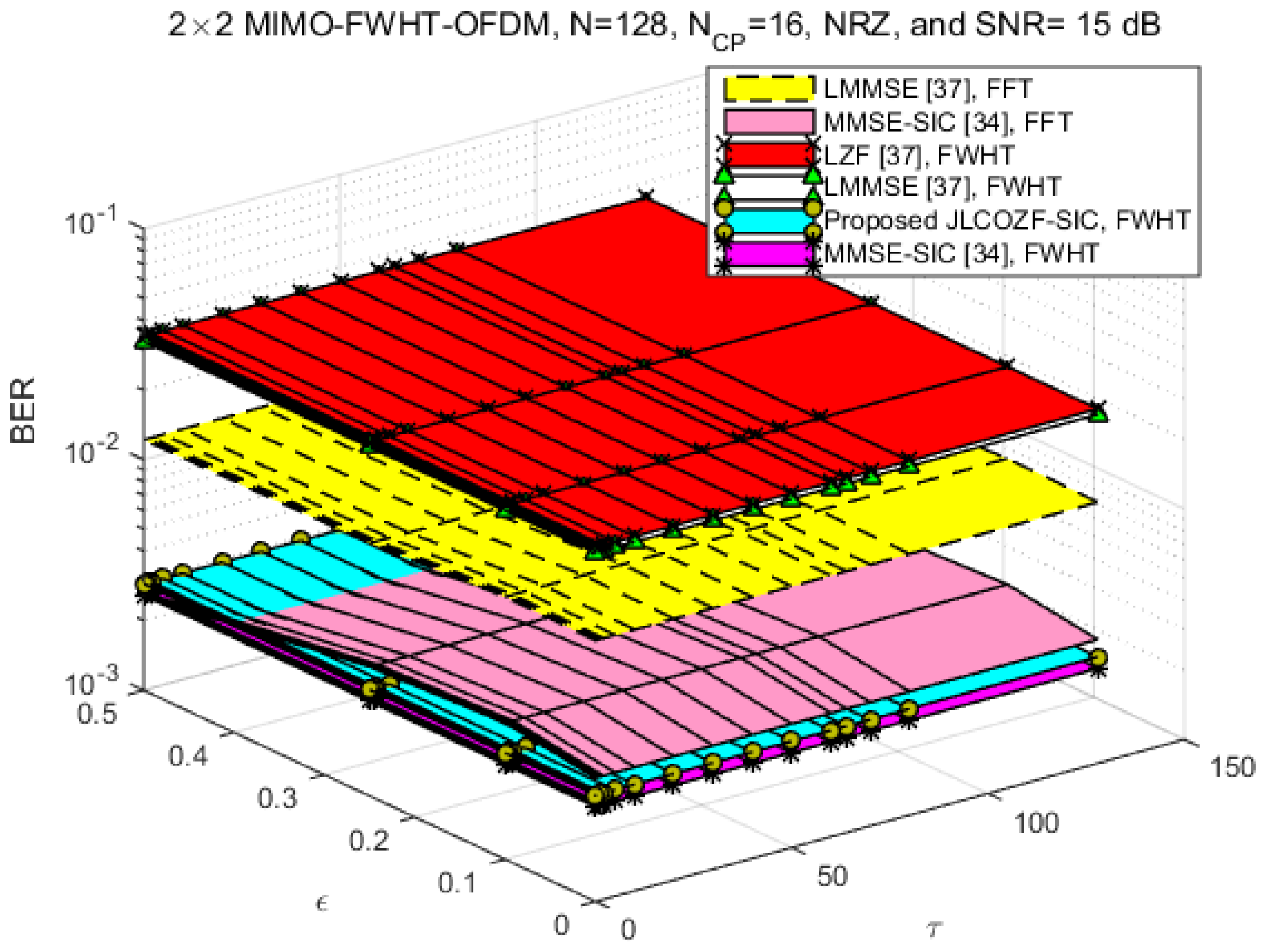

**Figure 13a**. The BER vs. normalized CFO at various compensation scenarios (i.e., τ).

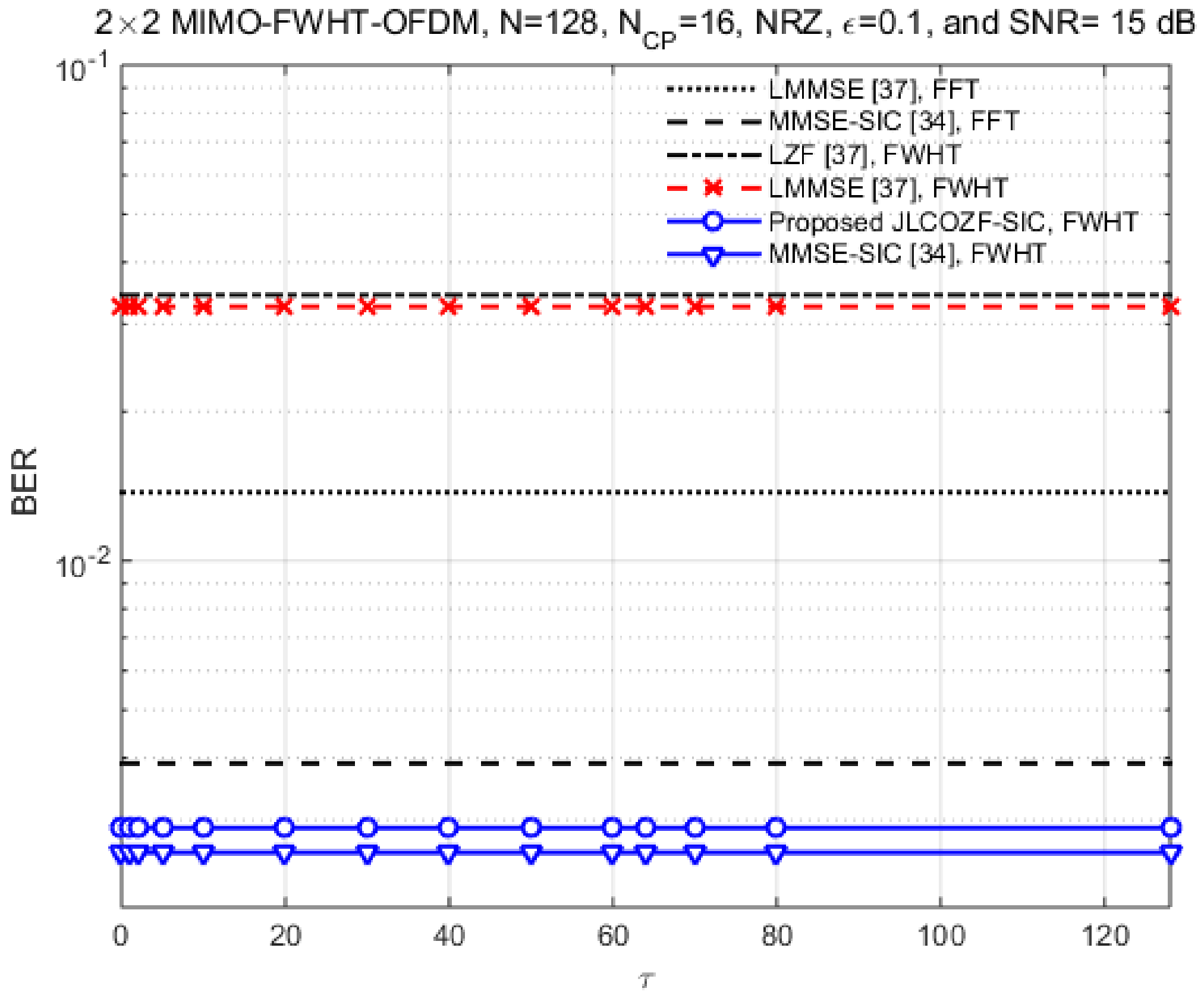

**Figure 13b**. The elevation view of Fig. 13a.

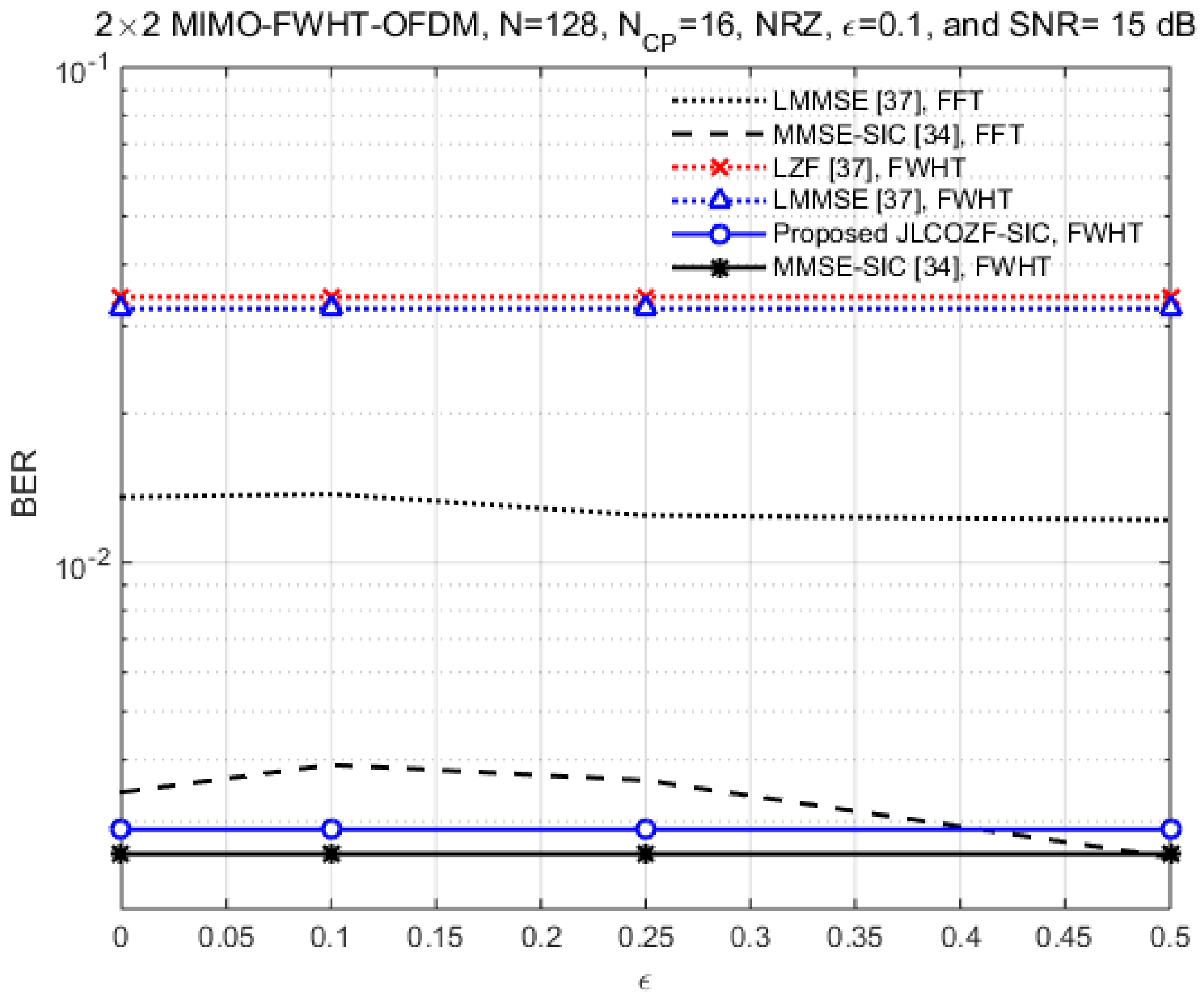

**Figure 13c**. The side view of Fig. 13a.

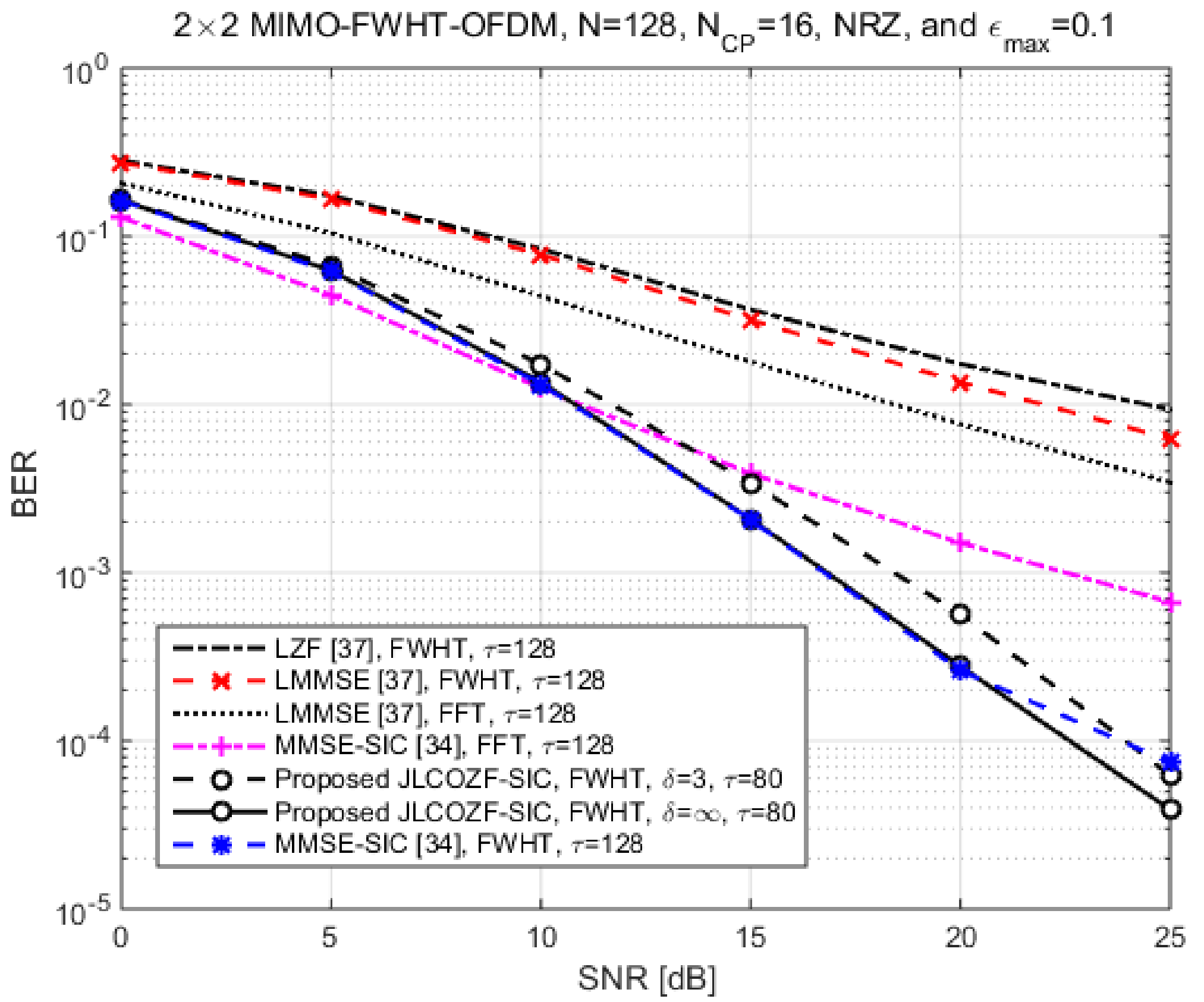

**Figure 14**. The BER vs. SNR at various equalizers.

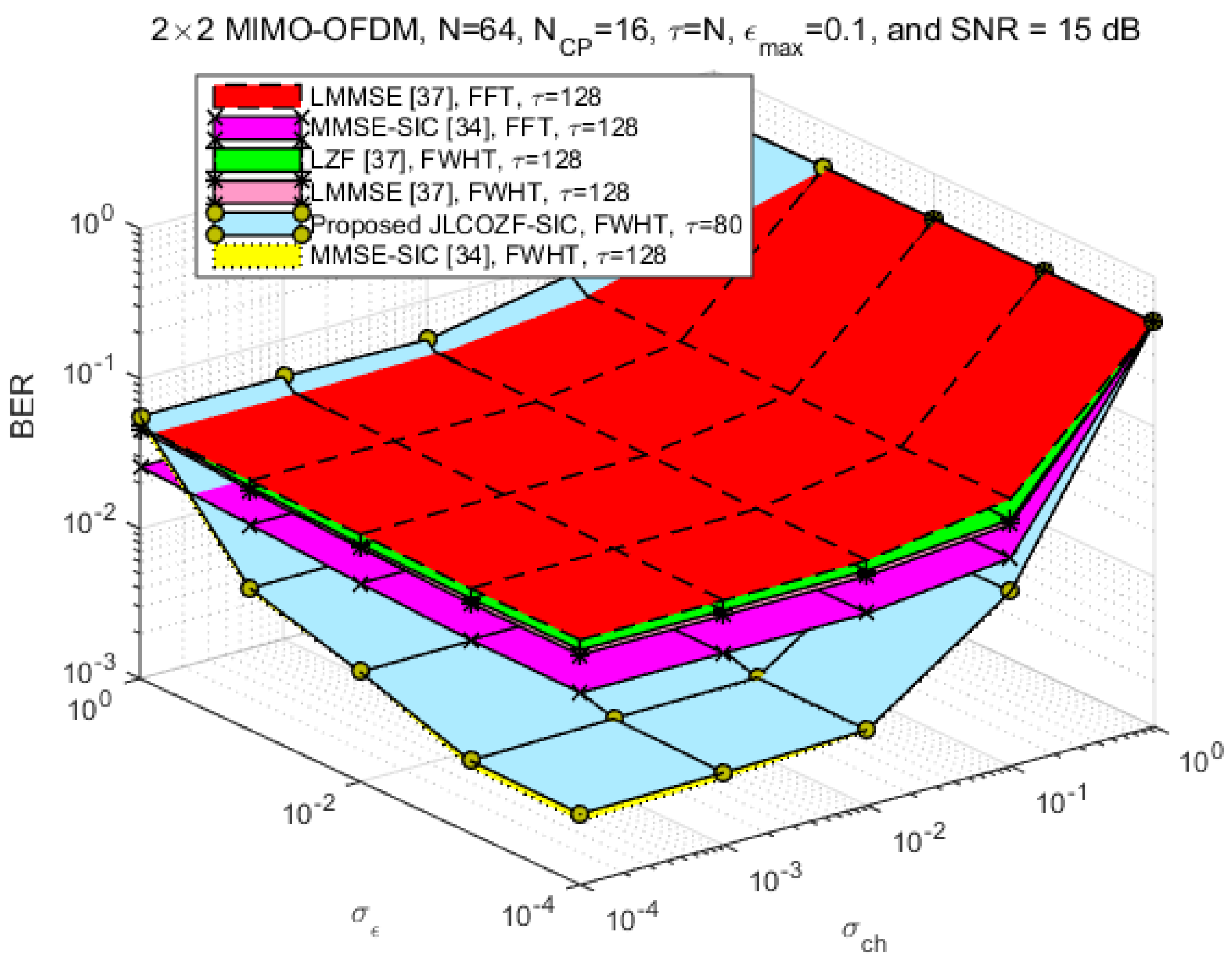

**Figure 15**. The BER performance in the case of estimation errors